\documentclass[useAMS,usenatbib,usegraphicx]{mn2e}

\def\deg{{$^{\circ}$}}
\def\kmsec{\mbox{km~s$^{\rm -1}$}}
\def\logg{\mbox{log~{\it g}}}
\def\msun{\mbox{M$_{\odot}$}}
\def\teff{\mbox{$T_{\rm eff}$}}
\def\vt{\mbox{$v_{\rm t}$}}

\def\rpro{\mbox{$r$-process}}
\def\spro{\mbox{$s$-process}}
\def\ncap{\mbox{$n$-capture}}
\def\loggf{$\log gf$}

\def\ngc{\mbox{NGC~5824}}
\def\apj{ApJ}
\def\aap{A\&A}
\def\aj{AJ}
\def\mnras{MNRAS}
\def\araa{ARA\&A}
\def\apjs{ApJS}
\def\apjl{ApJL}
\def\aaps{A\&AS}
\def\procspie{Proc.\ SPIE}
\def\jqsrt{J.\ Quant.\ Spectrosc.\ Rad.\ Trans.}
\def\pasp{PASP}
\def\pasj{PASJ}
\def\physscr{Phys.\ Scr.}
\def\nat{Nature}
\def\ar{Astron.\ Rep.}
\def\hsls{[$hs$/$ls$]}
\def\pbhs{[Pb/$hs$]}

\title[Abundances in NGC 5824]
{Detailed Chemical Abundances in NGC 5824:\ 
Another Metal-Poor Globular Cluster with Internal Heavy Element
Abundance Variations\thanks{
This paper includes data gathered with the 6.5 meter 
Magellan Telescopes located at Las Campanas Observatory, Chile.}
}

\author[Roederer et al.]{Ian U.\ Roederer$^{1,3}$\thanks{E-mail: iur@umich.edu},
Mario Mateo$^{1}$,
John I.\ Bailey III$^{1}$,
Meghin Spencer$^{1}$,\newauthor
Jeffrey D.\ Crane$^{2}$,
Stephen A.\ Shectman$^{2}$ \\
$^{1}$Department of Astronomy, University of Michigan,
1085 S.\ University Avenue, Ann Arbor, MI 48109, USA\\
$^{2}$Carnegie Observatories, 
813 Santa Barbara Street, Pasadena, CA 91101, USA\\
$^{3}$Joint Institute for Nuclear Astrophysics and Center for the
Evolution of the Elements, USA\\
}

\begin{document}

\pagerange{\pageref{firstpage}--\pageref{lastpage}} 
\pubyear{2015}
\maketitle
\label{firstpage}

\begin{abstract}

We present radial velocities,
stellar parameters, and detailed abundances of 39~elements
derived from high-resolution spectroscopic 
observations of red giant stars
in the luminous, metal-poor globular cluster NGC~5824.
We observe 26~stars in NGC~5824
using the Michigan/Magellan Fiber System (M2FS)
and two stars using the 
Magellan Inamori Kyocera Echelle (MIKE) spectrograph.
We derive a mean metallicity 
of [Fe/H]~$= -$1.94~$\pm$~0.02 (statistical) 
$\pm$~0.10 (systematic).
The metallicity dispersion
of this sample of stars,
0.08~dex,
is in agreement with previous work and
does not exceed the expected observational errors.
Previous work suggested an internal metallicity spread
only when fainter samples of stars were considered,
so we cannot exclude the possibility of an
intrinsic metallicity dispersion in NGC~5824.
The M2FS spectra reveal 
a large
internal dispersion
in [Mg/Fe], 0.28~dex,
which is found in a few other luminous, metal-poor clusters.
[Mg/Fe] is correlated with 
[O/Fe] and anti-correlated with [Na/Fe] and [Al/Fe].
There is no evidence for internal dispersion
among the other $\alpha$- or Fe-group abundance ratios.
Twenty-five of the 26~stars 
exhibit a $n$-capture enrichment pattern
dominated by $r$-process nucleosynthesis
($\langle$[Eu/Fe]$\rangle = +$0.11~$\pm$~0.12;
$\langle$[Ba/Eu]$\rangle = -$0.66~$\pm$~0.05).
Only
one star shows evidence of substantial
$s$-process enhancement 
([Ba/Fe]~$= +$0.56~$\pm$~0.12;
[Ba/Eu]~$= +$0.38~$\pm$~0.14), but
this star does not exhibit other characteristics
associated with $s$-process enhancement
via mass-transfer from a binary companion.
The Pb and other heavy elements produced by the $s$-process
suggest a timescale of no more than a few hundred Myr
for star formation and chemical enrichment, 
like the complex globular clusters 
M2, M22, and NGC~5286.

\end{abstract}


\section{Introduction}
\label{introduction}

The number of globular clusters
with spectroscopically-confirmed metallicity dispersions
has grown dramatically over the last decade.
These are generally among the most massive clusters,
and many are also metal-poor.
This list includes
$\omega$~Cen (e.g., \citealt{norris95,smith00,johnson10,marino11}),
M2 \citep{yong14b},
M19 \citep{johnson15b},
M22 (e.g., \citealt{dacosta09,marino09}),
M54 \citep{carretta10b},
\mbox{NGC~1851} \citep{yong08c,carretta10c,carretta11},
\mbox{NGC~3201} \citep{gonzalez98,simmerer13},
\mbox{NGC~5286} \citep{marino15}, and
Ter~5 \citep{origlia11,origlia13,massari14}.
Many of these clusters exhibit multiple sequences 
on the subgiant branch
on the color-magnitude diagram
when using broadband optical photometry
(e.g., \citealt{piotto12}),
but \mbox{NGC~3201} does not.
To this list we add M92,
where a high-precision differential analysis
by \citet{langer98} revealed one star
with a metallicity significantly higher
than other cluster members.
Clusters that are not traditionally
thought to contain internal metallicity spreads
may exhibit small variations ($<$~0.03~dex)
in Fe and other elements
when pushed to the limit of high-precision 
abundance techniques,
as in the case of \mbox{NGC~6752} \citep{yong13}.

The metallicity dispersions in M22 and \mbox{NGC~3201} have
recently been called into question by 
\citet{mucciarelli15b,mucciarelli15c}.
These authors report that the small dispersions
observed 
could be artifacts of the analysis,
and they have shown that spectroscopic 
gravities and Fe~\textsc{i} lines may yield spurious 
metallicity results.
%
Other contradicting results exist, too.
\citet{cohen10} reported a spread in
[Ca/H] in \mbox{NGC~2419}
based on the Ca~\textsc{ii} near-infrared triplet.
\citet{mucciarelli12} argued that this
spread was caused by changes in the continuous opacity
induced by severe Mg depletions in some stars,
rather than a dispersion in Ca abundances.
Intrinsic abundance dispersions
based on the Ca~\textsc{ii} triplet
have also been found in M22 by \citet{dacosta09}.
\citet{mucciarelli15c} reported that
no metallicity dispersion exists in M22,
but 
that result does not have a physical explanation
since Mg shows no depletions in M22.
%
While ongoing work seeks to re-examine
the small metallicity
dispersions ($\sim$~0.1~dex) in some clusters,
the range of metallicities is so large 
($\sim$~1~dex) 
in others
that a legitimate cosmic dispersion must exist.

Some of these clusters, and others, are embedded in
stellar halos extending beyond their
tidal radii \citep{olszewski09,correnti11,kunder14,marino14,navin15}.
The most striking example is
M54, which sits in the nucleus of the
Sagittarius dwarf galaxy (e.g., \citealt{ibata95,bellazzini08}).
The others could be the tidally-stripped
nuclei of former dwarf galaxies,
suggesting formation within individual dark matter halos.
It is important to 
establish which and how many clusters
belong to this class of complex objects.
This influences our understanding 
of their formation and evolution,
and growing the membership of this class
alleviates---slightly---the tension 
between cosmological simulations and the number
of observed Milky Way satellites,
also known as the ``missing satellites'' problem.

The elements heavier than the Fe group,
hereafter known as \ncap\ elements,
provide another perspective into the
complex formation histories of globular clusters.
In metal-poor clusters with no metallicity dispersion, 
the \ncap\ elements appear to have been
produced predominantly, if not exclusively,
by some form of rapid \ncap\ process (\rpro;
e.g., \citealt{gratton04,roederer10b}).
This indicates a rapid enrichment timescale
from core-collapse supernovae or neutron-star mergers.
In the class of complex clusters, the \ncap\ elements
show star-to-star variations in all cases 
where they have been studied
($\omega$~Cen, M2, M19, M22, \mbox{NGC~1851},
and \mbox{NGC~5286}).
The heavy elements in these clusters were produced by
each of
the slow \ncap\ process (\spro) and the \rpro,
and often the proportions of $r$- and \spro\ material
vary within each cluster.
The \rpro\ material
is thought to have been present in the gas
from which all stars formed.
The \spro\ material 
is thought to have been produced in other stars within the cluster,
ejected into the cluster ISM,
and incorporated into the stars observed today
(e.g., \citealt{smith00}).
However,
the chemical enrichment
scenarios to explain these signatures are neither simple
nor uniform from one cluster to another.

Recent observations by 
\citet{saviane12} and \citet{dacosta14}
revealed that \ngc\
may also be a member of the class of complex
globular clusters,
similar to M2, M22, or \mbox{NGC~5286}.
This southern cluster 
($\alpha =$~15:03:58.6, $\delta = -$33:04:05.6)
is the 14$^{\rm th}$ most-luminous cluster around the Milky Way
($M_{V} = -$8.85; Harris \citeyear{harris96}, 2010 edition),
with a stellar mass of $\sim$~6~$\times$~10$^{5}$~\msun\
\citep{mclaughlin05}.
\ngc\ is an outer halo cluster located
32.1~kpc from the Sun and 25.9~kpc from the Galactic center.
There is moderate reddening along the line of sight,
$E(B-V) =$~0.14,
and \ngc\ lies west of the Galactic bulge 
($\ell =$~332.6\deg)
and north of the Galactic plane
($b = +$22.1\deg).
\ngc\ does not show any split or broadened sequences
in broadband optical color-magnitude diagrams
\citep{piotto02},
but its blue horizontal branch is well-populated.
\citet{grillmair95} identified 
light beyond the tidal radius of \ngc, but
\citet{carballobello14} did not detect any photometric signatures
of an extended stellar halo around \ngc.
Low-resolution spectroscopy of the Ca~\textsc{ii} triplet
obtained by \citeauthor{dacosta14}\
revealed a possible metallicity spread at the $\approx$~0.1~dex level.
No high-resolution spectra have been 
collected previously to examine
the detailed kinematics and chemistry of this interesting cluster.

We present new
high-resolution and high signal-to-noise (S/N) 
spectroscopy of red giants in \ngc.
Our goals are
(1) to confirm or refute the existence 
of a metallicity dispersion in \ngc;
(2) to establish whether \ngc\
exhibits the light-element abundance variations
among O, Na, Mg, and Al that are common to globular clusters; and
(3) to characterize the chemical (in)homogeneities
of other $\alpha$-elements, Fe-group elements, and \ncap\ elements.
We describe our observations in Section~\ref{observations},
the radial velocity measurements in Section~\ref{rv},
the equivalent width (EW) measurements and line list 
in Sections~\ref{ew} and \ref{atomic},
and the details of the abundance analysis in Sections~\ref{atmosphere} and
\ref{analysis}.
We present our abundance results in Section~\ref{results},
weigh the evidence for a metallicity spread in Section~\ref{metallicity},
and discuss all other elements in Section~\ref{discussion}.
We summarize our conclusions in Section~\ref{summary}.

\section{Observations}
\label{observations}

\begin{table*}
\begin{minipage}{5.5in}  
\caption{Magnitudes, S/N Estimates, Velocities, and Model Atmosphere
Parameters for Stars Observed in NGC~5824
\label{obstab}}
\begin{tabular}{cccccccc}
\hline
Star &
$V$ &
S/N pix$^{-1}$ &
$V_{r}$ &
\teff\ &
\logg\ &
\vt\ &
[M/H] \\
 &
 &
(4570~\AA) &
(\kmsec) &
(K) &
 &
(\kmsec) &
 \\ 
\hline
11001198 & 16.35 & 51 & $-$24.5 (0.6)  & 4261 & 1.01 & 2.20 & $-$2.08 \\
12001300 & 16.24 & 40 & $-$28.0 (0.6)  & 4337 & 1.02 & 1.85 & $-$1.96 \\
21000267 & 15.61 & 62 & $-$28.7 (0.4)  & 4075 & 0.55 & 2.00 & $-$1.92 \\
21000688 & 15.75 & 47 & $-$42.5 (0.6)  & 4083 & 0.62 & 2.20 & $-$1.95 \\
21002944 & 15.39 & 54 & $-$25.8 (0.4)  &\ldots&\ldots&\ldots&\ldots \\
31003079 & 16.80 & 20 & $-$26.2 (0.7)  &\ldots&\ldots&\ldots&\ldots \\
32004229 & 16.52 & 27 & $-$29.6 (1.1)  &\ldots&\ldots&\ldots&\ldots \\
32004840 & 16.69 & 27 & $-$26.4 (1.2)  &\ldots&\ldots&\ldots&\ldots \\
41000940 & 16.47 & 40 & $-$26.7 (0.6)  & 4345 & 1.12 & 1.65 & $-$1.87 \\
41001607 & 16.15 & 50 & $-$32.9 (0.8)  & 4235 & 0.91 & 2.35 & $-$1.96 \\
41002449 & 17.07 & 28 & $-$20.8 (0.9)  &\ldots&\ldots&\ldots&\ldots \\
41003361 & 17.06 & 29 & $-$25.1 (0.8)  &\ldots&\ldots&\ldots&\ldots \\
42006307 & 16.34 & 46 & $-$23.5 (0.8)  & 4328 & 1.06 & 1.55 & $-$1.93 \\
42006712 & 17.27 & 25 & $-$36.5 $^{a}$ &\ldots&\ldots&\ldots&\ldots \\
42007331 & 16.82 & 28 & $-$18.0 (0.7)  &\ldots&\ldots&\ldots&\ldots \\
42007539 & 15.95 & 36 & $-$27.8 (0.8)  & 4188 & 0.79 & 1.90 & $-$1.90 \\
42007983 & 17.09 & 18 & $-$29.6 $^{a}$ &\ldots&\ldots&\ldots&\ldots \\
42008343 & 16.49 & 41 & $-$22.0 (0.6)  & 4392 & 1.16 & 1.55 & $-$1.98 \\
42008555 & 16.98 & 40 & $-$27.2 (0.9)  & 4488 & 1.42 & 1.45 & $-$1.88 \\
42009955 & 15.92 & 53 & $-$23.0 (0.8)  & 4200 & 0.79 & 2.20 & $-$1.93 \\
42010124 & 15.88 & 52 & $-$34.3 (0.9)  & 4146 & 0.73 & 2.10 & $-$1.92 \\
42010715 & 16.62 & 38 & $-$29.6 (0.7)  & 4378 & 1.20 & 1.90 & $-$2.00 \\
42011097 & 16.65 & 45 & $-$24.7 (0.8)  & 4383 & 1.22 & 1.75 & $-$1.93 \\
42011404 & 15.97 & 34 & $-$28.1 (1.1)  & 4190 & 0.80 & 1.95 & $-$1.82 \\
42011439 & 16.90 & 34 & $-$20.8 (0.5)  &\ldots&\ldots&\ldots&\ldots \\
42011588 & 16.42 & 31 & $-$33.9 (0.7)  &\ldots&\ldots&\ldots&\ldots \\
42012172 & 16.96 & 27 & $-$28.7 (0.8)  &\ldots&\ldots&\ldots&\ldots \\
42012437 & 17.06 & 34 & $-$23.0 (0.6)  &\ldots&\ldots&\ldots&\ldots \\
42012704 & 16.19 & 51 & $-$29.7 (0.7)  & 4248 & 0.94 & 2.20 & $-$1.99 \\
42013014 & 16.24 & 43 & $-$26.9 (0.6)  & 4211 & 0.93 & 2.30 & $-$1.89 \\
51000173 & 16.89 & 27 & $-$23.6 (0.4)  &\ldots&\ldots&\ldots&\ldots \\
52005710 & 16.96 & 30 & $-$34.8 (0.6)  & 4453 & 1.39 & 1.40 & $-$1.90 \\
52009395 & 17.37 & 29 & $-$25.5 (0.8)  &\ldots&\ldots&\ldots&\ldots \\
61000086 & 16.96 & 27 & $-$28.7 (0.7)  &\ldots&\ldots&\ldots&\ldots \\
61000142 & 17.04 & 31 & $-$29.0 (1.1)  &\ldots&\ldots&\ldots&\ldots \\
61000144 & 16.40 & 33 & $-$34.3 (1.0)  & 4293 & 1.06 & 1.85 & $-$1.90 \\
61000182 & 16.98 & 31 & $-$25.0 (0.4)  &\ldots&\ldots&\ldots&\ldots \\
61000397 & 16.31 & 39 & $-$23.9 (0.9)  & 4242 & 0.98 & 2.10 & $-$1.97 \\
61000826 & 16.27 & 43 & $-$34.8 (0.7)  & 4312 & 1.02 & 1.65 & $-$2.01 \\
61001052 & 16.79 & 29 & $-$25.9 (0.8)  &\ldots&\ldots&\ldots&\ldots \\
61001690 & 17.11 & 27 & $-$21.6 (0.8)  &\ldots&\ldots&\ldots&\ldots \\
61001836 & 16.88 & 36 & $-$24.1 (0.8)  &\ldots&\ldots&\ldots&\ldots \\
61003283 & 16.28 & 36 & $-$28.6 (0.4)  &\ldots&\ldots&\ldots&\ldots \\
61005163 & 15.91 & 52 & $-$21.1 (0.5)  & 4198 & 0.78 & 2.10 & $-$1.85 \\
61007408 & 16.17 & 51 & $-$34.6 (0.9)  & 4220 & 0.91 & 1.80 & $-$2.04 \\
61007565 & 15.96 & 58 & $-$28.0 (0.8)  & 4159 & 0.77 & 1.85 & $-$1.78 \\
61008247 & 16.43 & 42 & $-$34.4 (0.9)  & 4360 & 1.12 & 2.00 & $-$2.10 \\
62000027 & 16.43 & 33 & $-$27.7 (0.9)  &\ldots&\ldots&\ldots&\ldots \\
71000779 & 17.03 & 25 & $-$20.1 (1.0)  &\ldots&\ldots&\ldots&\ldots \\
71002877 & 15.81 & 55 & $-$27.3 (0.8)  & 4110 & 0.67 & 2.15 & $-$1.72 \\
\hline
\end{tabular} \\
$^{a}$ {Velocity derived from a single, noisy measurement only;
uncertainty is likely to be 1.5~\kmsec\ or greater.}
\end{minipage}
\end{table*}

Our first set of
observations were made using the Michigan/Magellan Fiber System
(M2FS)
and MSpec double spectrograph \citep{mateo12,bailey12}
mounted on the Nasmyth platform at the 6.5~m Landon Clay Telescope
(Magellan~II) at Las Campanas Observatory, Chile.
M2FS uses fiber plug plates to achieve high multiplexing capability
over a 30'-diameter field of view.
We observed 50~candidate members of \ngc\ and 11~blank sky positions.
Three observations were made on 2015 April 15, 2015 April 19, and
2015 April 20, 
with a total integration time of 11.5~h.

\begin{figure}
\centering
\includegraphics[angle=0,width=3.25in]{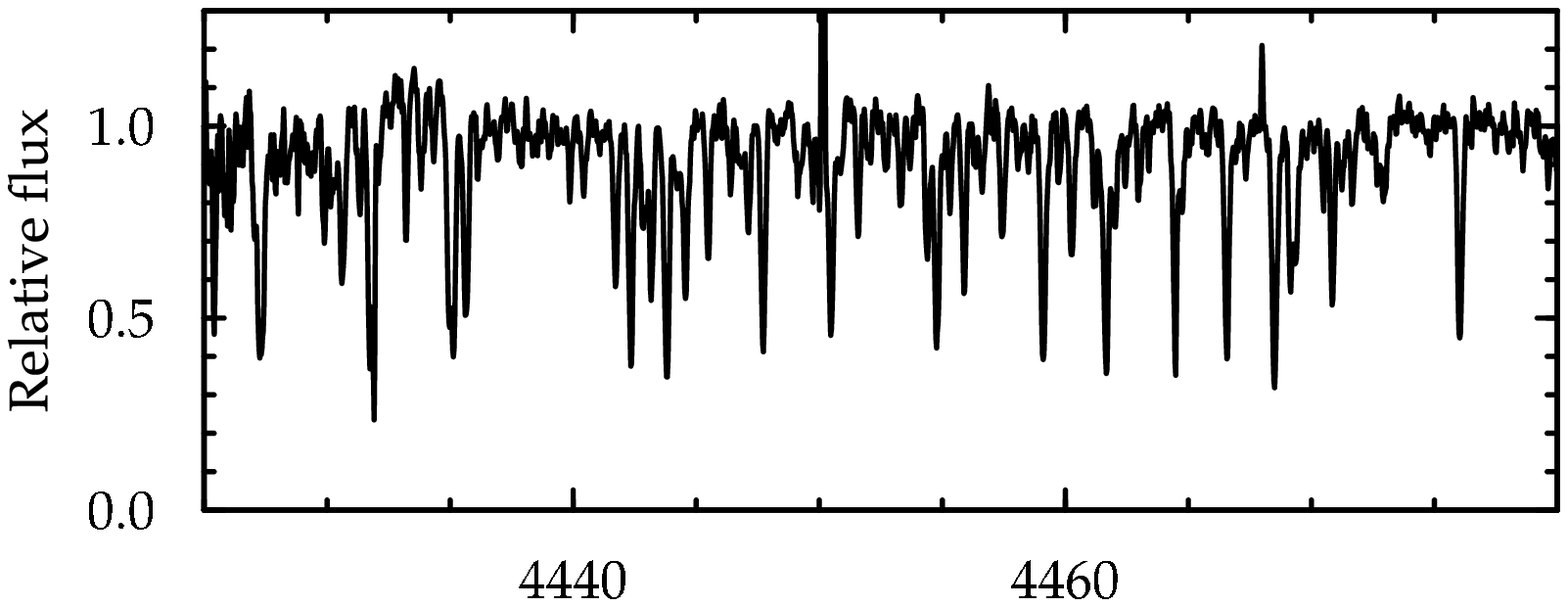} \\
\vspace*{0.1in}
\includegraphics[angle=0,width=3.25in]{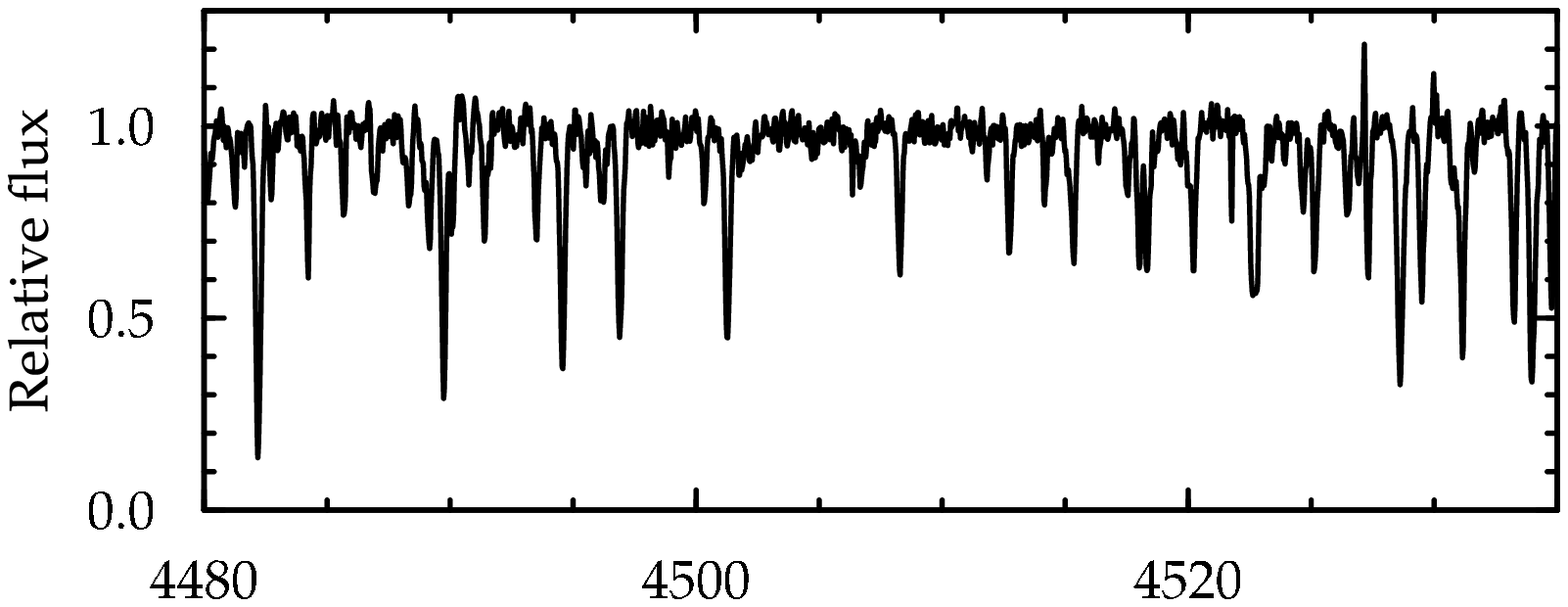} \\
\vspace*{0.1in}
\includegraphics[angle=0,width=3.25in]{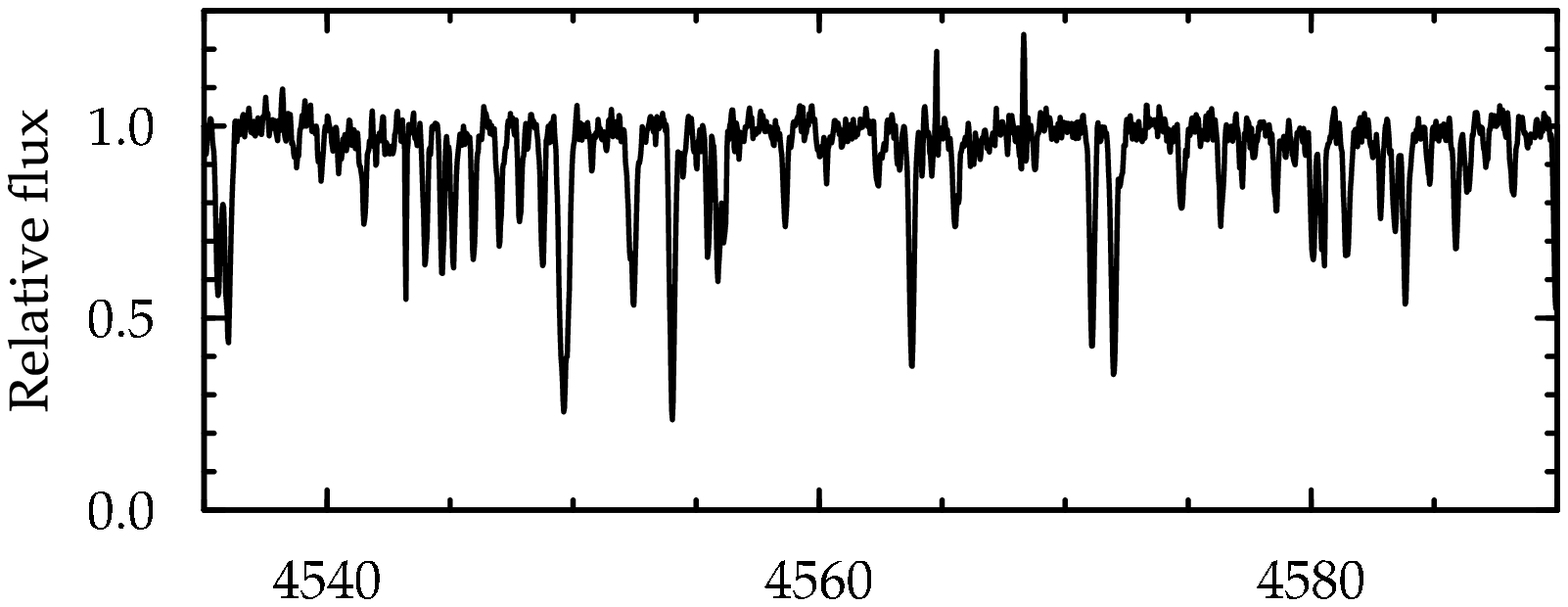} \\
\vspace*{0.1in}
\includegraphics[angle=0,width=3.25in]{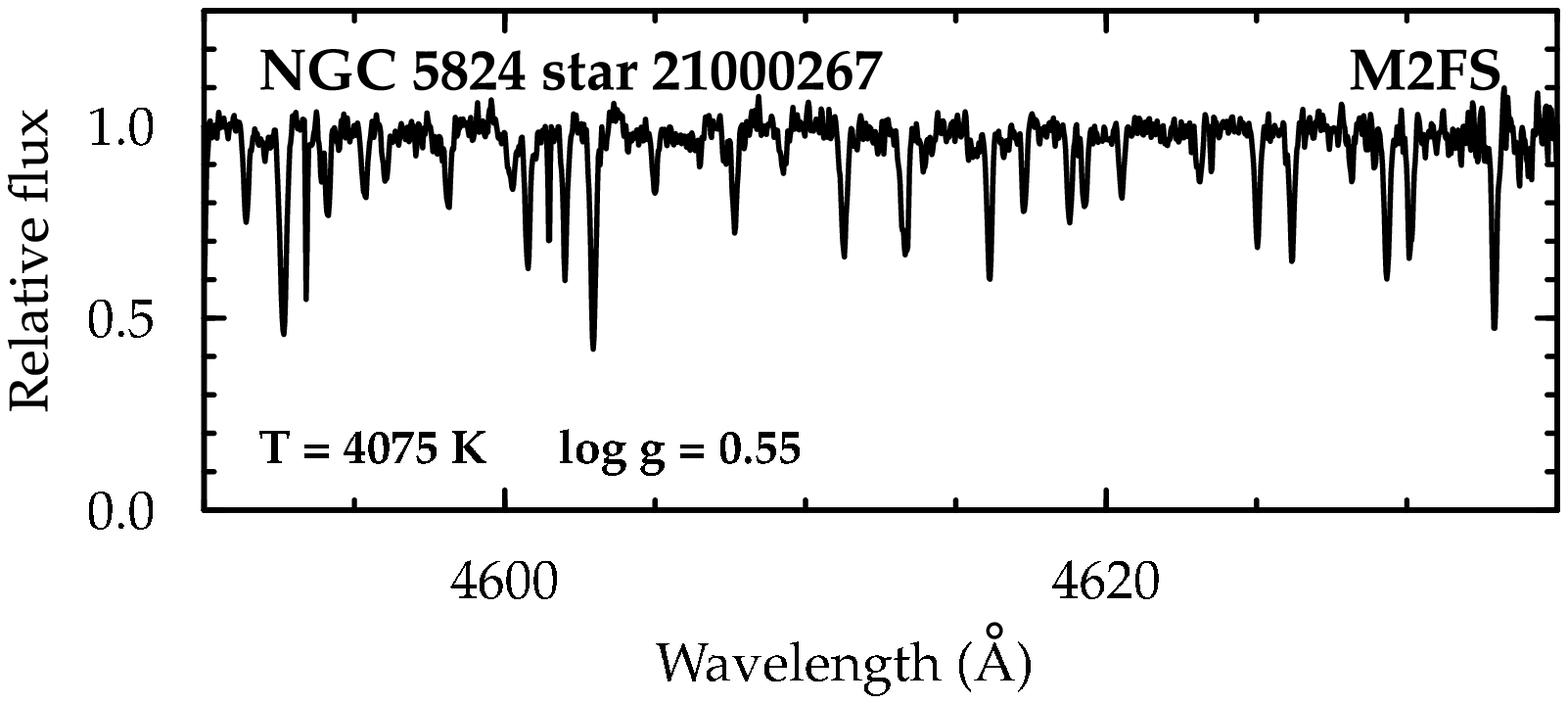} \\
\caption{
\label{m2fsrangeplot}
The M2FS spectrum of star 21000267.
}
\end{figure}

Our observations were made in HiRes mode with 
95~$\mu$m entrance slits.  
This setup delivers spectral resolving power
$R \equiv \lambda/\Delta\lambda \sim$~34,000,
as measured from isolated Th or Ar emission lines
in the comparison lamp spectra.
This corresponds to $\approx$~6.3~pixels per resolution element (RE).~
The spectral resolution varies by $\approx$~7~per cent 
from one fiber tetris to another 
due to the alignment of the fiber tetrises
with the slits.
We use a custom set of order-isolation filters to 
observe orders 77--80,
which covers roughly 4425~$\leq \lambda \leq$~4635~\AA\
for each target.
Figure~\ref{m2fsrangeplot} illustrates the entire M2FS 
spectrum of one star in our sample, 21000267.

Targets were chosen from the list of probable members given
by \citet{dacosta14}.
Stars observed with M2FS are on the upper red giant branch (RGB) in 
\ngc\ and are $\approx$~1 to 3~magnitudes brighter than the
horizontal branch.
Some data reduction (merging data from 
different CCD chip amplifiers, stacking images,
masking cosmic rays, 
and subtracting scattered light) 
was performed using \textsc{python} routines
written by J.~I.~B.
Standard \textsc{iraf} routines were used to perform
all other tasks (flatfielding, extraction, wavelength calibration, 
spectra co-addition, velocity shifting, and
continuum normalization).
The M2FS observations of \ngc\ were made during dark time.
Sky contamination was found to be negligible,
so no sky subtraction was performed.
Table~\ref{obstab} lists the $V$ magnitudes
(from \citeauthor{dacosta14})\ and
S/N ratios per pixel in the 
co-added M2FS spectra.
Our proposed M2FS integration times were shortened considerably
due to poor weather during the M2FS block run in April, 2015.
We measure radial velocities for all targets
observed (Section~\ref{rv}), but
a reliable chemical analysis (Section~\ref{analysis})
is only possible for the brightest targets.

After an initial analysis of the M2FS spectra,
two stars with nearly identical stellar parameters
but differing levels of \ncap\ elements were selected for
additional observations.
Stars 42009955 and 61005163 were observed 
on 2015 June 16
using the
Magellan Inamori Kyocera Echelle (MIKE)
spectrograph \citep{bernstein03},
which is also mounted on the
Nasmyth platform at Magellan~II
(but not simultaneously with M2FS).
These spectra were taken with the 0\farcs7\,$\times$\,5\farcs0 slit, 
yielding
a resolving power of $R \sim$~41,500 in the blue 
and $R \sim$~34,700 in the red as measured from isolated Th and Ar 
emission lines
in the comparison lamp images.
This corresponds to $\approx$~2.5 and 2.1~pixels RE$^{-1}$
on the blue and red arms, respectively.
The blue and red arms are split by a dichroic at $\approx$~4950~\AA.
This setup achieves complete wavelength coverage from 
3350--9150~\AA, although
only the spectra longward of $\approx$~3800~\AA\ 
have signal sufficient to perform a detailed abundance analysis.
Data reduction,
extraction, sky subtraction,
and wavelength calibration were performed using 
the \textsc{CarPy} MIKE data reduction pipeline
(see, e.g., \citealt{kelson03}).
Coaddition and continuum normalization 
were performed within
\textsc{iraf}\@.
The total integration times were 4.2~h and 3.3~h 
for stars 42009955 and 61005163, respectively.
The S/N ratios of the co-added
MIKE spectra range from $\approx$~25/1 near 3950~\AA\
to $\approx$~140/1 near 6750~\AA.~

\subsection{Comparing the M2FS and MIKE Spectra}
\label{comparespectra}

\begin{figure}
\centering
\includegraphics[angle=0,width=3.25in]{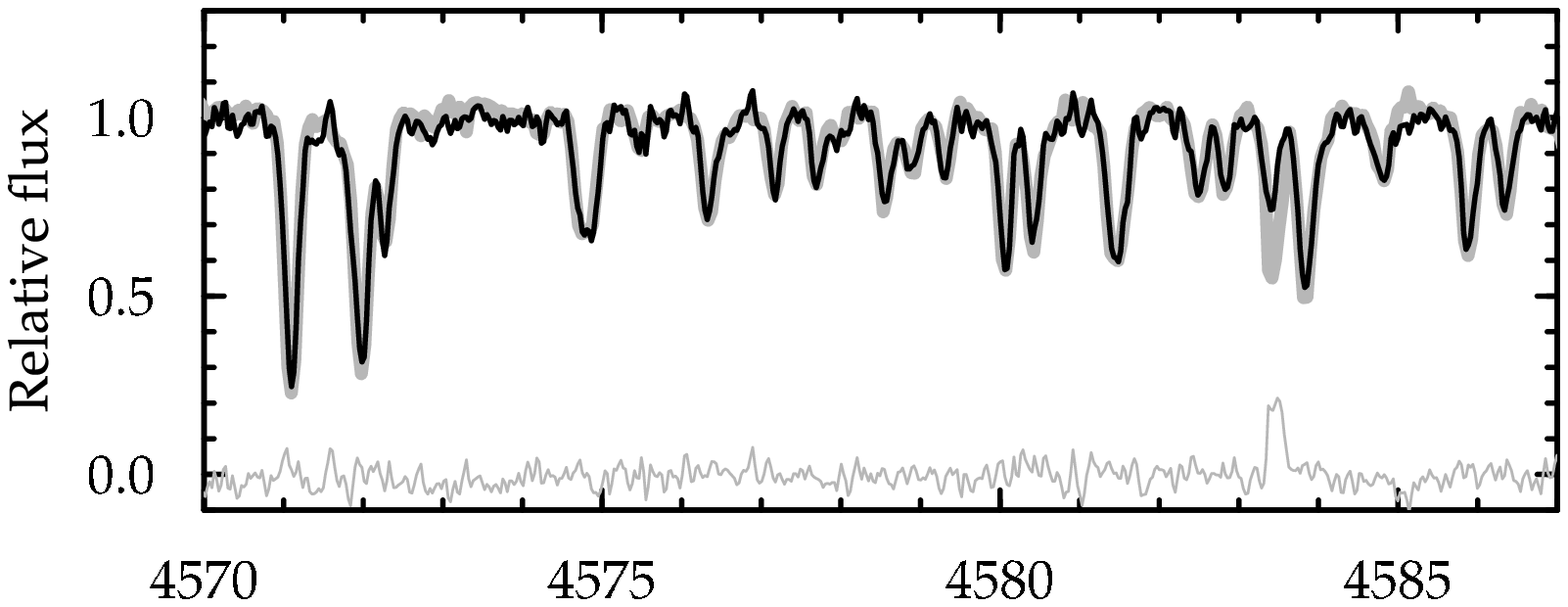} \\
\vspace*{0.1in}
\includegraphics[angle=0,width=3.25in]{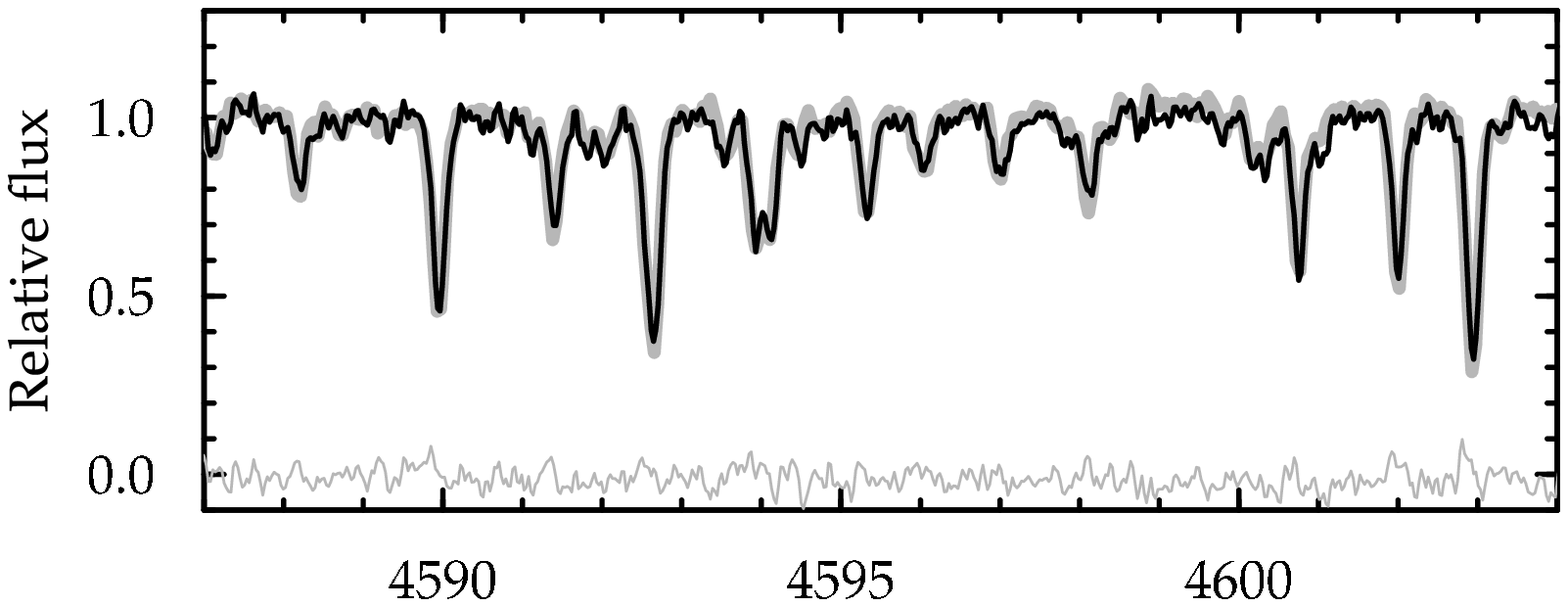} \\
\vspace*{0.1in}
\includegraphics[angle=0,width=3.25in]{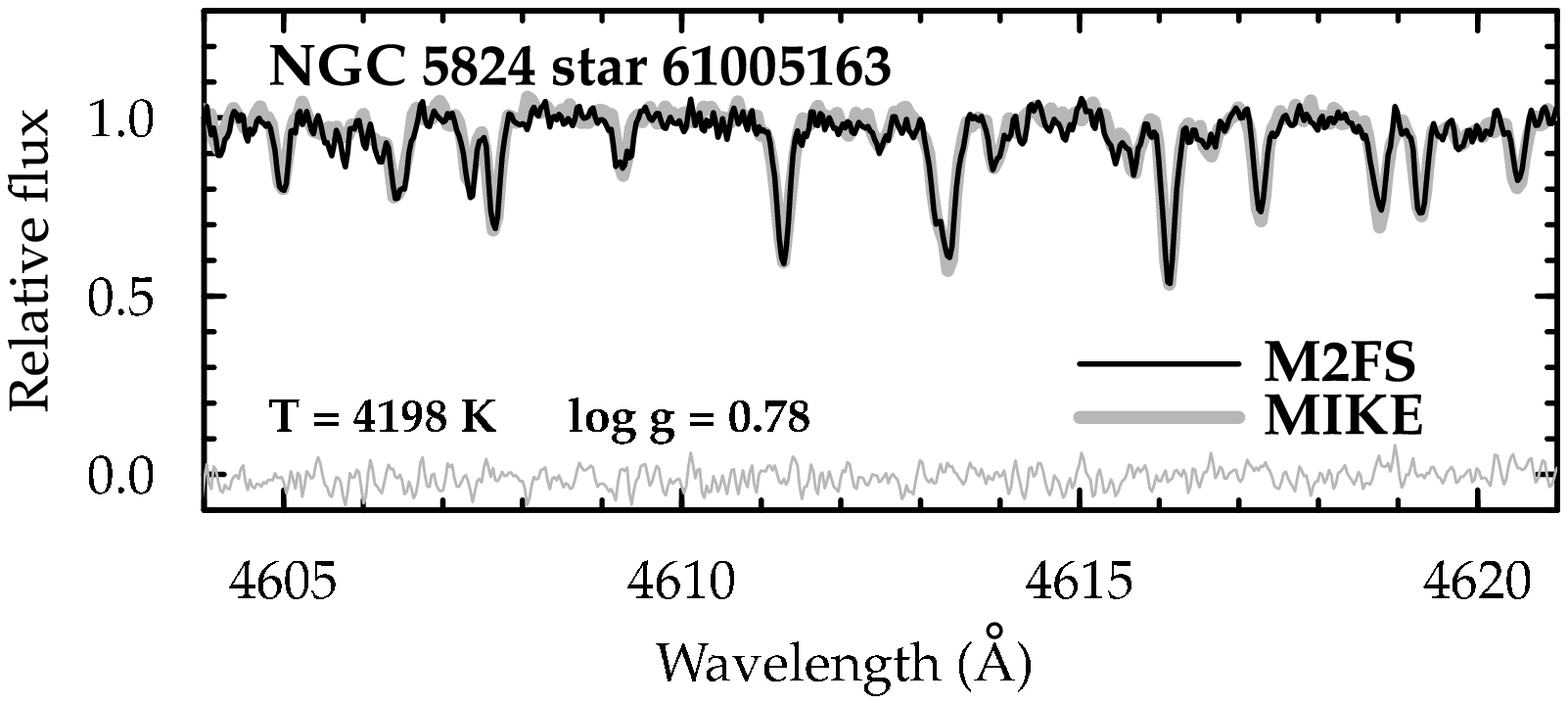} \\
\caption{
\label{m2fsmikeplot}
Portions of the M2FS and MIKE spectra of star 61005163.
The M2FS spectrum is shown by the black line, and
the MIKE spectrum is shown by the bold gray line.
The thin gray line at the bottom of each panel shows the 
residual between the two spectra on the same scale.
}
\end{figure}

Figure~\ref{m2fsmikeplot} illustrates a portion 
of the region of overlap between the M2FS and MIKE
spectra for one of the stars in common,
61005163.
The differences, resampled to the M2FS pixel spacing,
are shown in the bottom of each panel in Figure~\ref{m2fsmikeplot}.
Qualitatively,
the agreement between the spectra appears excellent,
and we now attempt to quantify the degree of similarity.

\begin{figure}
\centering
\includegraphics[angle=0,width=3.25in]{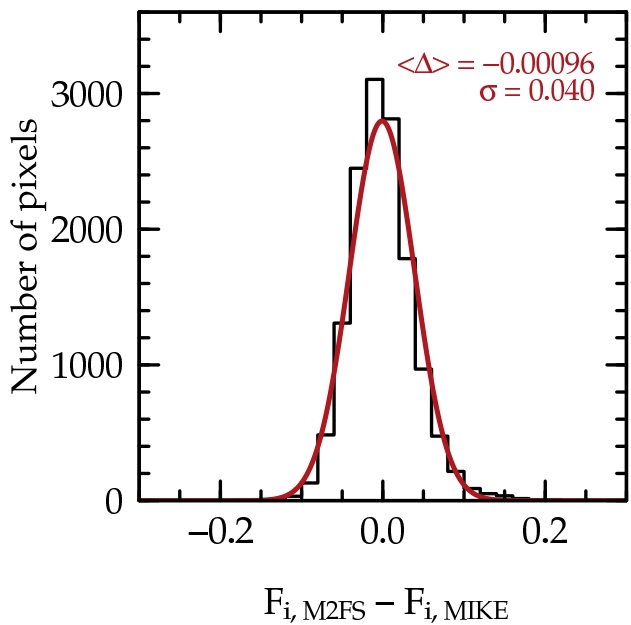}
\caption{
\label{specdeltaplot}
Residuals in the normalized flux between the M2FS and MIKE
spectra of stars 42009955 and 61005163.
The mean difference and standard deviation
are printed and illustrated by the red curve.
}
\end{figure}

Figure~\ref{specdeltaplot} shows a histogram of the
residuals of the normalized M2FS and MIKE spectra for
each of the two stars.
Only wavelengths from 4430 to 4630~\AA\ are included.
The mean difference is $-$0.096~per cent,
which reflects the relative continuum normalization
of the two spectra.
The standard deviation of the residuals is 4.0~per cent.
The S/N ratios of the M2FS spectra of these two stars
are $\approx$~52/1 per 0.0292~\AA\ pixel,
and the S/N ratios of the MIKE spectra 
are $\approx$~63/1 per 0.044~\AA\ pixel.
After converting to a common pixel size,
a residual of $\approx$~2.7~per cent can be accounted for
by photon noise alone.
The differing resolution of the two instruments
accounts for another $\approx$~0.7~per cent of the residuals,
if we assume that absorption lines account for $\approx$~20~per cent
of the spectrum.
This leaves $(0.040^{2} - 0.027^{2} - 0.007^{2})^{1/2} \approx$~2.9~per cent
unaccounted for, which we assume includes errors in the
flatfielding, cosmic ray removal, sky subtraction, order merging, 
etc.\
We conclude that the M2FS and MIKE spectra
are identical to within a few per cent.
In Sections~\ref{ew} and \ref{comparem2fsmike}, 
we compare EWs and abundances
derived from the two sets of spectra.

\section{Radial Velocity Measurements}
\label{rv}

We measure radial velocities for each M2FS observation
using the \textsc{iraf} FXCOR task to cross-correlate,
order-by-order, against a template.
We use one high-S/N spectrum of \mbox{21000267}
as the template.
We establish the zeropoint of this spectrum
by using SPLOT to measure the wavelengths of 51~Fe~\textsc{i} lines;
we then compare with the laboratory wavelengths 
presented by \citet{nave94}.
The velocity zeropoint of the template is 
accurate to better than 0.2~\kmsec.
The statistical uncertainties associated with
each cross-correlation relative to the template
vary with S/N, ranging from
0.4~\kmsec\ for the brightest targets to
1.3~\kmsec\ for the faintest ones.  
We estimate these uncertainties based on 
repeat observations of each star. 
We estimate the absolute velocity uncertainty to be $\approx$~0.2~\kmsec,
based on observations of the metal-poor standard star \mbox{HD~122563}
taken with the same M2FS configuration in February, 2015.

We report heliocentric radial velocities, $V_{r}$, for each star
in Table~\ref{obstab}.
We calculate heliocentric corrections 
using the \textsc{iraf} RVCORRECT task.
The velocities in Table~\ref{obstab} 
represent a weighted mean of our three observations.
The weights roughly correspond to the S/N obtained
in each set of observations,
with a weight of 1 for the first and third epochs
and a weight of 3 for the second epoch.
Our radial velocities differ by only 1.9~$\pm$~0.2~\kmsec\
($\sigma =$~9.1~\kmsec) from those reported by \citet{dacosta14}.

The mean systemic radial velocity for \ngc,
based on 50~stars, is
$-$27.4~$\pm$~0.7~\kmsec\ ($\sigma =$~4.9~\kmsec).
This agrees with the values measured by 
\citet{dacosta14},
$-$28.9~$\pm$~1.0~\kmsec,
and 
\citet{dubath97},
$-$26.0~$\pm$~1.6~\kmsec.
The dispersion of our velocity measurements,
4.9~\kmsec, is considerably smaller than the
core velocity dispersion 
derived from the integrated-light spectra of
\citeauthor{dubath97}, 11.1~$\pm$~1.6~\kmsec.
This may be expected since
our targets mostly sample the outer regions of the cluster
(0.9~$< r_{h} <$~14,
where $r_{h} =$~0.45~arcmin is the half-light radius given by
\citealt{harris96}).

We measure radial velocities from the MIKE spectra
of stars 42009955 and 61005163
by cross-correlating a late-type metal-poor template
against the echelle order containing the Mg~\textsc{i}~b lines,
as described in \citet{roederer14a}.
The MIKE velocity zeropoint is reliable to $\approx~$0.2~\kmsec\
(e.g., \citealt{roederer14c}).
The heliocentric radial velocities measured from the MIKE
spectra of stars 42009955 and 61005163,
$-$21.4~$\pm$~0.7~\kmsec\ and $-$20.8~$\pm$~0.7~\kmsec,
are in agreement with those measured two months 
earlier using M2FS,
$-$23.0~$\pm$~0.8~\kmsec\ and $-$21.1~$\pm$~0.5~\kmsec.
Our data offer no evidence for radial velocity variations
for either star.

\section{Equivalent Widths}
\label{ew}

We measure EWs from the spectra using a semi-automated 
routine, \textsc{ew.pro}, that fits Voigt (or Gaussian) 
line profiles to continuum-normalized spectra.
As discussed in \citet{roederer14a},
all fits are presented to the user for 
approval, modification, or rejection.
The EWs measured in the M2FS spectra are listed in
Table~\ref{m2fsewtab}, and the 
EWs measured in the MIKE spectra are
listed in Table~\ref{mikeewtab}.
The full list of lines examined in the M2FS spectra
is shown in Table~\ref{m2fslinelisttab}.

\begin{table}
\begin{minipage}{3.25in}
\caption{Equivalent Width Measurements for Lines
Detected in the M2FS Spectra
\label{m2fsewtab}}
\begin{tabular}{cccc}
\hline
Star &
Species &
$\lambda$ &
EW \\
 &
 & 
(\AA) &
(m\AA) \\
11001198 &  Mg~\textsc{i}  &   4571.10 &    142.5 \\
11001198 &  Ca~\textsc{i}  &   4526.94 &     36.2 \\
11001198 &  Ca~\textsc{i}  &   4578.55 &     39.1 \\
11001198 &  Ti~\textsc{i}  &   4453.70 &     29.7 \\
11001198 &  Ti~\textsc{i}  &   4481.26 &     43.7 \\
\hline
\end{tabular} \\
The complete version of Table~\ref{m2fsewtab} 
is available in machine-readable format in the
Supplementary Information found in the online edition of the journal.
Only a small portion is shown here to illustrate its form and content. \\
Note---
All abundances derived from lines with no EW
reported in Table~\ref{m2fsewtab}
have been derived by spectrum synthesis matching.
\end{minipage}
\end{table}

\begin{table}
\begin{minipage}{3.25in}
\caption{Equivalent Width Measurements for Lines
Detected in the MIKE Spectra
\label{mikeewtab}}
\begin{tabular}{ccccccc}
\hline
Species &
$\lambda$ &
E.P. &
\loggf &
Ref. &
EW &
EW \\
 &
 &
 &
 &
 &
42009955 &
61005163 \\
 &
(\AA) &
(eV) &
 &
 &
(m\AA) &
(m\AA) \\
\hline       
Li~\textsc{i}  &   6707.80 &  0.00 &    0.17 &  1 &  limit &   limit  \\
O~\textsc{i}   &   6300.30 &  0.00 & $-$9.78 &  2 &  limit &  \ldots  \\
O~\textsc{i}   &   7771.94 &  9.14 &    0.37 &  2 &  limit &     6.2  \\
O~\textsc{i}   &   7774.17 &  9.14 &    0.22 &  2 &  limit &     8.0  \\
O~\textsc{i}   &   7775.39 &  9.14 &    0.00 &  2 &  limit &     7.1  \\
\hline
\end{tabular} \\
The complete version of Table~\ref{mikeewtab} 
is available in machine-readable format in the
Supplementary Information found in the online edition of the journal.
Only a small portion is shown here to illustrate its form and content. \\
Notes---
Lines denoted ``limit'' were not detected,
and upper limits have been derived.
Lines denoted ``synth'' yielded
abundances via spectrum synthesis matching.\\
References---
 (1) \citealt{smith98}; 
 (2) NIST;
 (3) \citealt{aldenius09};
 (4) \citealt{lawler89}, using HFS from \citealt{kurucz95};
 (5) \citealt{lawler13};
 (6) \citealt{wood13};
 (7) \citealt{lawler14}, using HFS from \citealt{kurucz95};
 (8) \citealt{wood14v};
 (9) \citealt{sobeck07};
(10) \citealt{nilsson06};
(11) \citealt{booth84};
(12) \citealt{denhartog11} for both \loggf\ value and HFS;
(13) \citealt{ruffoni14};
(14) NIST, using HFS from \citealt{kurucz95};
(15) \citealt{wood14ni};
(16) \citealt{roederer12b};
(17) \citealt{biemont11};
(18) \citealt{ljung06};
(19) NIST, using HFS/IS from \citealt{mcwilliam98} when available;
(20) \citealt{lawler01la}, using HFS from \citealt{ivans06};
(21) This study;
(22) \citealt{roederer11c};  
(23) \citealt{lawler09}; 
(24) \citealt{li07}, using HFS from \citealt{sneden09};
(25) \citealt{ivarsson01}, using HFS from \citealt{sneden09};
(26) \citealt{denhartog03}, using HFS/IS from \citealt{roederer08} when available;
(27) \citealt{lawler06}, using HFS/IS from \citealt{roederer08} when available;
(28) \citealt{lawler01eu}, using HFS/IS from \citealt{ivans06};
(29) \citealt{roederer12d};
(30) \citealt{denhartog06};
(31) \citealt{lawler01tb};
(32) \citealt{wickliffe00}; 
(33) \citealt{lawler08};
(34) \citealt{lawler07};
(35) \citealt{biemont00}, using HFS/IS from \citealt{roederer12d}; 
(36) \citealt{nilsson02}. 
\end{minipage}
\end{table}

\begin{table}
\begin{minipage}{3.25in}
\caption{List of Lines Examined in M2FS Spectra
\label{m2fslinelisttab}}
\begin{tabular}{ccccc}
\hline
Species &
Wavelength &
E.P. &
\loggf &
Ref. \\
 &
(\AA) &
(eV) &
 &
 \\
\hline
Mg~\textsc{i}   &   4571.10  &  0.00  &  $-$5.62  &   1   \\
Ca~\textsc{i}   &   4526.94  &  2.71  &  $-$0.42  &   1   \\
Ca~\textsc{i}   &   4578.55  &  2.52  &  $-$0.56  &   1   \\
Sc~\textsc{ii}  &   4431.35  &  0.60  &  $-$1.97  &   2   \\
Ti~\textsc{i}   &   4449.14  &  1.89  &  $+$0.47  &   3   \\
\hline
\end{tabular} \\
The complete version of Table~\ref{m2fslinelisttab} 
is available in machine-readable format in the
Supplementary Information found in the online edition of the journal.
Only a small portion is shown here to illustrate its form and content. \\
References---
(1)  NIST;
(2)  \citealt{lawler89}, using HFS from \citealt{kurucz95};
(3)  \citealt{lawler13};
(4)  \citealt{wood13};
(5)  \citealt{lawler14}, using HFS from \citealt{kurucz95};
(6)  \citealt{sobeck07};
(7)  \citealt{booth84};
(8)  \citealt{ruffoni14};
(9)  \citealt{wood14ni};
(10) NIST, using HFS/IS from \citealt{mcwilliam98};
(11) This work;
(12) \citealt{lawler09};
(13) \citealt{denhartog03}, using HFS/IS from \citealt{roederer08} when 
        available;
(14) \citealt{lawler06}, using HFS/IS from \citealt{roederer08} when available;
(15) \citealt{lawler01eu}, using HFS/IS from \citealt{ivans06} when available;
(16) \citealt{wickliffe00}.
\end{minipage}
\end{table}

\begin{figure}
\centering
\includegraphics[angle=0,width=3.25in]{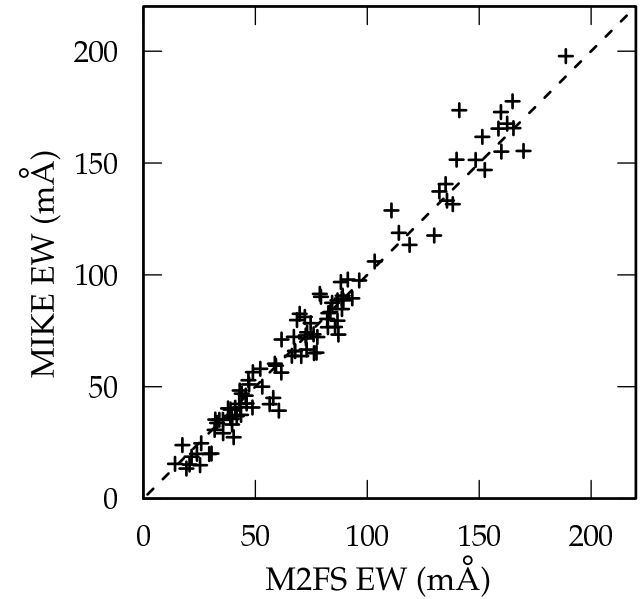}
\caption{
\label{ewplot}
Comparison of EWs measured in M2FS and MIKE spectra
of stars 42009955 and 61005163.
The dashed line marks the one-to-one correspondence.
}
\end{figure}

EWs measured from MIKE spectra are useful 
to demonstrate the integrity of EWs measured from M2FS spectra.
\citet{roederer14a} found that EWs measured using \textsc{ew.pro}
were statistically identical
among spectra obtained with three 
different spectrographs---MIKE, 
the Tull Coud\'{e} Spectrograph, and the
High Resolution Spectrograph (HRS).
(The Tull and HRS instruments are at McDonald Observatory, Texas).
\citet{bedell14} performed an analysis of EWs measured from
high-quality asteroid-reflected solar spectra 
taken with MIKE and ESPaDOnS.
(ESPaDOnS is at the the Canada-France-Hawaii Telescope.)
They found abundance differences at the $\approx$~0.04~dex level.
This comprised the largest component of the 
error budget in their differential analysis,
but it is sufficiently small for our purposes.
Figure~\ref{ewplot} compares the EWs measured with MIKE and M2FS
for 103 lines in common in the spectra
of stars 42009955 and 61005163.
We find a difference of 0.25~$\pm$~0.78~m\AA\ 
($\sigma =$~7.9~m\AA), which is not significant,
indicating that the M2FS EWs are also consistent with
external scales.

\section{Atomic Data}
\label{atomic}

Tables~\ref{mikeewtab} and \ref{m2fslinelisttab}
include references for the atomic data.
We privilege \loggf\ values from recent laboratory
studies whenever possible,
since these investigations frequently deliver
$\approx$~5~per cent precision ($\approx$~0.02~dex) or better.
We also attempt to use a single source for the \loggf\ values
of a given species whenever possible,
to minimize any systematic offsets from one study to another.
Our goal in making these choices is to 
minimize the impact of systematics arising from
the set of lines examined.
As we show in Section~\ref{comparem2fsmike}, this goal can 
be achieved in most cases.

Hyperfine splitting (HFS) structure and
isotope shifts (IS) are included
in the spectrum synthesis when these data are available.
We adopt an isotopic ratio for $^{12}$C/$^{13}$C of 4.
For Cu, we adopt the solar isotopic ratio, 
$^{63}$Cu/$^{65}$Cu~$=$~2.24.
We adopt \rpro\ isotopic fractions for Ba, Nd, Sm, Eu, and Pb
using the values presented in \citet{sneden08}.
For star 61005163, we instead adopt
an appropriate mixture of $r$- and \spro\ isotopic fractions
(see Section~\ref{ncapture}).
At the request of the referee,
we have included the HFS and IS component patterns
for lines of \ncap\ elements studied in this work.
These are presented in Table~\ref{hfstab}.
We emphasize that these data simply echo those given in the references to
Table~\ref{mikeewtab}.

\begin{table}
\begin{minipage}{3.25in}
\caption{List of Hyperfine Component Patterns for Heavy Elements Sorted by 
Species and Wavelength
\label{hfstab}}
\begin{tabular}{ccccc}
\hline
Species &
Isotope &
Wavelength &
E.P. &
\loggf\footnote{The \loggf\ values of each isotope are normalized to the total
  \loggf\ value for the transition.} \\
 & 
 &
(\AA) &
(eV) &
 \\
\hline
Ba~\textsc{ii} & 137 & 4553.9980 & 0.000 & $-$0.6660 \\
Ba~\textsc{ii} & 137 & 4554.0000 & 0.000 & $-$0.6660 \\
Ba~\textsc{ii} & 137 & 4554.0010 & 0.000 & $-$1.0640 \\
Ba~\textsc{ii} & 135 & 4554.0020 & 0.000 & $-$0.6660 \\
Ba~\textsc{ii} & 135 & 4554.0030 & 0.000 & $-$0.6660 \\
\hline
\end{tabular} \\
The complete version of Table~\ref{hfstab}
is available in machine-readable format in the
Supplementary Information found in the online edition of the journal.
Only a small portion is shown here to illustrate its form and content. 
\end{minipage}
\end{table}

The La~\textsc{ii} line at 4522.37~\AA\ was not covered 
by the laboratory study of \citet{lawler01la},
but it is the only reliable La~\textsc{ii} line that
appears in our M2FS spectra.
We use a high-resolution spectrum of the \rpro-rich 
standard star \mbox{BD$+$17~3248},
a red giant with [Fe/H]~$= -$2.1 \citep{cowan02},
to derive an empirical \loggf\ value for this line.
Our examination of 
five other La~\textsc{ii} lines in \mbox{BD$+$17~3248}
yields \loggf~$= -$1.22~$\pm$~0.10 for the line at 4522.37~\AA.~
No HFS data are available for this line.
It falls on the linear part of the 
curve of growth in the spectra of \ngc\ red giants
and the spectrum of \mbox{BD$+$17~3248},
so errors resulting from the neglect of HFS
in our spectrum synthesis should be minimal.

\section{Model Atmosphere Parameters}
\label{atmosphere}

We calculate effective temperatures, \teff,
using the de-reddened $V-I$ color-\teff\ relation
from \citet{alonso99b},
assuming $E(B-V) =$~0.14 \citep{zinn80,reed88}
and the \citet{cardelli89} extinction coefficients.
We transform the \citet{dacosta14} $V-I$ colors,
measured in the standard Johnson-Cousins system,
to the Johnson system using the relation 
given by \citet{bessell79}.
A second step is frequently employed by
fitting a relationship between a magnitude (often $K$)
and \teff, which reduces the impact of photometric errors
and differential reddening
(e.g., \citealt{carretta07}; \citealt{roederer15}).
This approach assumes that the RGB
has no intrinsic spread at a given color.
This assumption may be invalid for \ngc\
since an internal metallicity spread may be present,
so we do not employ this second step 
when calculating \teff.
Monte Carlo error propagation calculations
indicate that the statistical uncertainties on \teff\ from
all sources of error 
(intrinsic scatter in the color-\teff\ relation,
photometric error, and
uncertainty in the reddening)
are $\approx$~145~K.
The 
optional second step---fitting and interpolating 
a relationship
between $K$ and \teff---yields statistical uncertainties
$\approx$~40--50~K, which is considerably better.
This demonstrates why the second step is preferable
when applicable.

We calculate the log of the surface gravity, \logg, 
using the relation
$\log g = 4\log(T_{\rm eff,\star})
+\log(M/M_{\odot})
-0.4(M_{\rm bol,\sun} - M_{\rm bol,\star})
-10.61$.
The constant 10.61 is derived from the 
solar values given in \citet{cox00},
the quantity $M_{\rm bol,\star}$ is given by
${\rm BC}_{V} + m_{V} - (m-M)$, and
$BC_{V}$ is given by eq.\ 18 of \citet{alonso99b}.
We assume masses of 0.8~$\pm$~0.15~M$_{\odot}$ for all stars.
We adopt the distance modulus given 
by \citet{harris96},
$m-M =$~17.94~$\pm$~0.05,
which is based on the magnitude of the 
horizontal branch measured by \citet{brocato96}.
Our Monte Carlo calculations indicate
that statistical uncertainties in the \logg\ values
are $\approx$~0.14~dex.
This quantity is dominated by uncertainties in \teff\ and
mass estimates.

We interpolate model atmospheres from the
\textsc{atlas9} grid of $\alpha$-enhanced models
\citep{castelli03}, using an interpolator 
provided by A.\ McWilliam (2009, private communication).
We derive abundances 
using a recent version of the \textsc{moog} line analysis
software \citep{sneden73}, with updates described in 
\citet{sobeck11}.
We determine microturbulent velocities, \vt, 
by requiring no trend between the line strength,
parameterized by $\log$(EW/$\lambda$),
and abundance derived from Fe~\textsc{i} lines.
We estimate a statistical uncertainty of 
$\approx$~0.1~\kmsec\ in \vt\ 
for a fixed set of \teff\ and \logg.
We set the model metallicity, [M/H], equal to the 
Fe abundance derived from Fe~\textsc{ii} lines
and adjust the microturbulent velocity and metallicity
iteratively.
We also iteratively cull 
Fe~\textsc{i} and \textsc{ii} lines
giving abundances more than 2~$\sigma$
from the mean.
We consider the model parameters to have converged
when subsequent adjustment of
\vt\ is less than 0.05~\kmsec,
adjustment of the model [M/H] is less than 0.03~dex,
and all Fe~\textsc{i} and \textsc{ii} lines
agree to within 2~$\sigma$.

Systematic sources of uncertainty 
are undoubtedly larger than the statistical 
errors noted here, 
but they have little effect on the 
relative abundances of stars 
in the same evolutionary state.
We discuss the impact of these uncertainties
in Section~\ref{spread}.

The adopted model atmosphere parameters are listed
in Table~\ref{obstab} for stars
with moderate or high S/N levels 
and \teff~$>$~4000~K.
We examine the detailed abundance patterns of these
26~stars.
Star 21002944 
also has superb S/N, but
it may be on the AGB.~
Assuming a mass of 0.65~M$_{\odot}$ \citep{gratton10},
we estimate a temperature of 3890~K,
which is nearly 200~K cooler than the
next-coolest giant.
Our method may not be applicable to stars
this cool without additional modification,
and we aim to preserve a differential 
quality to the analysis.
We do not analyze the abundances in star~21002944
and will not discuss it further.

\section{Analysis}
\label{analysis}

Most abundances are derived using an adaptation of the
batch mode capabilities of \textsc{moog},
forcing theoretical EWs
to match the observed ones by adjusting the abundance.
Other abundances are derived by matching
synthetic to observed spectra.
We adopt the latter method when lines are
too blended to measure reliable EWs 
or when HFS or IS could affect the derived abundance.
In a few cases, we report 3~$\sigma$ upper limits on the abundance
using a version of the formula presented on p.\ 590 of
\citet{frebel08}, which is derived from
equation A8 of \citet{bohlin83}.

We assume local thermodynamic equilibrium (LTE)
holds for the line-forming layers of the atmosphere
for all species except 
Na~\textsc{i} \citep{lind11} and 
K~\textsc{i} \citep{takeda02}.
Both of these species---and O,
discussed below---can only be detected in the MIKE spectra.
The non-LTE corrections for the Na~\textsc{i} 
lines at 5682, 5688, 6154, and 6160~\AA\ 
range from $-$0.02 to $-$0.08~dex.
Non-LTE corrections for the 
K~\textsc{i} line at 7698~\AA\ are $\approx -$0.48~dex.
We include an additional 0.1~dex statistical uncertainty
in our error estimates to account for uncertainties
in the corrections.
Non-LTE corrections for each of the
three high-excitation O~\textsc{i} 
7771, 7774, and 7775~\AA\ triplet lines
are expected to be substantial,
but the stellar parameters of stars 42009955 and 61005163
are far from the edge of the 
grid of corrections presented by \citet{fabbian09},
so we do not employ them.
We caution that the LTE O abundances derived
from these lines are likely overestimated,
and we only recommend their use in
a differential sense.
The [O~\textsc{i}] line at 6300~\AA,
which is expected to form under LTE,
is weak and cannot be reliably measured 
even after removing the telluric lines.

No C or N lines are detected in our M2FS spectra.
In the two MIKE spectra, 
C abundances are derived from the
CH $A^2\Delta - X^2\Pi$ G band ($\approx$~4290--4330~\AA)
using a line list
provided by B.\ Plez (2007, private communication).
N abundances are derived from the
CN $B^2\Sigma - X^2\Sigma$ violet band ($\approx$~3875--3885~\AA)
using the line list from \citet{kurucz95}
after setting the C abundance using the CH G band.
We use the spectrum synthesis matching technique 
to derive the C and N abundances.

\section{Results}
\label{results}

Table~\ref{m2fsabundtab} lists the mean abundances
derived for each element in each star observed with M2FS.~
Tables~\ref{9955mikemeantab} and \ref{5163mikemeantab}
list the mean abundances derived from the MIKE spectra
of stars 42009955 and 61005163.
The abundances in Tables~\ref{9955mikemeantab} and \ref{5163mikemeantab}
reflect the non-LTE corrections discussed in Section~\ref{analysis}.
We use standard definitions of elemental abundances and ratios
in these tables.
For element X, the logarithmic abundance is defined
as the number of X atoms per 10$^{12}$ hydrogen atoms,
$\log\epsilon$(X)~$\equiv \log_{10}(N_{\rm X}/N_{\rm H}) +$~12.0.
For elements X and Y, [X/Y] is 
the logarithmic abundance ratio relative to the solar ratio
on the \citet{asplund09} abundance scale,
defined as $\log_{10} (N_{\rm X}/N_{\rm Y}) -
\log_{10} (N_{\rm X}/N_{\rm Y})_{\odot}$, 
using like ionization states;
i.e., neutrals with neutrals and ions with ions.
Abundances or ratios denoted with the ionization state
indicate the total elemental abundance as derived from transitions of
that particular state.

\begin{table*}
\begin{minipage}{4.2in}
\caption{Mean Abundances in Individual Stars,
As Derived from the M2FS Spectra
\label{m2fsabundtab}}
\begin{tabular}{ccccccccc}
\hline
Species & 
Star & 
$N_{\rm lines}$ &
$\log\epsilon$ &
{[Fe~\textsc{i}/H]} &
$\sigma_{\rm stat}$ &
$\sigma_{\rm tot}$ &
$\sigma_{\rm neut}$ &
$\sigma_{\rm ions}$ \\
\hline
Fe~\textsc{i} & 11001198 & 18 &   5.11 & $-$2.39 &  0.04 &  0.19 & \ldots& \ldots \\
Fe~\textsc{i} & 12001300 & 18 &   5.28 & $-$2.22 &  0.04 &  0.19 & \ldots& \ldots \\
Fe~\textsc{i} & 21000267 & 22 &   5.16 & $-$2.34 &  0.04 &  0.19 & \ldots& \ldots \\
Fe~\textsc{i} & 21000688 & 17 &   5.11 & $-$2.39 &  0.04 &  0.19 & \ldots& \ldots \\
Fe~\textsc{i} & 41000940 & 16 &   5.09 & $-$2.41 &  0.05 &  0.19 & \ldots& \ldots \\
\hline
\end{tabular} \\
The complete version of Table~\ref{m2fsabundtab} 
is available in machine-readable format in the
Supplementary Information found in the online edition of the journal.
Only a small portion is shown here to illustrate its form and content. \\
\end{minipage}
\end{table*}

\begin{table*}
\begin{minipage}{3.8in}  
\caption{Mean Abundances in Star 42009955, As Derived from the MIKE Spectrum
\label{9955mikemeantab}}
\begin{tabular}{cccccccc}
\hline
Species &
$N_{\rm lines}$ &
$\log\epsilon$ &
[X/Fe]\footnote{[Fe/H] is given for Fe~\textsc{i} and Fe~\textsc{ii}} &
$\sigma_{\rm stat}$ &
$\sigma_{\rm tot}$ &
$\sigma_{\rm neut}$ &
$\sigma_{\rm ions}$ \\
\hline
 Fe~\textsc{i}  & 117 &  $+$5.11 & $-$2.39 &  0.06 &  0.19 &  0.00 &  0.00  \\
 Fe~\textsc{ii} &  10 &  $+$5.48 & $-$2.02 &  0.07 &  0.10 &  0.00 &  0.00  \\
 Li~\textsc{i}  &   1 & $<-$0.75 &$<-$1.62 & \ldots& \ldots& \ldots& \ldots \\
 C~(CH)         &   1 &  $+$5.25 & $-$1.16 &  0.15 &  0.25 &  0.18 &  0.18  \\ 
 N~(CN)         &   1 &  $+$7.20 & $+$1.39 &  0.25 &  0.32 &  0.27 &  0.27  \\
 O~\textsc{i}   &   3 & $<+$7.00 &$<+$0.70 & \ldots& \ldots& \ldots& \ldots \\
 Na~\textsc{i}  &   4 &  $+$4.40 & $+$0.55 &  0.13 &  0.22 &  0.14 &  0.20  \\
 Mg~\textsc{i}  &   4 &  $+$5.28 & $+$0.07 &  0.06 &  0.21 &  0.09 &  0.18  \\
 Al~\textsc{i}  &   4 &  $+$5.14 & $+$1.08 &  0.09 &  0.20 &  0.11 &  0.18  \\
 Si~\textsc{i}  &   3 &  $+$5.95 & $+$0.83 &  0.17 &  0.25 &  0.19 &  0.24  \\
 K~\textsc{i}   &   1 &  $+$2.90 & $+$0.26 &  0.12 &  0.25 &  0.15 &  0.23  \\
 Ca~\textsc{i}  &  12 &  $+$4.21 & $+$0.26 &  0.11 &  0.22 &  0.12 &  0.20  \\
 Sc~\textsc{ii} &   7 &  $+$1.27 & $+$0.14 &  0.05 &  0.09 &  0.17 &  0.09  \\
 Ti~\textsc{i}  &  17 &  $+$2.53 & $-$0.03 &  0.04 &  0.19 &  0.07 &  0.17  \\
 Ti~\textsc{ii} &  13 &  $+$3.27 & $+$0.33 &  0.05 &  0.10 &  0.16 &  0.09  \\
 V~\textsc{i}   &  10 &  $+$1.38 & $-$0.16 &  0.11 &  0.21 &  0.13 &  0.20  \\
 V~\textsc{ii}  &   1 &  $+$2.07 & $+$0.16 &  0.19 &  0.23 &  0.26 &  0.23  \\
 Cr~\textsc{i}  &  14 &  $+$2.96 & $-$0.29 &  0.05 &  0.20 &  0.08 &  0.18  \\
 Cr~\textsc{ii} &   4 &  $+$3.80 & $+$0.18 &  0.05 &  0.10 &  0.18 &  0.09  \\
 Mn~\textsc{i}  &   4 &  $+$2.64 & $-$0.40 &  0.07 &  0.19 &  0.09 &  0.17  \\
 Co~\textsc{i}  &   3 &  $+$2.74 & $+$0.14 &  0.17 &  0.25 &  0.18 &  0.24  \\
 Ni~\textsc{i}  &   7 &  $+$3.81 & $-$0.02 &  0.04 &  0.18 &  0.07 &  0.17  \\
 Cu~\textsc{i}  &   1 &  $+$1.28 & $-$0.52 &  0.08 &  0.20 &  0.10 &  0.18  \\
 Zn~\textsc{i}  &   2 &  $+$2.58 & $+$0.41 &  0.04 &  0.18 &  0.07 &  0.16  \\
 Rb~\textsc{i}  &   0 & $<+$1.45 &$<+$1.32 & \ldots& \ldots& \ldots& \ldots \\
 Sr~\textsc{ii} &   2 &  $+$0.89 & $+$0.04 &  0.17 &  0.20 &  0.19 &  0.20  \\
 Y~\textsc{ii}  &   7 &  $-$0.16 & $-$0.35 &  0.06 &  0.09 &  0.17 &  0.09  \\
 Zr~\textsc{ii} &   6 &  $+$0.77 & $+$0.21 &  0.06 &  0.10 &  0.17 &  0.09  \\
 Mo~\textsc{i}  &   1 &  $-$0.30 & $+$0.21 &  0.14 &  0.23 &  0.15 &  0.22  \\
 Ba~\textsc{ii} &   3 &  $-$0.14 & $-$0.31 &  0.06 &  0.11 &  0.16 &  0.10  \\
 La~\textsc{ii} &  11 &  $-$0.94 & $-$0.03 &  0.06 &  0.10 &  0.18 &  0.10  \\
 Ce~\textsc{ii} &  14 &  $-$0.63 & $-$0.19 &  0.05 &  0.10 &  0.18 &  0.09  \\
 Pr~\textsc{ii} &   3 &  $-$1.31 & $-$0.02 &  0.05 &  0.09 &  0.18 &  0.09  \\
 Nd~\textsc{ii} &  23 &  $-$0.63 & $-$0.03 &  0.06 &  0.10 &  0.18 &  0.10  \\
 Sm~\textsc{ii} &  15 &  $-$0.89 & $+$0.17 &  0.05 &  0.10 &  0.18 &  0.09  \\
 Eu~\textsc{ii} &   5 &  $-$1.26 & $+$0.24 &  0.08 &  0.11 &  0.17 &  0.10  \\
 Gd~\textsc{ii} &   1 &  $-$0.66 & $+$0.29 &  0.06 &  0.10 &  0.18 &  0.10  \\
 Tb~\textsc{ii} &   1 &  $-$1.70 & $+$0.02 &  0.11 &  0.14 &  0.21 &  0.14  \\
 Dy~\textsc{ii} &   3 &  $-$0.63 & $+$0.29 &  0.06 &  0.10 &  0.18 &  0.09  \\
 Er~\textsc{ii} &   1 &  $-$0.84 & $+$0.26 &  0.28 &  0.29 &  0.33 &  0.29  \\
 Hf~\textsc{ii} &   2 &  $-$0.76 & $+$0.41 &  0.13 &  0.16 &  0.22 &  0.15  \\
 Pb~\textsc{i}  &   1 &  $-$0.23 & $+$0.12 &  0.18 &  0.25 &  0.19 &  0.24  \\
 Th~\textsc{ii} &   1 & $<-$1.60 & $<+$0.36 & \ldots& \ldots& \ldots& \ldots \\
\hline
\end{tabular}
\end{minipage}
\end{table*}

\begin{table*}
\begin{minipage}{3.8in}  
\caption{Mean Abundances in Star 61005163, As Derived from the MIKE Spectrum
\label{5163mikemeantab}}
\begin{tabular}{cccccccc}
\hline
Species &
$N_{\rm lines}$ &
$\log\epsilon$ &
[X/Fe]\footnote{[Fe/H] is given for Fe~\textsc{i} and Fe~\textsc{ii}} &
$\sigma_{\rm stat}$ &
$\sigma_{\rm tot}$ &
$\sigma_{\rm neut}$ &
$\sigma_{\rm ions}$ \\
\hline
 Fe~\textsc{i}  & 124 &  $+$5.21 & $-$2.29 &  0.06 &  0.20 &  0.00 &  0.00  \\
 Fe~\textsc{ii} &  10 &  $+$5.58 & $-$1.92 &  0.07 &  0.10 &  0.00 &  0.00  \\
 Li~\textsc{i}  &   1 & $<-$0.48 &$<-$1.45 & \ldots& \ldots& \ldots& \ldots \\
 C~(CH)         &   1 &  $+$6.40 & $-$0.11 &  0.15 &  0.25 &  0.18 &  0.18  \\
 N~(CN)         &   1 &  $+$6.70 & $+$0.79 &  0.25 &  0.32 &  0.27 &  0.27  \\
 O~\textsc{i}   &   3 &  $+$7.89 & $+$1.50 &  0.11 &  0.21 &  0.13 &  0.19  \\
 Na~\textsc{i}  &   1 &  $+$3.92 & $-$0.03 &  0.13 &  0.22 &  0.15 &  0.21  \\
 Mg~\textsc{i}  &   4 &  $+$5.77 & $+$0.47 &  0.07 &  0.21 &  0.09 &  0.19  \\
 Al~\textsc{i}  &   3 & $<+$4.37 &$<+$0.21 & \ldots& \ldots& \ldots& \ldots \\
 Si~\textsc{i}  &   3 &  $+$6.10 & $+$0.89 &  0.17 &  0.24 &  0.18 &  0.23  \\
 K~\textsc{i}   &   1 &  $+$2.95 & $+$0.21 &  0.12 &  0.25 &  0.15 &  0.23  \\
 Ca~\textsc{i}  &  12 &  $+$4.32 & $+$0.27 &  0.11 &  0.22 &  0.12 &  0.20  \\
 Sc~\textsc{ii} &   7 &  $+$1.25 & $+$0.01 &  0.05 &  0.09 &  0.18 &  0.09  \\
 Ti~\textsc{i}  &  15 &  $+$2.62 & $-$0.04 &  0.04 &  0.20 &  0.08 &  0.18  \\
 Ti~\textsc{ii} &  13 &  $+$3.35 & $+$0.32 &  0.05 &  0.10 &  0.17 &  0.09  \\
 V~\textsc{i}   &  10 &  $+$1.50 & $-$0.13 &  0.11 &  0.21 &  0.13 &  0.20  \\
 V~\textsc{ii}  &   0 &   \ldots &   \ldots& \ldots& \ldots& \ldots& \ldots \\
 Cr~\textsc{i}  &  14 &  $+$3.11 & $-$0.24 &  0.04 &  0.20 &  0.08 &  0.18  \\
 Cr~\textsc{ii} &   3 &  $+$3.82 & $+$0.10 &  0.05 &  0.09 &  0.18 &  0.09  \\
 Mn~\textsc{i}  &   4 &  $+$2.70 & $-$0.44 &  0.06 &  0.19 &  0.09 &  0.17  \\
 Co~\textsc{i}  &   3 &  $+$2.71 & $+$0.02 &  0.18 &  0.25 &  0.19 &  0.24  \\
 Ni~\textsc{i}  &   7 &  $+$3.92 & $-$0.00 &  0.04 &  0.18 &  0.08 &  0.16  \\
 Cu~\textsc{i}  &   1 &  $+$1.31 & $-$0.59 &  0.08 &  0.21 &  0.10 &  0.19  \\
 Zn~\textsc{i}  &   2 &  $+$2.71 & $+$0.45 &  0.04 &  0.18 &  0.07 &  0.16  \\
 Rb~\textsc{i}  &   0 & $<+$1.25 &$<+$1.02 & \ldots& \ldots& \ldots& \ldots \\
 Sr~\textsc{ii} &   1 &  $+$1.08 & $+$0.13 &  0.16 &  0.19 &  0.19 &  0.19  \\
 Y~\textsc{ii}  &   8 &  $+$0.13 & $-$0.16 &  0.06 &  0.09 &  0.17 &  0.09  \\
 Zr~\textsc{ii} &   6 &  $+$0.87 & $+$0.21 &  0.06 &  0.09 &  0.17 &  0.09  \\
 Mo~\textsc{i}  &   0 &   \ldots & \ldots  & \ldots& \ldots& \ldots& \ldots \\
 Ba~\textsc{ii} &   3 &  $+$0.83 & $+$0.56 &  0.05 &  0.12 &  0.14 &  0.12  \\
 La~\textsc{ii} &  17 &  $-$0.35 & $+$0.46 &  0.05 &  0.09 &  0.17 &  0.08  \\
 Ce~\textsc{ii} &  18 &  $+$0.13 & $+$0.47 &  0.05 &  0.09 &  0.17 &  0.08  \\
 Pr~\textsc{ii} &   4 &  $-$0.78 & $+$0.42 &  0.07 &  0.10 &  0.18 &  0.09  \\
 Nd~\textsc{ii} &  26 &  $-$0.02 & $+$0.47 &  0.05 &  0.09 &  0.17 &  0.08  \\
 Sm~\textsc{ii} &  16 &  $-$0.59 & $+$0.37 &  0.04 &  0.09 &  0.18 &  0.09  \\
 Eu~\textsc{ii} &   5 &  $-$1.21 & $+$0.18 &  0.08 &  0.11 &  0.18 &  0.10  \\
 Gd~\textsc{ii} &   2 &  $-$0.37 & $+$0.48 &  0.05 &  0.09 &  0.18 &  0.09  \\
 Tb~\textsc{ii} &   1 &  $-$1.55 & $+$0.07 &  0.11 &  0.14 &  0.21 &  0.14  \\
 Dy~\textsc{ii} &   4 &  $-$0.31 & $+$0.51 &  0.07 &  0.10 &  0.18 &  0.10  \\
 Er~\textsc{ii} &   1 &  $-$0.45 & $+$0.55 &  0.28 &  0.29 &  0.33 &  0.29  \\
 Hf~\textsc{ii} &   2 &  $-$0.58 & $+$0.49 &  0.10 &  0.13 &  0.20 &  0.12  \\
 Pb~\textsc{i}  &   1 &  $+$0.76 & $+$1.01 &  0.06 &  0.23 &  0.12 &  0.21  \\
 Th~\textsc{ii} &   0 &   \ldots & \ldots  & \ldots& \ldots& \ldots& \ldots \\
\hline
\end{tabular}
\end{minipage}
\end{table*}

Four sets of uncertainties are listed in these tables.
The statistical uncertainty, $\sigma_{\rm stat}$, 
is given by equation~A17 of \citet{mcwilliam95}.
This includes uncertainties in the EW,
line profile fitting, \loggf\ values, and non-LTE corrections, if any.
The total uncertainty, $\sigma_{\rm tot}$, 
is given by equation~A16 of \citeauthor{mcwilliam95} 
This includes the statistical uncertainty and 
reflects uncertainties in the model atmosphere parameters.
The other two uncertainties,
$\sigma_{\rm neut}$ and $\sigma_{\rm ions}$,
are useful when constructing abundance ratios among different elements.
We recommend that $\sigma_{\rm neut}$ for element A 
be added in quadrature with $\sigma_{\rm stat}$ for
element B when computing the ratio [A/B] when B is 
derived from lines of the neutral species.
Similarly, we recommend using
$\sigma_{\rm ions}$ instead of $\sigma_{\rm neut}$
when element B 
is derived from lines of the ionized species.
These latter two sets of uncertainties are
omitted for the [Fe/H] ratios.

Table~\ref{meantab} reports the weighted mean abundances
of all elements studied in the M2FS spectra of \ngc.
Star 61005163, whose anomalous abundance pattern
will be discussed in great detail in later sections, 
is omitted from the means presented in Table~\ref{meantab}.

\begin{table}
\begin{minipage}{3.25in}  
\caption{Mean Abundances in NGC~5824 Determined from M2FS Spectra
\label{meantab}}
\begin{tabular}{cccccc}
\hline
Ratio &
Species &
Mean &
Std.\ err.\ &
Std.\ dev.\ &
N$_{\rm stars}$ \\
\hline
{[Fe/H]}  &  \textsc{i} & $-$2.38 & 0.02 & 0.08 & 25 \\
{[Fe/H]}  & \textsc{ii} & $-$1.94 & 0.02 & 0.08 & 25 \\
{[Mg/Fe]} &  \textsc{i} & $-$0.20 & 0.06 & 0.28 & 25 \\
{[Ca/Fe]} &  \textsc{i} & $+$0.20 & 0.02 & 0.10 & 25 \\
{[Sc/Fe]} & \textsc{ii} & $-$0.02 & 0.04 & 0.15 & 15 \\
{[Ti/Fe]} &  \textsc{i} & $-$0.04 & 0.02 & 0.09 & 25 \\
{[Ti/Fe]} & \textsc{ii} & $+$0.29 & 0.02 & 0.08 & 25 \\
{[V/Fe]}  &  \textsc{i} & $-$0.26 & 0.03 & 0.14 & 24 \\
{[Cr/Fe]} &  \textsc{i} & $-$0.25 & 0.02 & 0.08 & 25 \\
{[Cr/Fe]} & \textsc{ii} & $+$0.13 & 0.03 & 0.15 & 25 \\
{[Mn/Fe]} &  \textsc{i} & $-$0.38 & 0.06 & 0.16 &  7 \\
{[Ni/Fe]} &  \textsc{i} & $-$0.11 & 0.03 & 0.14 & 25 \\
{[Sr/Fe]} &  \textsc{i} & $-$0.36 & 0.04 & 0.17 & 22 \\
{[Ba/Fe]} & \textsc{ii} & $-$0.55 & 0.05 & 0.24 & 25 \\
{[La/Fe]} & \textsc{ii} & $-$0.14 & 0.03 & 0.11 & 12 \\
{[Ce/Fe]} & \textsc{ii} & $-$0.19 & 0.02 & 0.10 & 25 \\
{[Nd/Fe]} & \textsc{ii} & $-$0.04 & 0.03 & 0.15 & 25 \\
{[Sm/Fe]} & \textsc{ii} & $+$0.12 & 0.03 & 0.16 & 24 \\
{[Eu/Fe]} & \textsc{ii} & $+$0.11 & 0.02 & 0.12 & 25 \\
{[Dy/Fe]} & \textsc{ii} & $+$0.16 & 0.04 & 0.17 & 22 \\
\hline
\end{tabular} \\
Note---Star \mbox{61005163} has been excluded 
from the data presented in this table.
\end{minipage}
\end{table}

\subsection{Comparing Abundances Derived from M2FS and MIKE Spectra}
\label{comparem2fsmike}

\begin{figure}
\centering
\includegraphics[angle=0,width=3.25in]{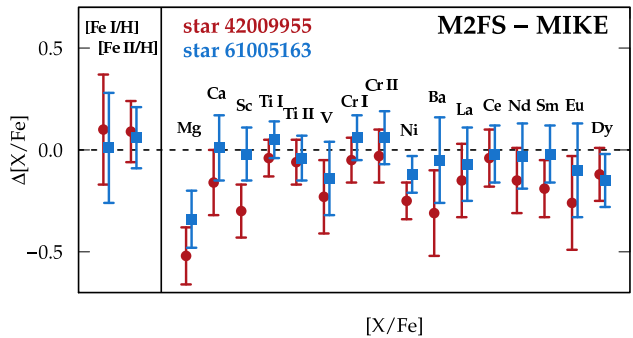}
\caption{
\label{comparemplot}
Comparison of abundance ratios derived from M2FS and MIKE
spectra for the two stars in common.
Red circles indicate star 42009955, and 
blue squares indicate star 61005163.
The dotted line marks a difference of zero.
}
\end{figure}

Figure~\ref{comparemplot} compares the abundance
ratios derived from M2FS and MIKE spectra for the 
two stars in common, 42009955 and 61005163.
In general, there is superb agreement.
The most significant disagreement occurs for the [Mg/Fe]
ratio, where the M2FS ratio is lower by 0.43~$\pm$~0.16~dex
than the MIKE ratio, on average.
Only one Mg~\textsc{i} line, at 4571.10~\AA, 
is covered in the M2FS spectra.
This line originates from the ground level of the Mg~\textsc{i} atom,
whereas all other Mg~\textsc{i} lines typically studied
in metal-poor giants originate from levels
at $\approx$~2.7 or $\approx$~4.3~eV
(see, e.g., Table~8 of \citealt{roederer14a}).
The direction of the offset we identify is consistent with
expectations that the ground level of Mg~\textsc{i} 
may experience substantial overionization relative to the 
LTE level populations \citep{asplund05}.
The non-LTE calculations of \citet{shimanskaya00} 
confirm this, although 
they do not extend to stars cool enough to 
justify applying corrections to our data.
We derive Mg abundances from three other Mg~\textsc{i} lines
in the MIKE spectra.
The Mg~\textsc{i} line at 4571~\AA\ also gives abundances
lower than these other three lines by 0.41~dex, on average,
confirming that our M2FS measurements are not uniquely in error.
We caution that the [Mg/Fe] ratios reported in Tables~\ref{m2fsabundtab} and
\ref{meantab} are likely to be lower by $\approx$~0.4~dex
than [Mg/Fe] ratios derived from other Mg~\textsc{i} lines.

The [Ni/Fe] ratios are both mildly different in these two 
sets of spectra, where the M2FS ratio is lower than the MIKE
ratio by 0.19~$\pm$~0.12~dex.
Two Ni~\textsc{i} lines, at 4470.48 and 4604.99~\AA,
have been used in both sets of spectra 
of both stars.
The abundances
derived from them agree to 0.07~dex or better.
Abundances derived from these lines in the MIKE spectra
are lower by 0.18~dex, on average, than the abundances derived from
the five other Ni~\textsc{i} lines in each star.
This indicates that our M2FS Ni abundance determinations are themselves
not in error, just as we have found in the case of Mg.
No non-LTE calculations are available for Ni lines,
to the best of our knowledge;
however, the results of \citet{wood14ni} 
suggest that any non-LTE effects on neutral Ni are minimal,
at least in metal-poor subgiants.
The cause of the Ni offset is unclear at present.
We caution that the [Ni/Fe] ratios reported in 
Tables~\ref{m2fsabundtab} and \ref{meantab} 
are likely to be lower by $\approx$~0.2~dex
than [Ni/Fe] ratios derived from other Ni~\textsc{i} lines.

\section{The Metallicity of NGC~5824 and the
Possibility of Internal Dispersion}
\label{metallicity}

\subsection{Comparison to Da Costa et al.\ Metallicities}
\label{comparedacosta}

We compare our derived metallicities, inferred from
individual Fe~\textsc{ii} lines in 26 stars in \ngc,
with those calculated by \citet{dacosta14},
inferred from a calibration of the Ca~\textsc{ii} 
near-infrared triplet
EWs.
\citeauthor{dacosta14}\ did not report metallicities for individual
stars, so we follow their procedure 
using the calibration given by \citet{saviane12}.
We estimate uncertainties on the \citeauthor{dacosta14}\ metallicities
using a Monte Carlo approach.

\begin{figure}
\centering
\includegraphics[angle=0,width=3.25in]{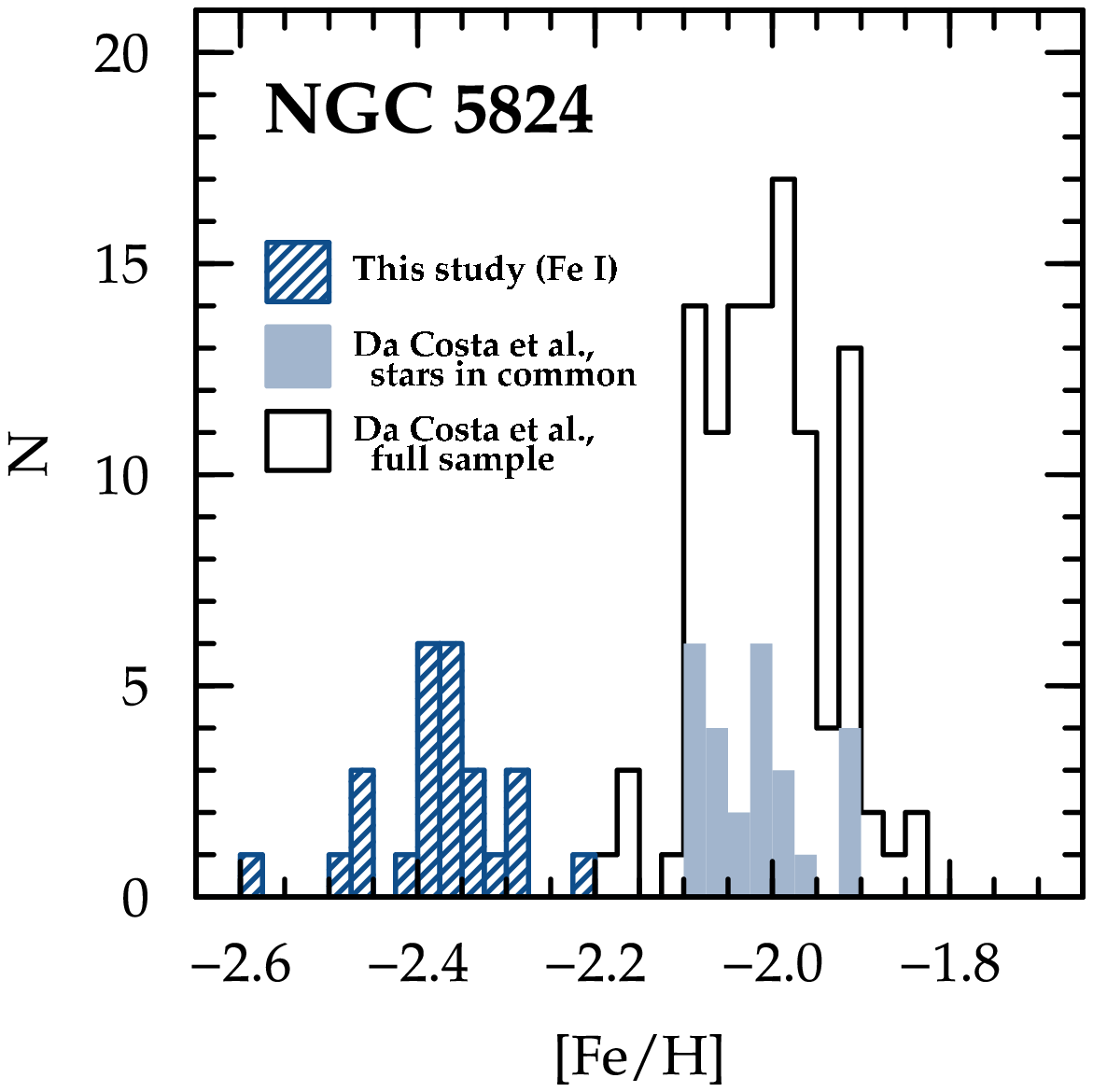} \\
\vspace*{0.1in}
\includegraphics[angle=0,width=3.25in]{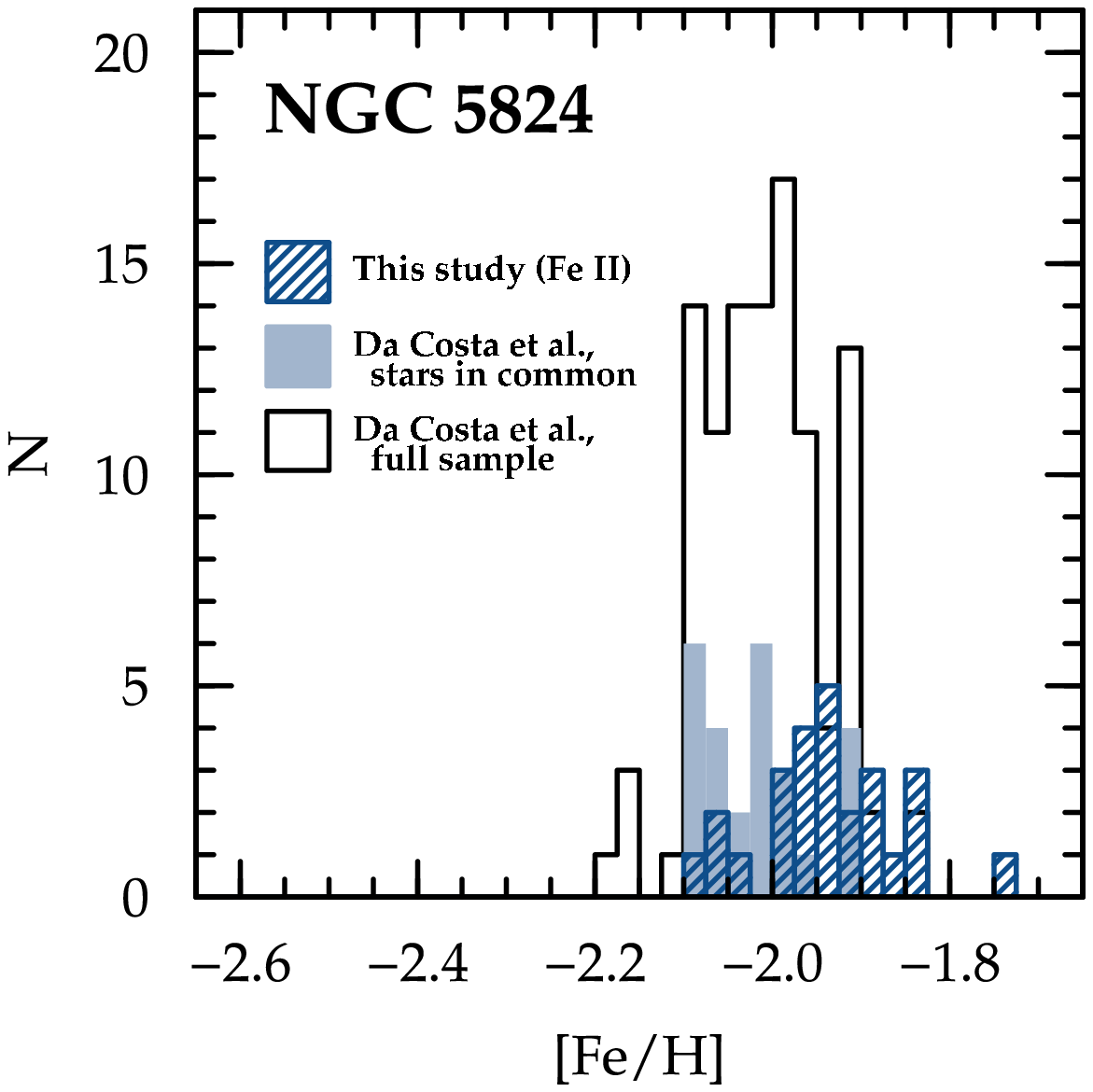}
\caption{
\label{mdfplot}
The metallicity distribution function of \ngc.
The dark blue hatched histogram represents metallicities
derived by us using 
Fe~\textsc{i} lines (top panel) or
Fe~\textsc{ii} lines (bottom panel) in 26~stars.
The unfilled black histogram represents
metallicities for the full sample of 108~stars
examined by \citet{dacosta14}.
The filled light blue histogram represents
the subset of 26~stars from \citeauthor{dacosta14}\
that are common to our study.
}
\end{figure}

Figure~\ref{mdfplot} illustrates the metallicity distribution
for stars in \ngc.
The hatched dark blue histogram represents metallicities
derived by us.
The black line represents the full sample of 108~stars
examined by \citet{dacosta14},
and the light blue histogram represents the subset of 26~stars
in common to both studies.
Our mean metallicities derived from Fe~\textsc{i} and Fe~\textsc{ii} are
different by 0.44~dex, but both have a
standard deviation of 0.08~dex.
The metallicities derived from Fe~\textsc{ii} lines
are in much better agreement with those of 
\citeauthor{dacosta14}\ and an external metallicity scale
(Section~\ref{koch}), so we prefer these values.

A small offset in [Fe/H] is present between our work and 
that of \citet{dacosta14}.
For the 26~stars in common, our [Fe/H] values 
derived from Fe~\textsc{ii} lines
are higher
by 0.08~$\pm$~0.02~dex.
Part of this offset may be explained by the 
slightly lower value of [Ca/Fe]
that we have derived for \ngc\
relative to the other metal-poor clusters that
\citet{saviane12} used to derive their calibration.
We find [Ca/Fe]~$= +$0.20~$\pm$~0.02,
while Gratton et al. (\citeyear{gratton04}) reported a mean
of [Ca/Fe]~$= +$0.25~$\pm$~0.02 for 28 clusters
with [Fe/H]~$< -$~1, including many 
used by \citeauthor{saviane12} 
\citeauthor{dacosta14}\ note that the uncertainty in the
mean metallicity derived
for \ngc\ using their Ca triplet calibration,
[Fe/H]~$= -$2.01~$\pm$~0.13, 
is dominated by the calibration relation itself.

\subsection{Tests of Internal Metallicity Dispersion}
\label{spread}

\begin{figure}
\centering
\includegraphics[angle=0,width=3.25in]{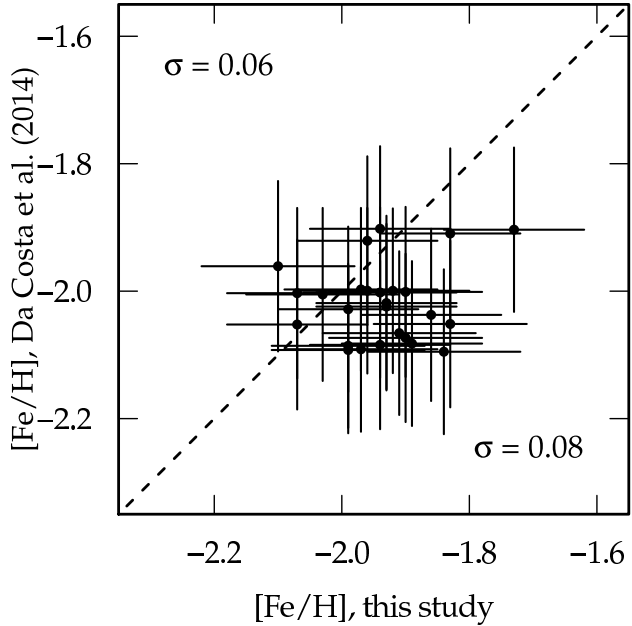}
\caption{
\label{dacostafeplot}
Comparison of metallicities derived in our study
(using Fe~\textsc{ii} lines) with those
inferred from Ca~\textsc{ii} triplet EWs
measured by \citet{dacosta14}.
The dashed line represents a one-to-one correlation.
}
\end{figure}

Figure~\ref{dacostafeplot} compares our results
with those of \citet{dacosta14}.
The metallicities derived by \citeauthor{dacosta14}\ and us
are inferred from independent spectral features.
We might expect to find a correlation in Figure~\ref{dacostafeplot}
if [Ca/Fe] is relatively constant 
and there is an intrinsic metallicity dispersion within \ngc\
that exceeds the observational uncertainties.
The $p$-value for the linear correlation coefficient 
is 0.52, affirming our visual 
impression that no significant correlation is found.

\begin{figure}
\centering
\includegraphics[angle=0,width=3.25in]{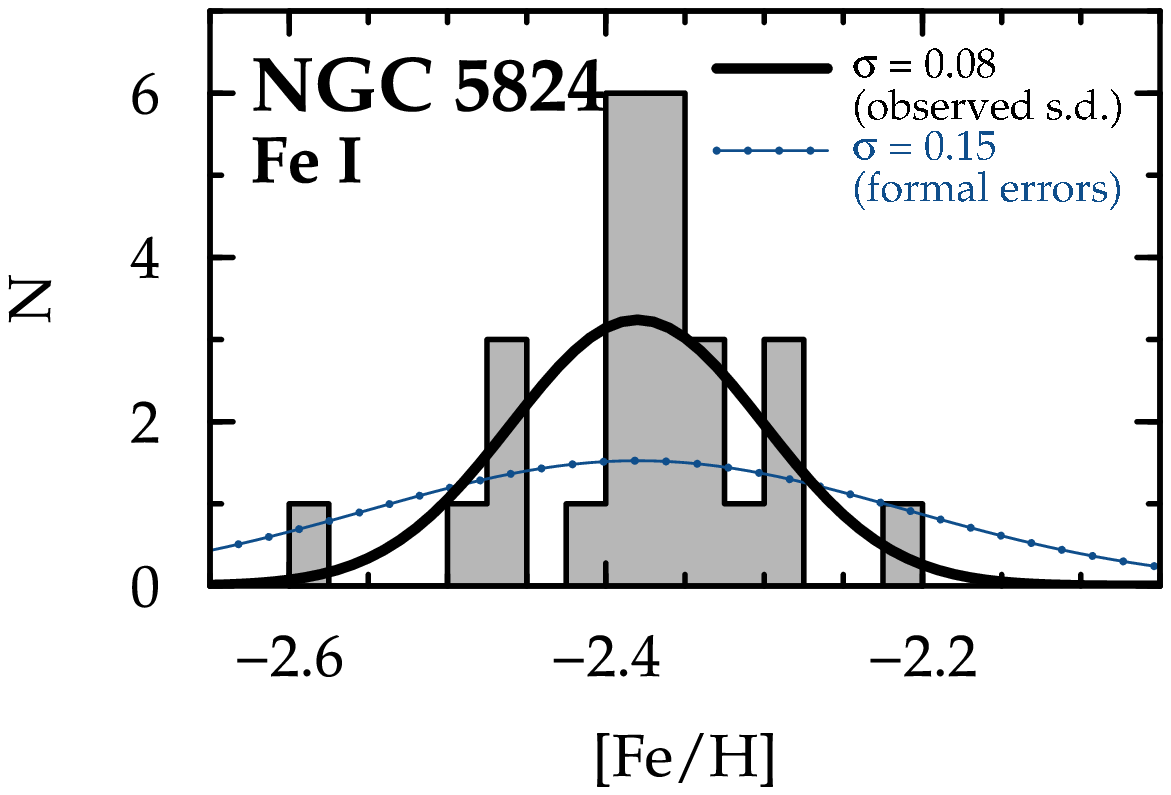} \\
\vspace*{0.1in}
\includegraphics[angle=0,width=3.25in]{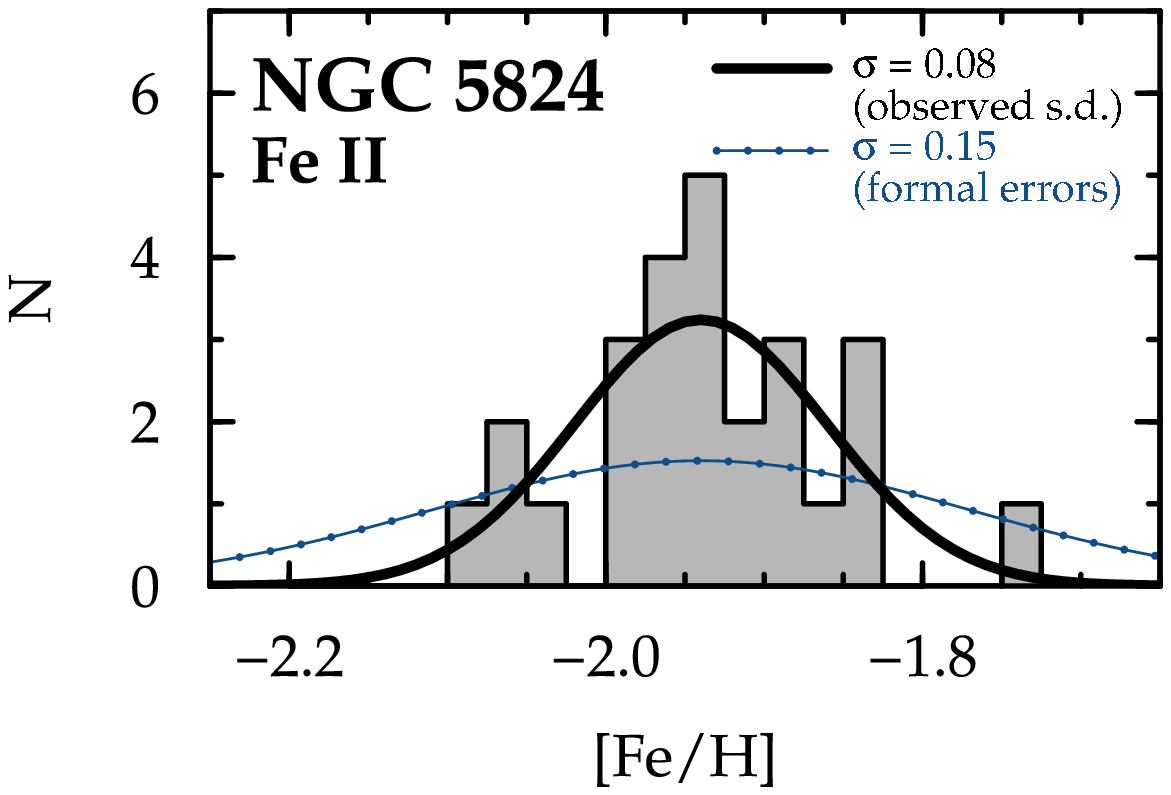}
\caption{
\label{altmdfplot}
Observed metallicity distributions (gray shaded histograms) in \ngc.
The top panel shows the results from Fe~\textsc{i} lines, and 
the bottom panel shows the results from Fe~\textsc{ii} lines. 
The bold black line represents a normal distribution
with mean 
[Fe/H]~$= -$2.38 (Fe~\textsc{i}) or
[Fe/H]~$= -$1.94 (Fe~\textsc{ii})
and standard deviation 0.08,
which is equivalent to the standard deviation of the
derived metallicities for both Fe~\textsc{i} and \textsc{ii}.
The studded blue line represents a normal distribution
with mean 
[Fe/H]~$= -$2.38 (Fe~\textsc{i}) or
[Fe/H]~$= -$1.94 (Fe~\textsc{ii}) 
and standard deviation 0.15,
which is equivalent to the formal error distributions
predicted for our data.
All distributions are normalized to the areas of the observed histograms.
}
\end{figure}

Figure~\ref{altmdfplot} illustrates our observed metallicity distributions
in \ngc, which are equivalent to the hatched blue histograms
in Figure~\ref{mdfplot}.
Our metallicities
have standard deviations of 
0.08~dex derived from either 
Fe~\textsc{i} or Fe~\textsc{ii} lines.
The black curves in Figure~\ref{altmdfplot} illustrate
normal distributions with a mean metallicity of $-$2.38 (for Fe~\textsc{i}) 
or $-$1.94 (for Fe~\textsc{ii})
and standard deviations of 0.08~dex.
We estimate the formal widths of the [Fe/H]
distributions that would be expected
given errors in the input quantities used to calculate
\teff\ and \logg.
We assume $\sigma_{\teff} =$~145~K,
$\sigma_{\logg} =$~0.14~dex, and
$\sigma_{\vt} =$~0.1~\kmsec,
as estimated in Section~\ref{atmosphere}.
Most of the errors on the other input quantities in the \logg\ calculation in
Section~\ref{atmosphere} are systematic,
so they affect all stars similarly.
We ignore the errors in these quantities for now.
For a given star,
we use 1000 realizations of model atmosphere parameters
drawn from normal distributions with these standard deviations,
and we rederive the [Fe/H] ratios. 
The standard deviations of the resulting [Fe/H] distributions
vary somewhat with \teff, but the median values are $\approx$~0.15~dex
for both Fe~\textsc{i} and \textsc{ii}.
This value is illustrated by the blue curves in Figure~\ref{altmdfplot}.

It is apparent from Figure~\ref{altmdfplot} that the
expected distributions overestimate the observed widths
by about a factor of two.
This likely signals that we have overestimated the 
errors on the model atmosphere parameters.
The largest source of error that contributes to the 
width in the predicted [Fe/H] distribution is $\sigma_{\teff}$,
which enters the calculation twice:\
once through \teff\ itself, and once through 4$\log \teff$
in the \logg\ formula.
The \citet{alonso99b} $V-I$ color-\teff\ calibration has an intrinsic
scatter of 125~K.
If we arbitrarily assume zero scatter in this calibration,
$\sigma_{\teff}$ decreases to 74~K
(reflecting quantities related to the photometry),
and $\sigma_{\rm [Fe/H]}$
decreases to 0.09~dex,
close to the observed value.

We compare the results with another \teff\ scale
to assess whether this arbitrary assumption is plausible.
The $V-I_{c}$ color-\teff\ relation of \citet{casagrande10}
has an intrinsic scatter of only 59~K.
This scale was calibrated using dwarfs,
but \citet{casagrande14} have shown that it is also 
applicable to giants.
The \citeauthor{casagrande10}\ calibration
predicts \teff\ values warmer than the \citeauthor{alonso99b}\
scale by 85~K for these stars.
The standard deviation
of the residuals between the two scales is only 3~K, however.
The excellent agreement of these two scales---zeropoint 
aside---offers reassurance that 
the statistical errors on \teff\ may be smaller
than assumed.

Our tests indicate that
no reasonable assumptions for the errors in \teff, \logg, or \vt\
can produce a predicted [Fe/H] 
distribution 
for either Fe~\textsc{i} or \textsc{ii}
narrower than the observed one.
Furthermore, we have assumed during these calculations
that the measurement errors on the EWs
are zero, which also would lead us to underestimate
the widths of the predicted [Fe/H] distributions.
We conclude that there is no internal
metallicity dispersion among the stars observed in \ngc.
We estimate a systematic uncertainty of 0.10~dex
in the mean metallicity
when sources of systematic uncertainty are considered.

We do not find evidence in our data
for the metallicity spread reported by \citet{dacosta14},
but that study would also not have found evidence for
a metallicity spread if they were restricted to the
26~stars studied by us.
The dispersion in their metallicities of these 26~stars,
0.06~dex, is equivalent to the expected error distribution.
This value is
smaller than the
FWHM of the distribution shown in Figure~12 of \citeauthor{dacosta14},
0.16~dex, 
or the inner-quartile range, 0.10~dex,
reported based on their full sample of 108~stars.

\begin{figure}
\centering
\includegraphics[angle=0,width=3.25in]{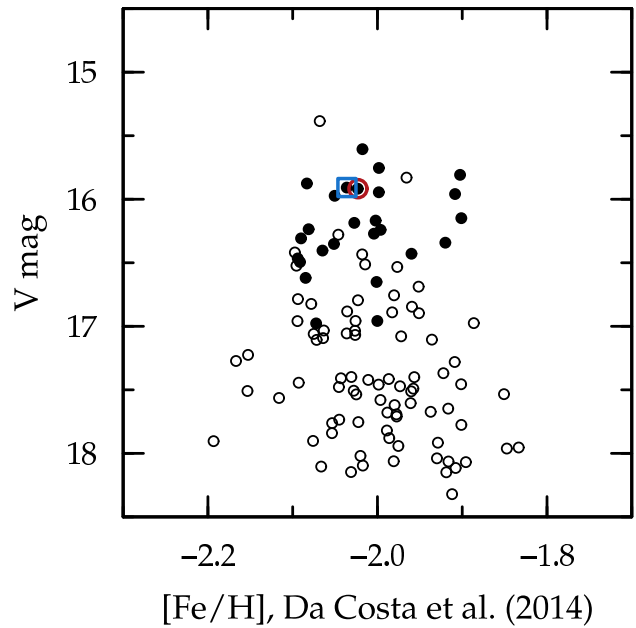}
\caption{
\label{otherdacostaplot}
Metallicities for the \citet{dacosta14} sample
inferred from the Ca~\textsc{ii} triplet EWs.
Open circles denote all 108~stars in their sample,
and filled circles mark the 26~stars in common with us.
The red circle marks star 42009955, and
the blue square marks star 61005163.
}
\end{figure}

Figure~\ref{otherdacostaplot} illustrates
the \citeauthor{dacosta14}\ metallicities
as a function of $V$ magnitude.
Open circles represent their full sample,
and filled circles represent stars in
common with our sample.
Stars that contribute to the 
tails of their metallicity distribution 
have $V >$~17,
where we have no stars in common.
One explanation is that this simply reflects
the increasing measurement errors for fainter stars.
If, however, \ngc\ has an intrinsic metallicity spread, 
the stars with higher metallicities may not 
occupy the upper RGB, and
our observations would be biased against them.
Stars with lower metallicities would generally not be fainter
and should be found in our sample, so
this explanation may only tell part of the story.
The radial distributions of our sample and the full \citeauthor{dacosta14}\
sample are only mildly different, with the \citeauthor{dacosta14}\
sample being slightly more extended than ours.
The two-sided Kolmogorov-Smirnov test returns a $p$-value of
0.12
when applied to the cumulative radial distributions of these samples,
and the Cramer-von Mises test returns a $p$-value of 0.044.
If lower metallicity stars are found at greater radii, on average,
this could explain the tail to lower metallicities.

\begin{figure}
\centering
\includegraphics[angle=0,width=3.25in]{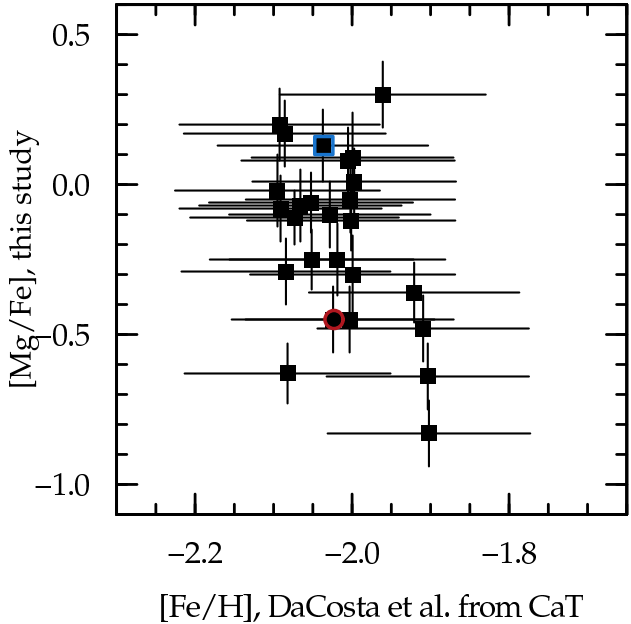}
\caption{
\label{camgplot}
Our [Mg/Fe] ratios as a function of the
[Fe/H] inferred from the Ca~\textsc{ii} triplet (CaT)
measurements of \citet{dacosta14}.
The red circle marks star 42009955, and
the blue square marks star 61005163.
}
\end{figure}

Another partial
explanation for the range of Ca~\textsc{ii}
triplet strengths found by \citet{dacosta14}
could be an opacity effect 
caused by the Mg-deficient stars.
\citet{mucciarelli12} noted that the most 
Mg-deficient stars in \mbox{NGC~2419}
were preferentially those with the highest [Ca/H] ratios.
Mg is one of the dominant donors of electrons 
to form the H$^{-}$ ion, the primary source of
continuous opacity in late-type metal-poor stars.
\citeauthor{mucciarelli12}\ contend that
the continuous opacity decreases when Mg is depleted, 
leading to stronger Ca~\textsc{ii} lines
even though the Ca abundance is constant.
The saturated Ca~\textsc{ii} triplet lines
are partially formed in layers of the atmosphere
most sensitive to the electron pressure (see Figure~7 of
\citeauthor{mucciarelli12}),
whereas lines on the weak
part of the curve of growth are minimally-affected.
Figure~\ref{camgplot} compares the \citeauthor{dacosta14}\
metallicities derived from Ca~\textsc{ii} triplet
lines with the [Mg/Fe] ratios derived by us.
We estimate the significance of the correlation in 
Figure~\ref{camgplot} by calculating the
$p$-values for the Spearman and Pearson coefficients
from 1000 resamplings of the [Mg/Fe] and 
Ca~\textsc{ii} triplet metallicity distributions 
given their observed errors.
The median $p$-values are 0.40 and 0.37, respectively,
indicating that the correlation is not significant.
Regardless, depletions in Mg could not fully explain the situation,
as noted in Section~\ref{introduction}.
\citet{dacosta09} discovered an intrinsic
dispersion in the strengths of the Ca~\textsc{ii} triplet
in M22 red giants,
similar to that found in \ngc.
However, no significant depletions
in [Mg/Fe] or other electron donors
are found in M22 \citep{marino11}.
No satisfactory explanation has been found
for the \citeauthor{dacosta09}\ result in M22,
and the possibility of a metallicity spread in \ngc\
is unresolved by our data.

\subsection{A Differential Analysis of Stars 42009955 and 61005163}
\label{diff}

We use a differential approach to examine whether there may be
abundance variations between the two stars observed with MIKE,
42009955 and 61005163.
These two stars are highlighted in
Figures~\ref{otherdacostaplot} and \ref{camgplot}.
There are 101~Fe~\textsc{i} lines in common,
10~Fe~\textsc{ii} lines, 
12~Ca~\textsc{i} lines,
14~Ti~\textsc{i} lines,
13~Ti~\textsc{ii} lines,
14~Cr~\textsc{i} lines,
and
7~Ni~\textsc{i} lines.
The line-by-line differentials, in the sense of
star 42009955 minus star 61005163, are
$\delta$[Ca/Fe]~$=-$0.020~$\pm$~0.027,
$\delta$[Ti~\textsc{i}/Fe]~$=-$0.036~$\pm$~0.039,
$\delta$[Ti~\textsc{ii}/Fe]~$=-$0.038~$\pm$~0.061,
$\delta$[Cr/Fe]~$=+$0.008~$\pm$~0.023
$\delta$[Ni/Fe]~$=+$0.007~$\pm$~0.021.
All of the errors reported here are statistical only, and
none of these differences are significant.
These ratios are also robust and consistent
with no difference when Fe is removed from the construction:\
$\delta$[Ca/Ni]~$= -$0.028~$\pm$~0.031, for example.
We conclude that the abundances 
of the Fe-group elements (including Ca and Ti)
are constant in these two stars.

The overall metallicities are significantly different
when considering Fe~\textsc{i} lines,
$\delta$[Fe~\textsc{i}/H]~$= -$0.129~$\pm$~0.010, 
but they are not significantly different when considering 
Fe~\textsc{ii} lines,
$\delta$[Fe~\textsc{ii}/H]~$= -$0.059~$\pm$~0.057.
As noted previously, other studies
(e.g., \citealt{mucciarelli15b})
have pointed out that Fe~\textsc{i} lines 
may be a poor representation of the 
Fe abundance in cool, metal-poor giants
in globular clusters.
Whatever the cause of this effect,
\citeauthor{mucciarelli15b}\ have shown that
it may be most pronounced
in stars that show the \spro\ enhancement,
like star 61005163 (see Section~\ref{ncapture}).
Fe~\textsc{ii} lines may be more reliable.
We conclude that this differential analysis
does not offer compelling evidence for
metallicity differences between stars
42009955 and 61005163.

Our referee notes, correctly, that this technique may not
be strictly valid if the He abundance or 
total abundance of C, N, and O
differs between these two stars.
\citet{sbordone11} have shown that
stars with the same color but differing He or 
total C, N, and O abundances
will separate into different sequences using 
broadband optical photometry.
Our method for deriving \logg\ would then need to be
refined for the additional sequence.
We cannot evaluate the He abundances
or total abundance of C, N, and O in these two stars, and
broadband optical photometry has not yet
revealed multiple sequences on the red giant branch of
\ngc\ \citep{piotto02}.
Evidence suggests, however, that \ngc\ may be a good
candidate to search for the presence of He variations
(Section~\ref{light}), 
which have been detected in every cluster investigated \citep{milone15a},
so this matter is unresolved at present.

\subsection{The Metallicity of NGC 5824 on the Koch \& McWilliam Scale}
\label{koch}

\citet{koch08} have established a globular cluster metallicity
scale based on a differential abundance analysis
with the K-giant Arcturus ($\alpha$~Boo),
whose temperature is known to better than 30~K.
\citet{koch11} placed the metal-poor cluster
\mbox{NGC~6397} on this scale, finding
[Fe/H]~$= -$2.10~$\pm$~0.02 (statistical) $\pm$~0.07 (systematic).
For three unblended 
Fe~\textsc{ii} lines in common between three stars in \mbox{NGC~6397}
(stars 7230, 8958, and 13414) and our MIKE spectra of
stars 42009955 and 61005163,
we find a mean line-by-line offset of $+$0.23~$\pm$~0.04~dex
in the sense of \ngc\ minus \mbox{NGC~6397}.
Relative to the \citeauthor{koch11} metallicity of \mbox{NGC~6397},
\ngc\ has a metallicity of 
[Fe/H]~$= -$1.87~$\pm$~0.04 (statistical) $\pm$~0.07 (systematic).
This is in reasonable agreement with the metallicity we
have derived,
[Fe/H]~$= -$1.94~$\pm$~0.02 (statistical) $\pm$~0.10 (systematic).

\section{Discussion}
\label{discussion}

\subsection{Light Elements}
\label{light}

Mg is the only light element that is sometimes found
to vary within globular clusters that 
is covered by our M2FS spectra.
Nevertheless, the dispersion in the [Mg/Fe] ratios,
0.28~dex, far exceeds that of any other $\alpha$- or Fe-group element.
The inner quartile range of [Mg/Fe] ratios is 0.35~dex,
and the full range is 1.1~dex.
Recall (Section~\ref{comparem2fsmike})
that the [Mg/Fe] ratios derived from the M2FS spectra
may be systematically underestimated by $\approx$~0.4~dex,
but the star-to-star dispersion should be 
reliable.

Large star-to-star variations in [Mg/Fe] 
have been found in some luminous, metal-poor clusters, including
\mbox{$\omega$~Cen} (e.g., \citealt{norris95,smith00}),
M3 \citep{sneden04,johnson05},
M13 (\citealt{pilachowski96,shetrone96,kraft97}),
M15 \citep{sneden97,carretta09b},
\mbox{NGC~2419} \citep{cohen12,mucciarelli12},
\mbox{NGC~2808} \citep{carretta09b,carretta14b},
\mbox{NGC~4833} \citep{carretta14,roederer15},
\mbox{NGC~6752} \citep{gratton01,yong03},
and possibly
M62 \citep{yong14a} and
M92 \citep{shetrone96}.
Some of these clusters are also suspected to contain
groups of stars with substantial He enhancements
(\mbox{$\omega$~Cen}:\ \citealt{piotto05}, \citealt{king12};
M62:\ \citealt{milone15a};
\mbox{NGC~2419}:\ \citealt{dicriscienzo11};
\mbox{NGC~2808}:\ \citealt{piotto07}, \citealt{milone15b};
\mbox{NGC~6752}:\ \citealt{milone13}).
\ngc, too, is luminous, metal-poor, and shows a large
internal [Mg/Fe] dispersion.
The relatively large dispersion in [Mg/Fe] 
distinguishes \ngc\ from other 
clusters with complex heavy-element
abundance patterns,
where the [Mg/Fe] dispersions range from
0.04~dex to 0.15~dex
(M2, M19, M22, \mbox{NGC~1851}, \mbox{NGC~5286};
\citealt{carretta11,marino11,marino15,yong14b,johnson15b}).
\ngc\ may be a good candidate to search for 
the presence of internal He variations.

\begin{figure*}
\centering
\includegraphics[angle=0,width=2.16in]{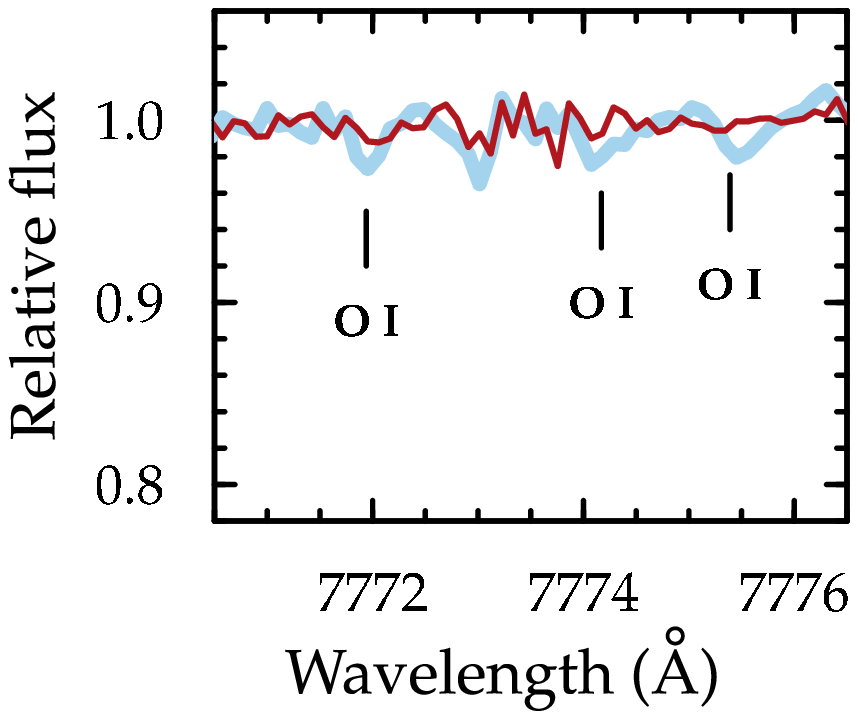}
\includegraphics[angle=0,width=2.16in]{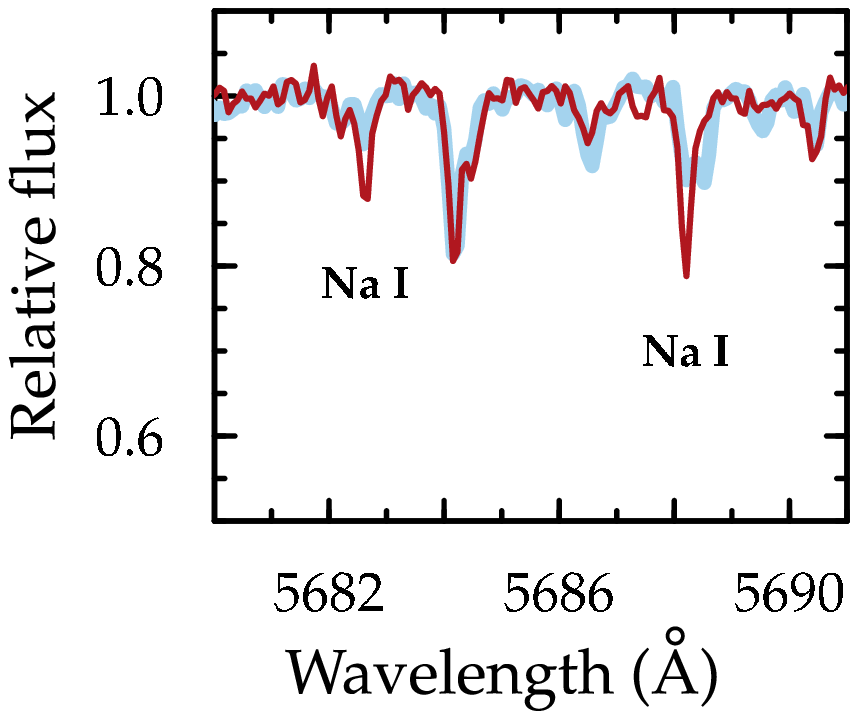}
\includegraphics[angle=0,width=2.16in]{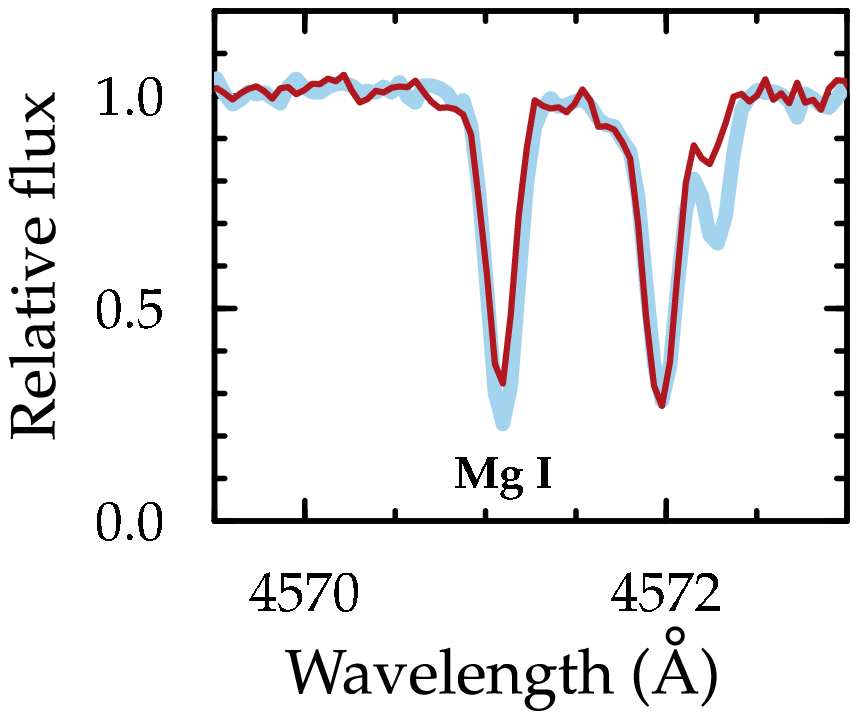} \\
\vspace*{0.05in}
\includegraphics[angle=0,width=2.16in]{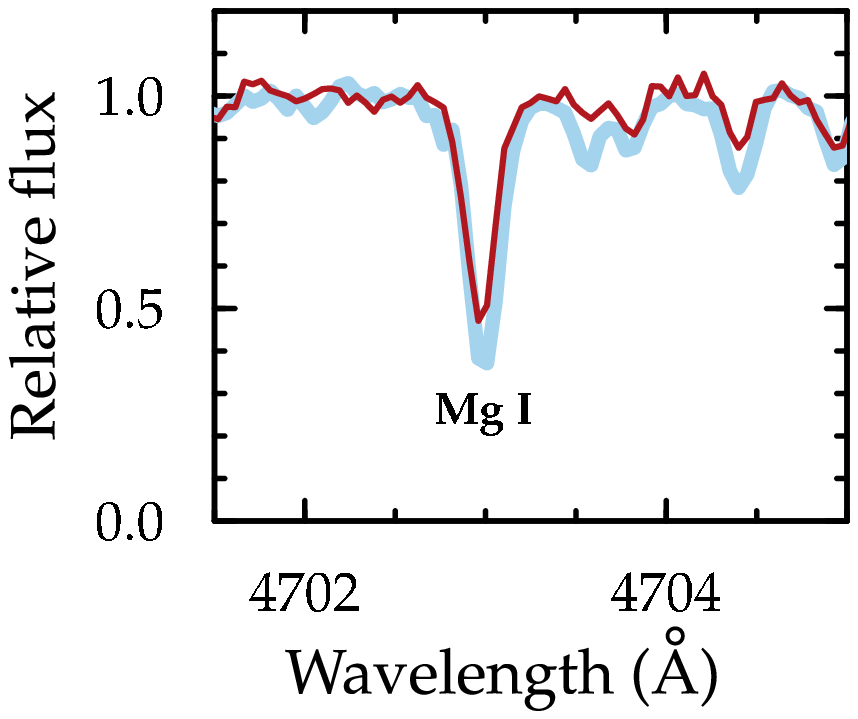}
\includegraphics[angle=0,width=2.16in]{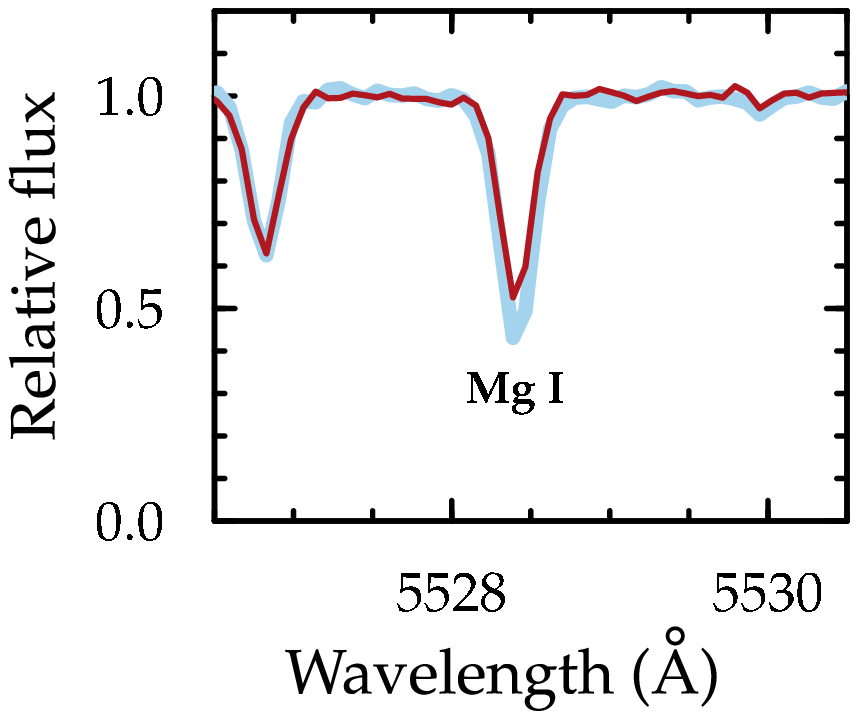}
\includegraphics[angle=0,width=2.16in]{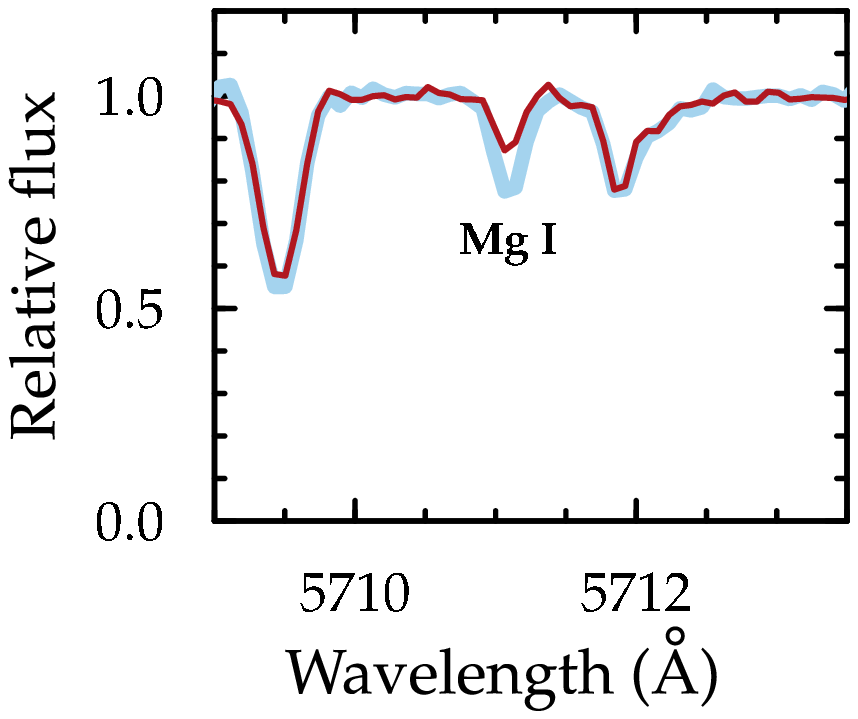} \\
\vspace*{0.05in}
\hspace*{-2.21in}
\includegraphics[angle=0,width=2.16in]{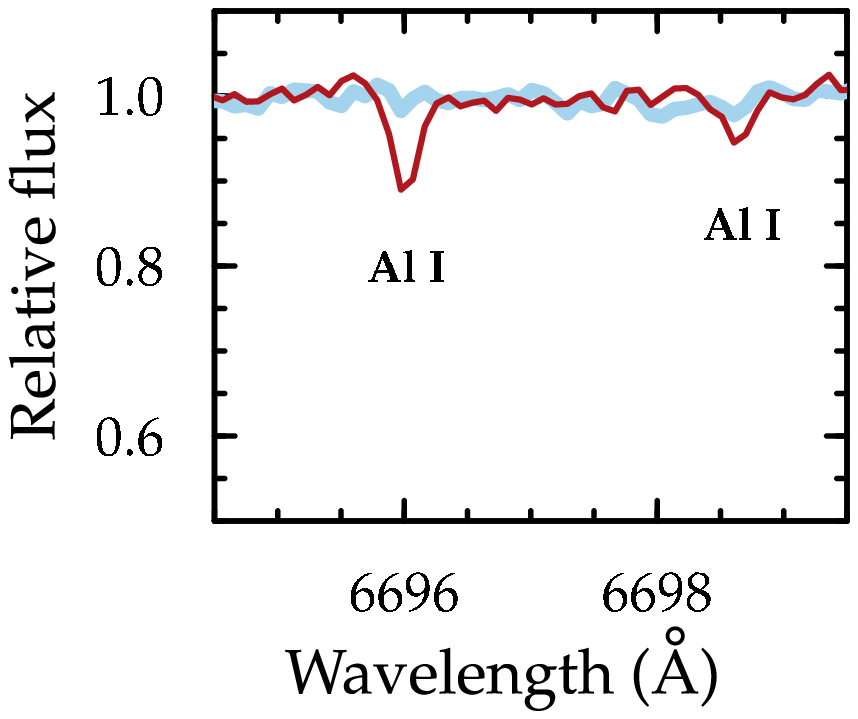}
\includegraphics[angle=0,width=2.16in]{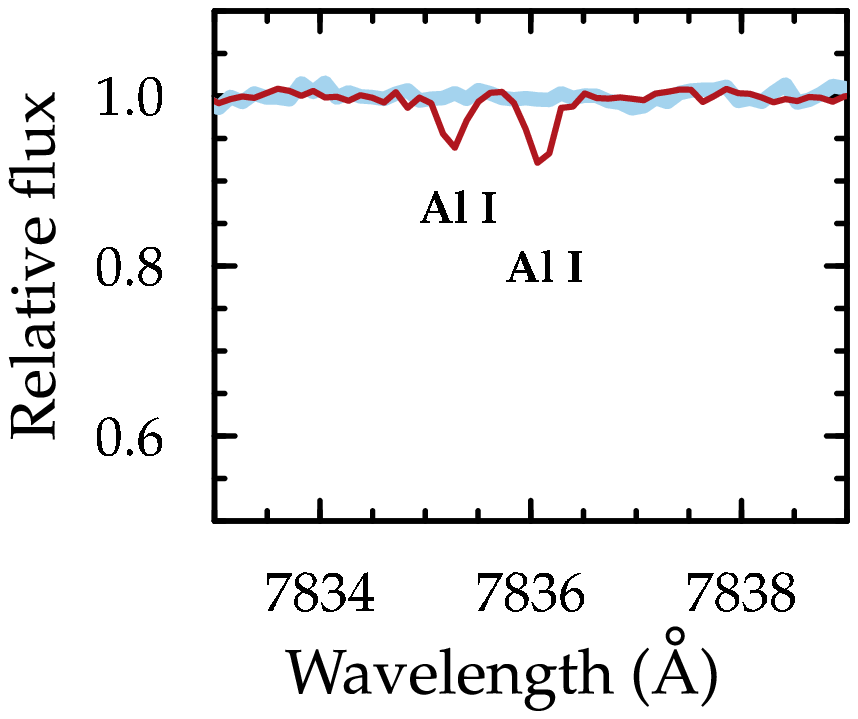}
\caption{
\label{lightspecplot}
Selections of the MIKE spectra of stars 42009955 (thin red line)
and 61005163 (bold blue line) around several  O~\textsc{i}, Na~\textsc{i}, 
Mg~\textsc{i}, and Al~\textsc{i} lines.
}
\end{figure*}

Our MIKE spectra 
of stars 42009955 and 61005163
include lines of O~\textsc{i}, Na~\textsc{i}, 
Mg~\textsc{i}, and Al~\textsc{i}.
These are illustrated in Figure~\ref{lightspecplot}.
Examination of the spectra reveal the 
usual light element abundance variations found
within globular clusters.
Star 61005163 has relatively 
strong O~\textsc{i} lines,
weak Na~\textsc{i} lines, 
strong Mg~\textsc{i} lines,
and non-detected Al~\textsc{i} lines.
Star 42009955, in contrast, has 
non-detected O~\textsc{i} lines, 
relatively strong Na~\textsc{i} lines,
weak Mg~\textsc{i} lines, and 
strong Al~\textsc{i} lines.
The abundances reported in Tables~\ref{9955mikemeantab} and
\ref{5163mikemeantab} quantitatively confirm these impressions.

Si and K have also been shown to occasionally participate
in the high-temperature $p$-capture reactions that 
produce the light element abundance variations in 
globular clusters
\citep{yong05,carretta09b,carretta13b,cohen12,mucciarelli12,mucciarelli15a,
ventura12,meszaros15}.
These two elements are only covered in the MIKE spectra.
In both cases, the [Si/Fe] and [K/Fe] ratios
are in agreement.
These two stars show variations 
in the [O/Fe], [Na/Fe], [Mg/Fe], and [Al/Fe] ratios,
so [Si/Fe] and [K/Fe] may not
vary substantially within \ngc.

CH and CN molecular features are detected in our two MIKE spectra.
[C/Fe] is subsolar and [N/Fe] is supersolar in both stars,
which is consistent with the scenario
where C is depleted and converted to N 
as stars evolve up the RGB.
We estimate the natal [C/Fe] ratios in these two stars
using the results of 
\citet{placco14}, who provide a set of corrections
based on stellar evolution models 
that relate the current [C/Fe] ratio 
of a star on the RGB
to the initial [C/Fe] ratio.
For star 42009955, the
observed [C/Fe]~$= -$1.16
maps to [C/Fe]$_{\rm init} \approx -$0.55.
For star 61005163, the
observed [C/Fe]~$= -$0.11 
maps to [C/Fe]$_{\rm init} \approx +$0.44.
Neither of these initial [C/Fe] ratios
would be considered ``carbon-enhanced'' by 
modern definitions
\citep{aoki07},
which require [C/Fe]$_{\rm init} \geq +$0.7.

For completeness, we note that the Li~\textsc{i} line at 6707~\AA\
is covered in our MIKE spectra.
Li~\textsc{i} is not detected in either star.
The upper limits are consistent with
the levels of Li-depletion commonly found in other
cluster and field stars on the upper RGB
(e.g., \citealt{gratton00,lind09}).

\subsection{Elements in the Iron Group}
\label{fegroup}

\begin{figure}
\centering
\includegraphics[angle=0,width=3.25in]{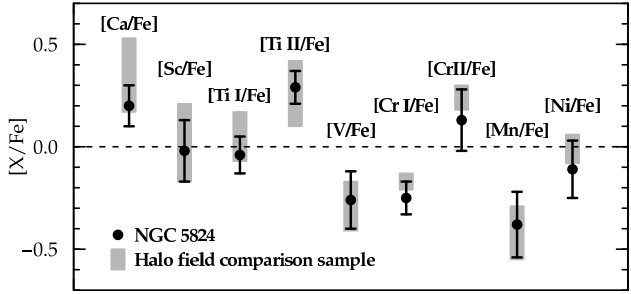}
\caption{
\label{haloplot}
Comparison of abundance ratios in \ngc\ and a
sample of field red giants with similar metallicity.
The black points represent the \ngc\ ratios, derived from
the M2FS spectra, and the
gray shaded boxes represent the median~$\pm$~one standard deviation
for the field sample.
}
\end{figure}

Figure~\ref{haloplot} compares the mean abundance ratios
in \ngc\ (excluding star 61005163)
with those in halo stars of similar metallicity.
The gray boxes represent the median~$\pm$~one
standard deviation
of 14 red giant stars in the solar neighborhood
with $-$2.5~$<$~[Fe/H]~$< -$1.5
\citep{roederer14a}.
There is superb agreement between the Fe-group elements in \ngc\
the halo field sample for the nine ratios shown in Figure~\ref{haloplot}.
We conclude that the overall composition of \ngc\ 
for elements with 20~$\leq Z \leq$~28 (Ca to Ni) 
is indistinguishable from that of the Galactic halo
in the solar neighborhood.

Cu has been found to correlate with the
\spro\ enhancement found in other complex, low-metallicity clusters,
like M2, M22, and \mbox{NGC~5286}.
Zn also correlates with the \spro\ enhancement in M22.
Cu~\textsc{i} and Zn~\textsc{i} lines are only 
covered in our MIKE spectra.
One star, 42009955 shows no \spro\ enhancement, while the other,
61005163, does (see Section~\ref{ncapture}).
We find no
significant differences among the [Cu/Fe] or [Zn/Fe]
ratios in these two stars.

\subsection{Neutron-Capture Elements}
\label{ncapture}

\begin{figure*}
\centering
\includegraphics[angle=0,width=1.625in]{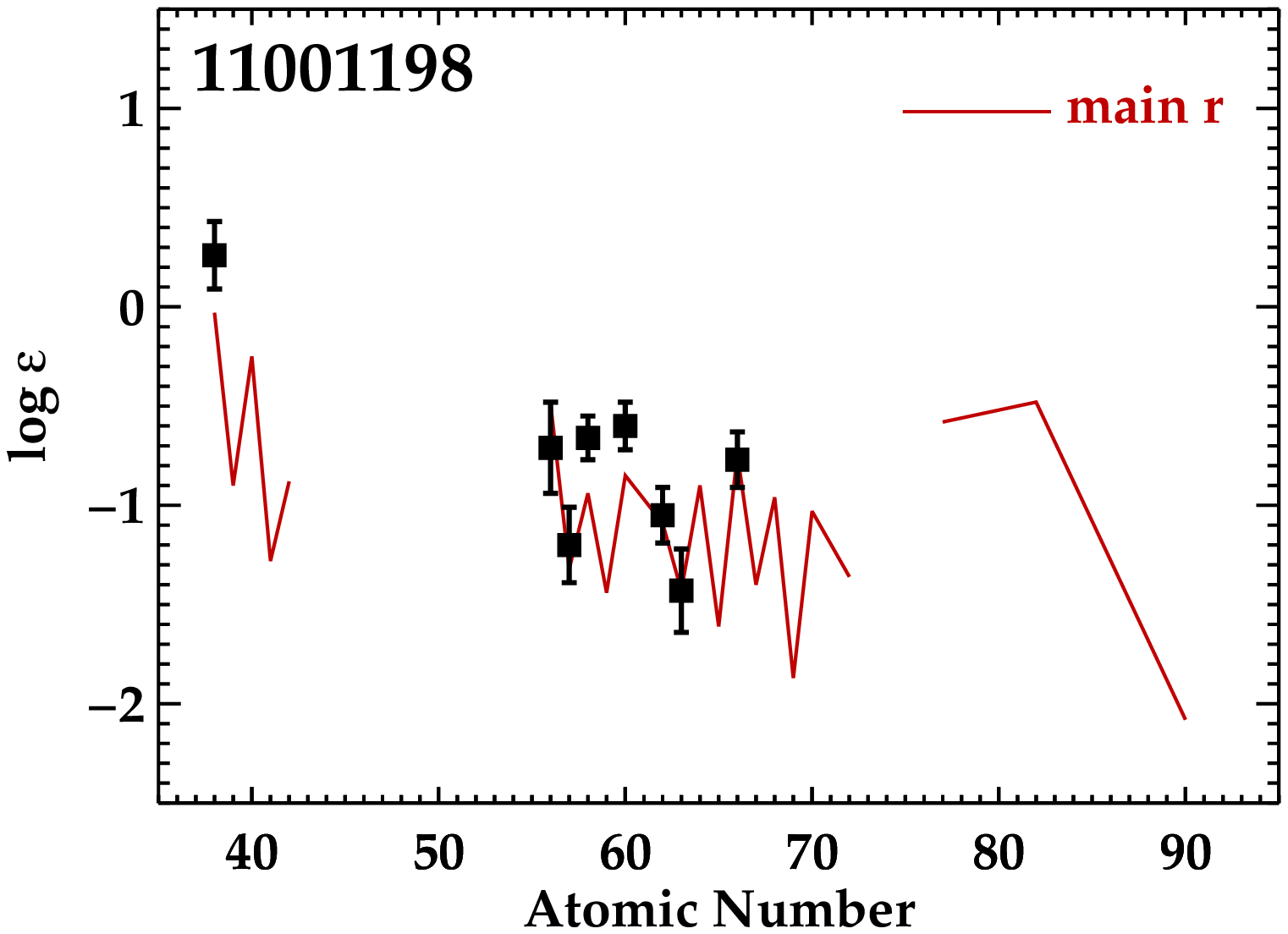}
\includegraphics[angle=0,width=1.625in]{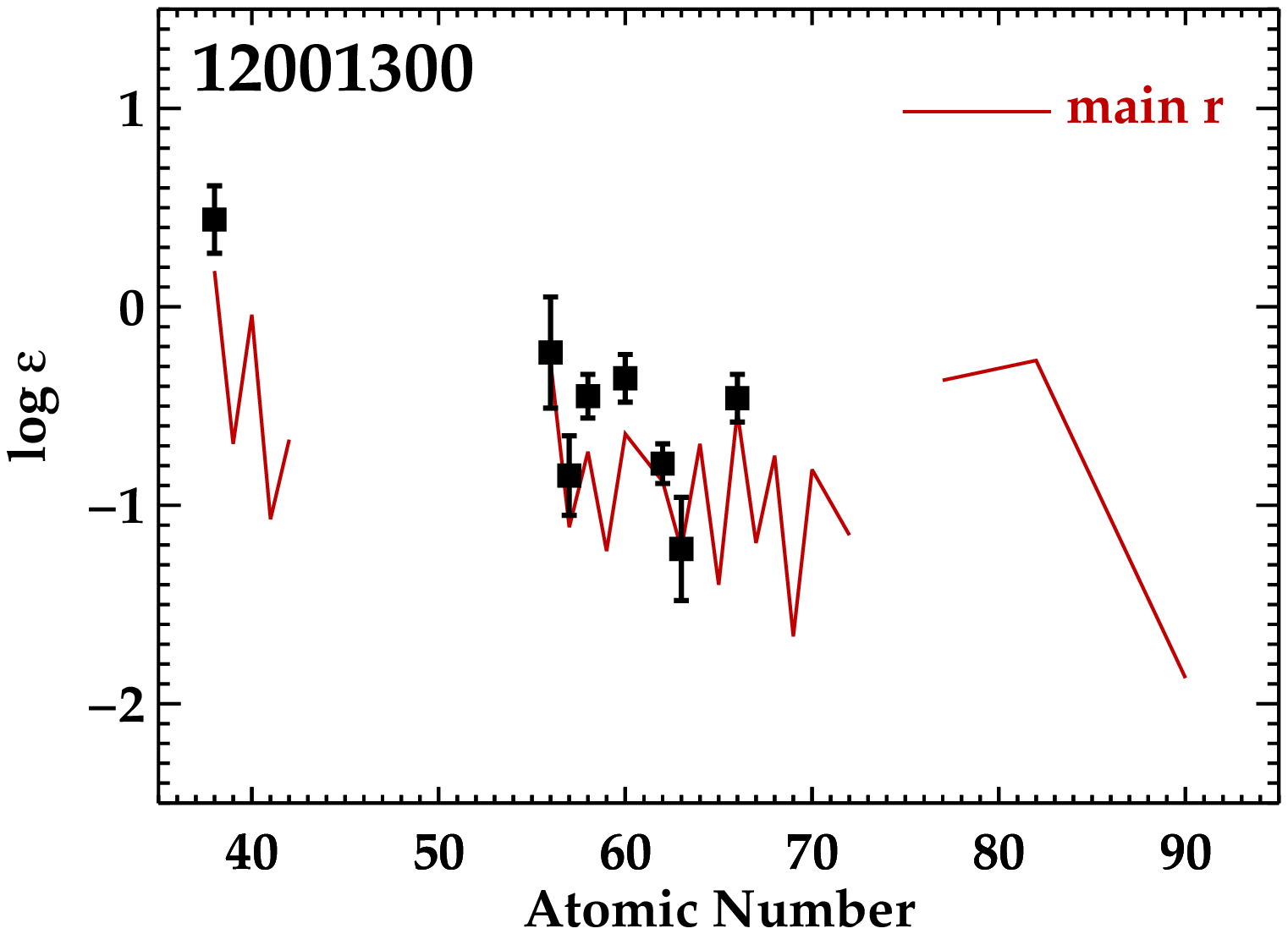}
\includegraphics[angle=0,width=1.625in]{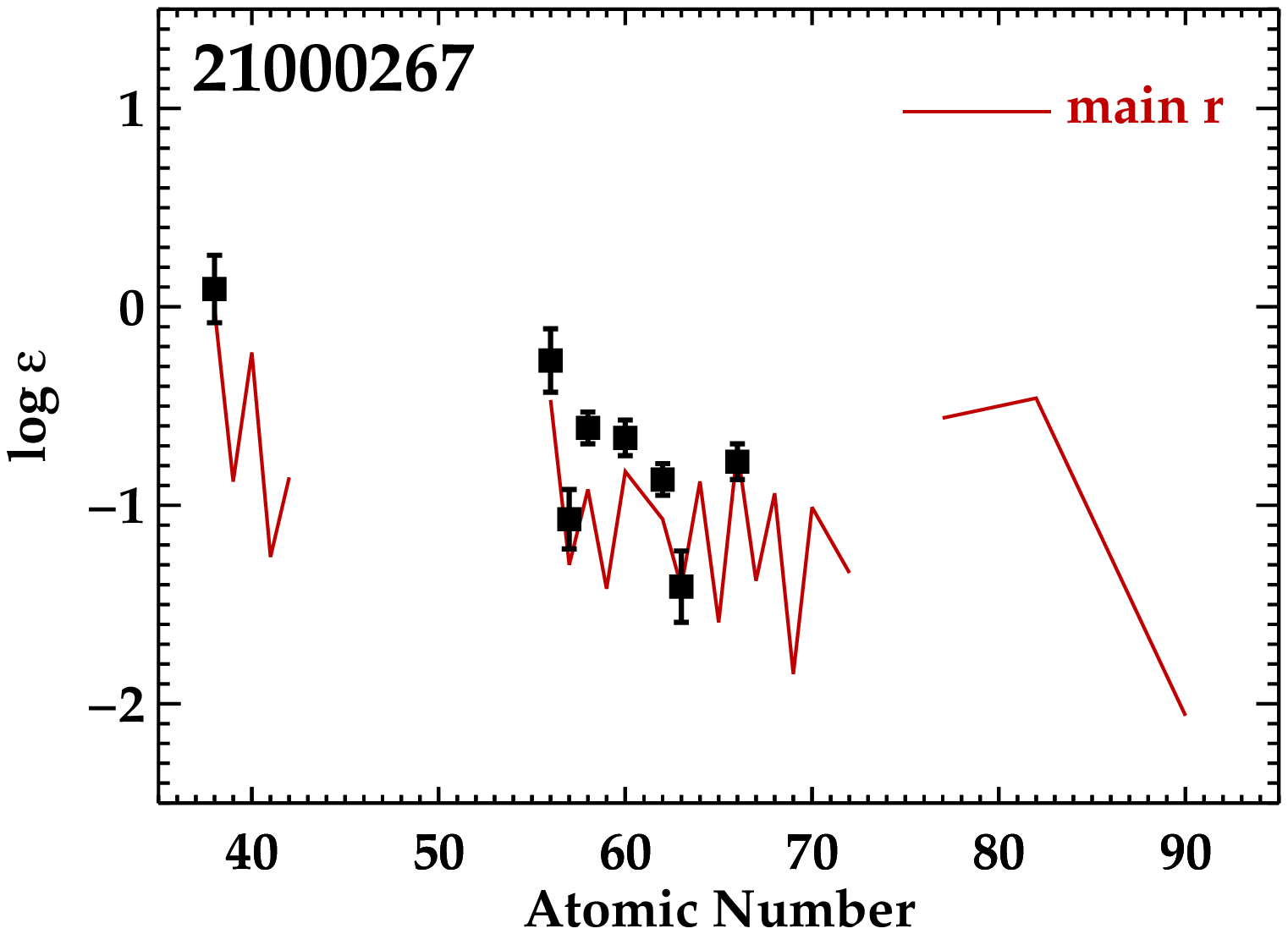}
\includegraphics[angle=0,width=1.625in]{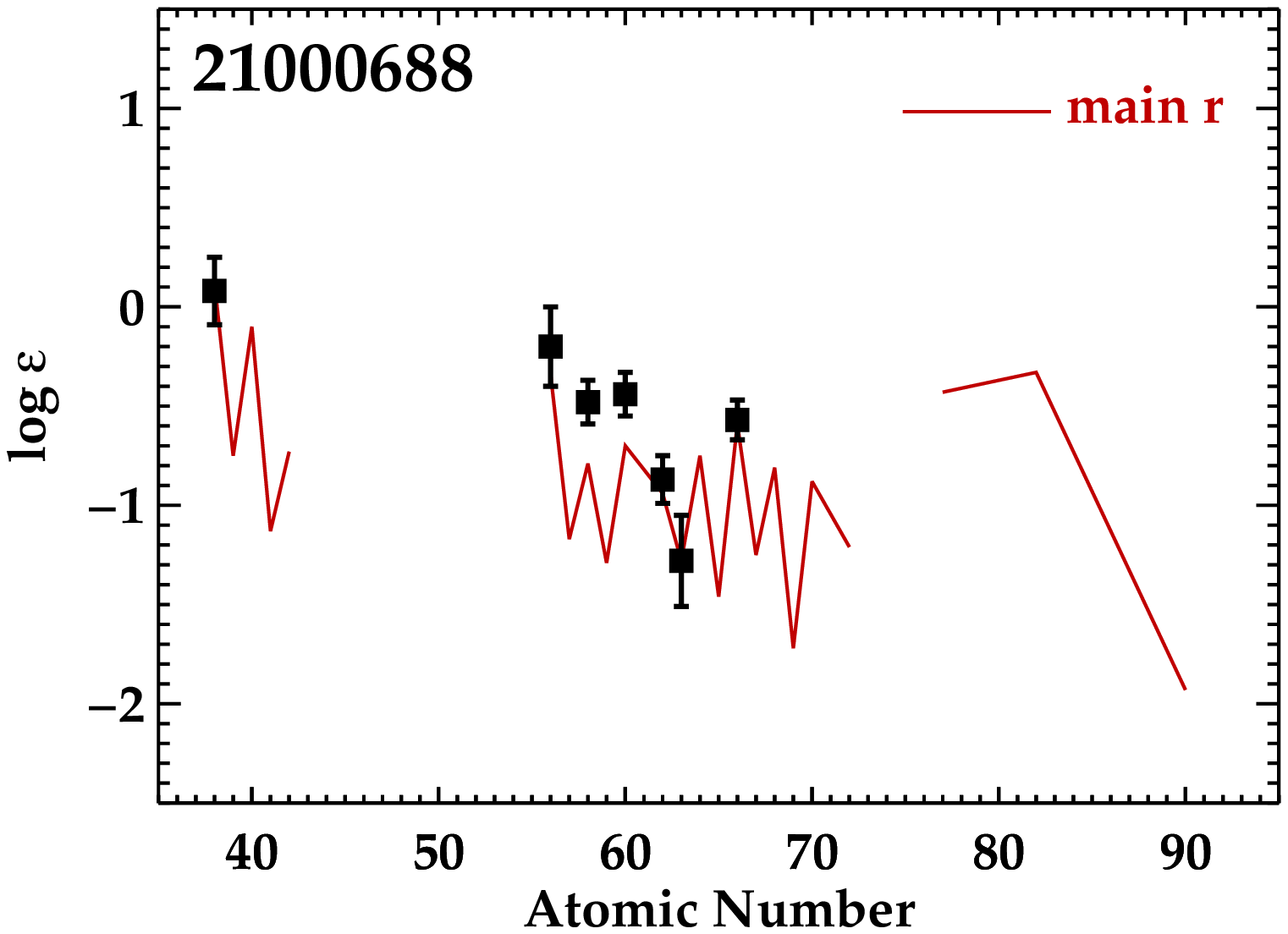} \\
\vspace*{0.05in}
\includegraphics[angle=0,width=1.625in]{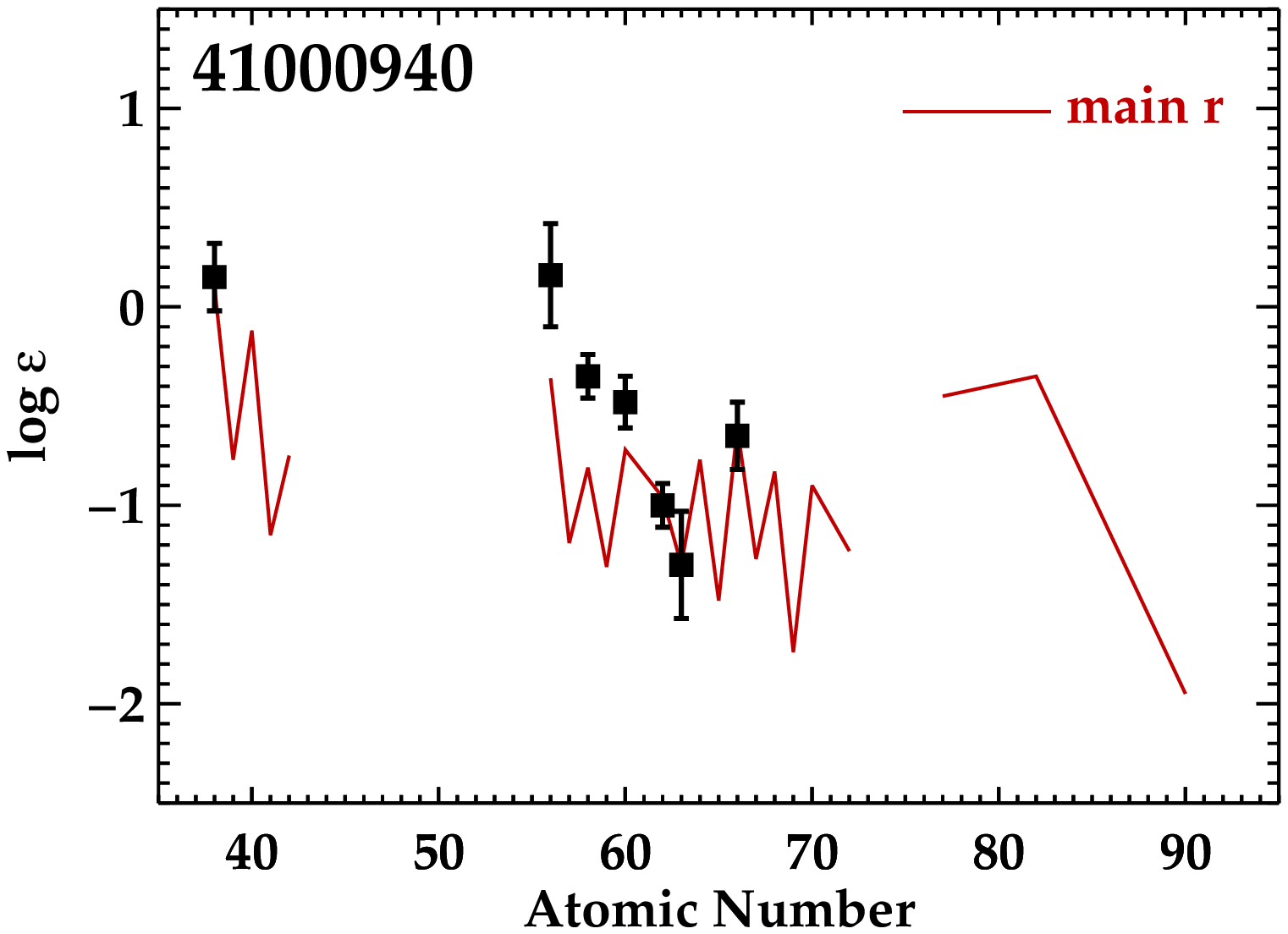}
\includegraphics[angle=0,width=1.625in]{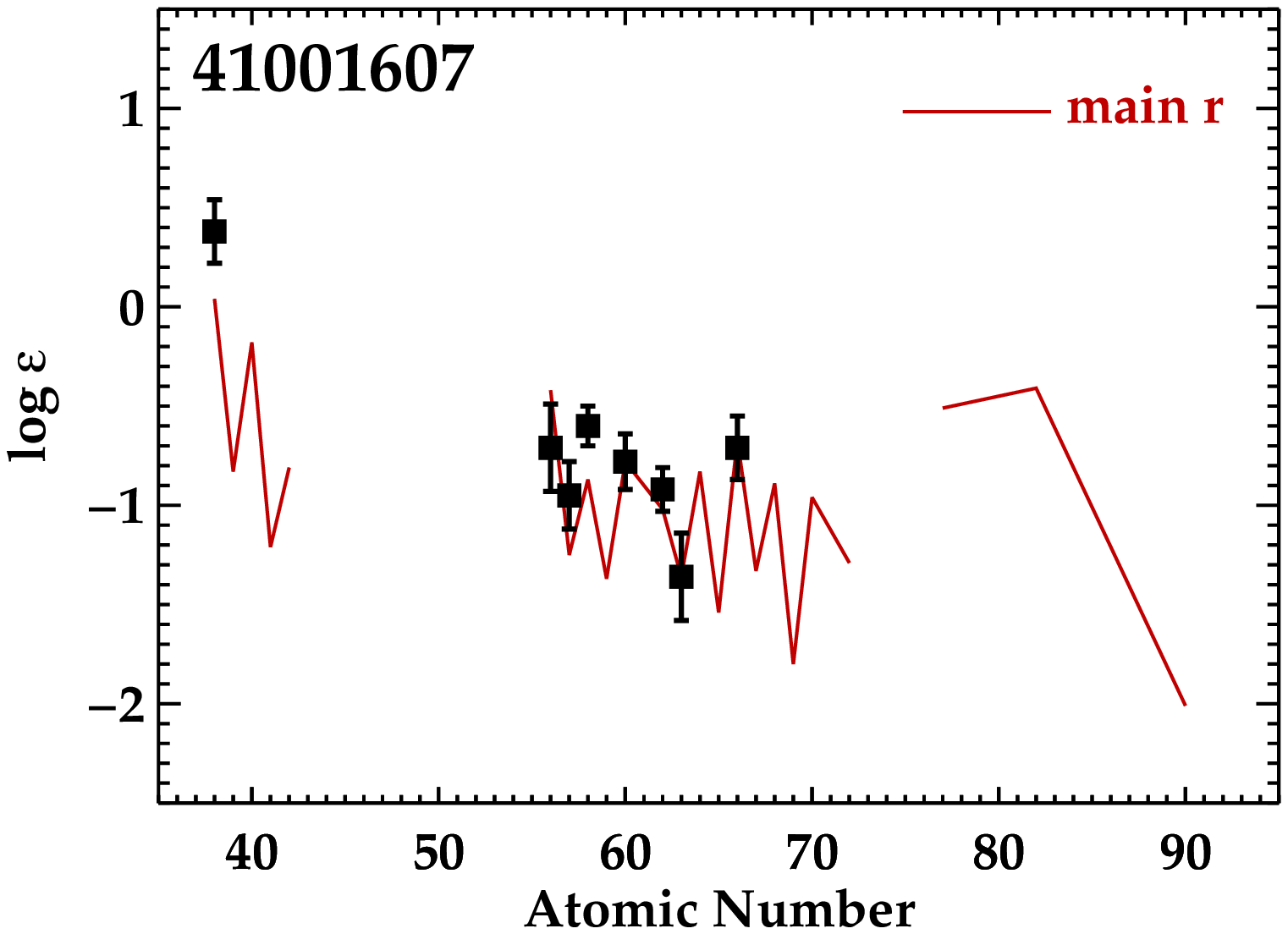}
\includegraphics[angle=0,width=1.625in]{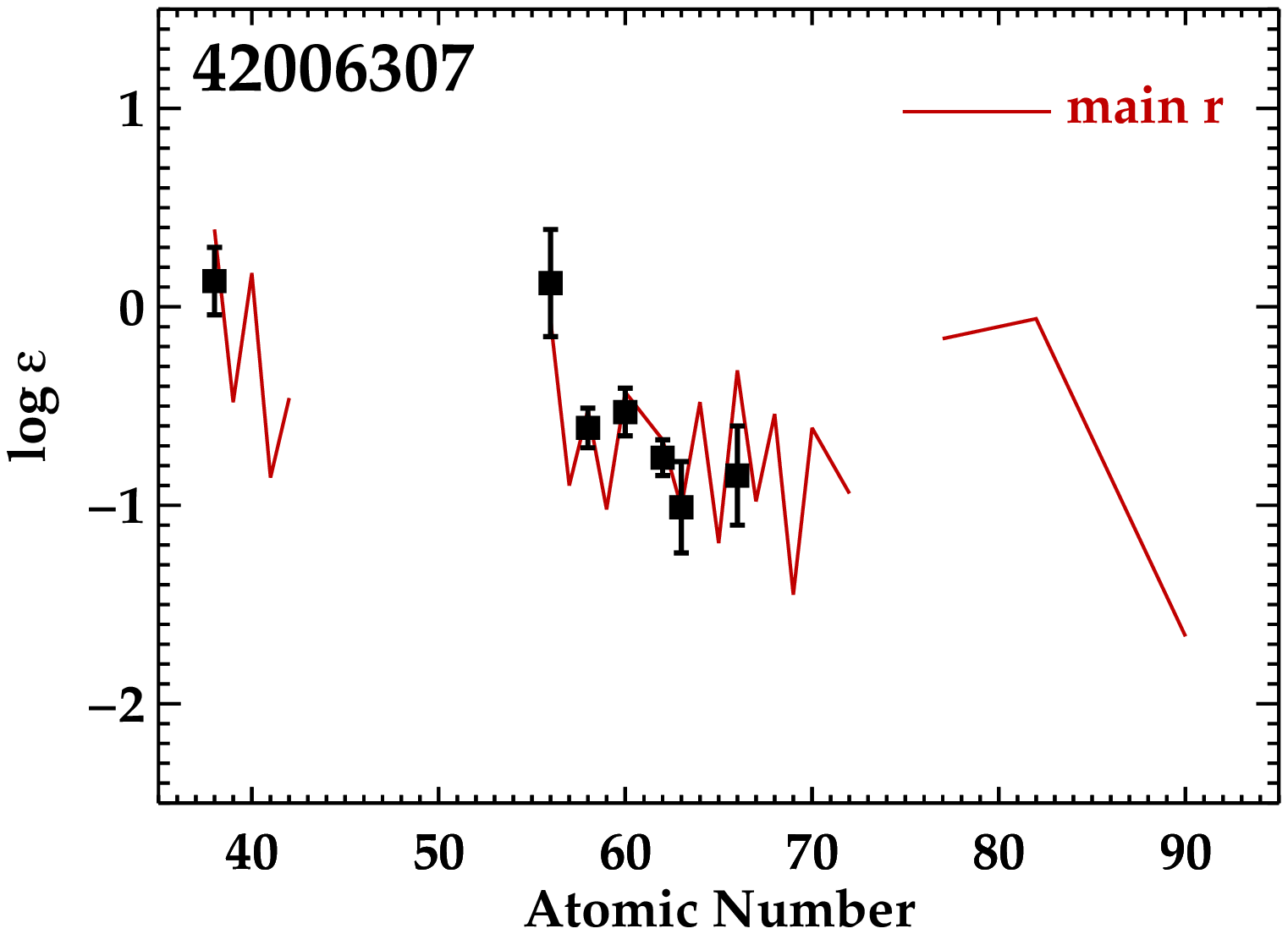}
\includegraphics[angle=0,width=1.625in]{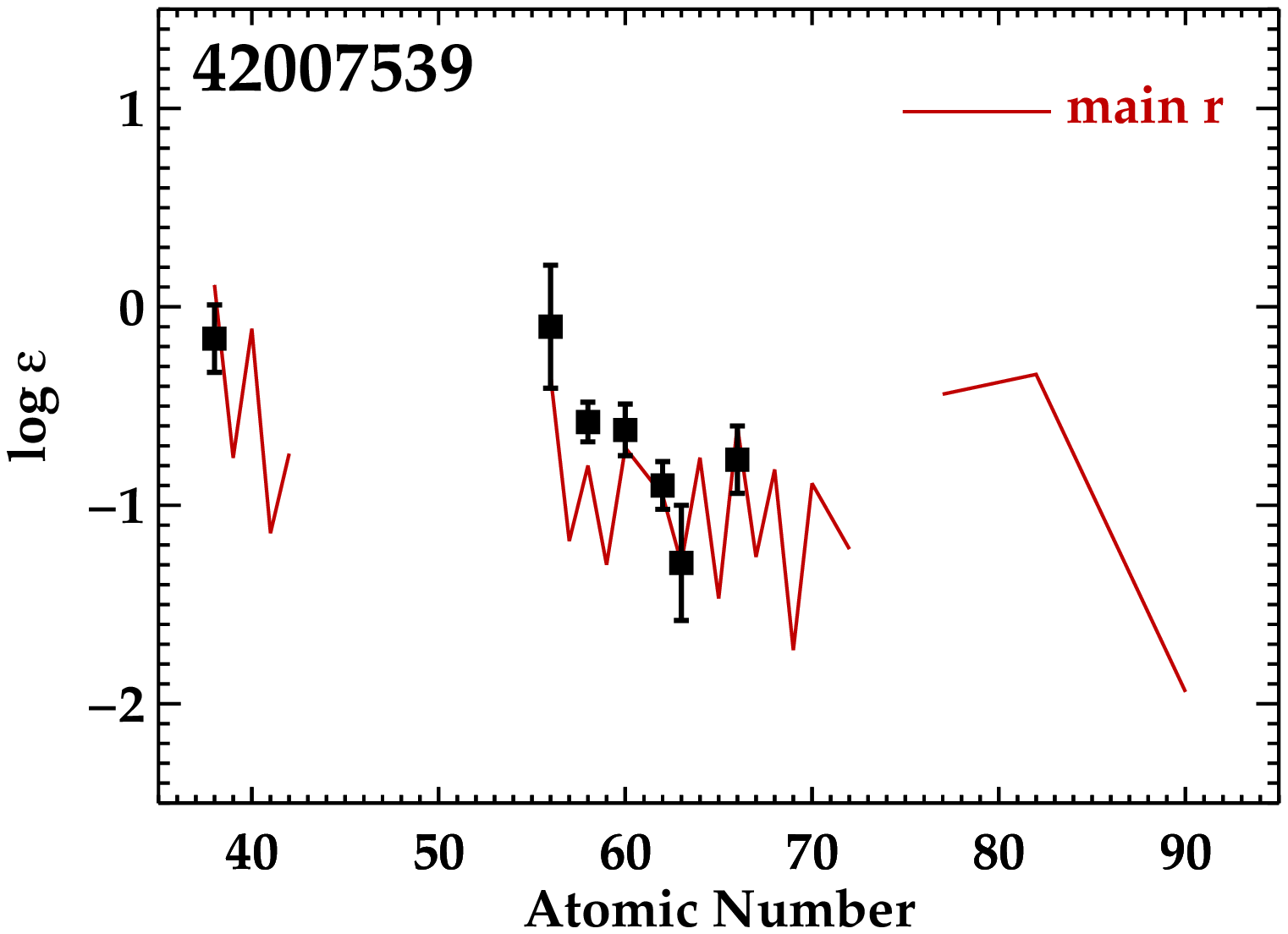} \\
\vspace*{0.05in}
\includegraphics[angle=0,width=1.625in]{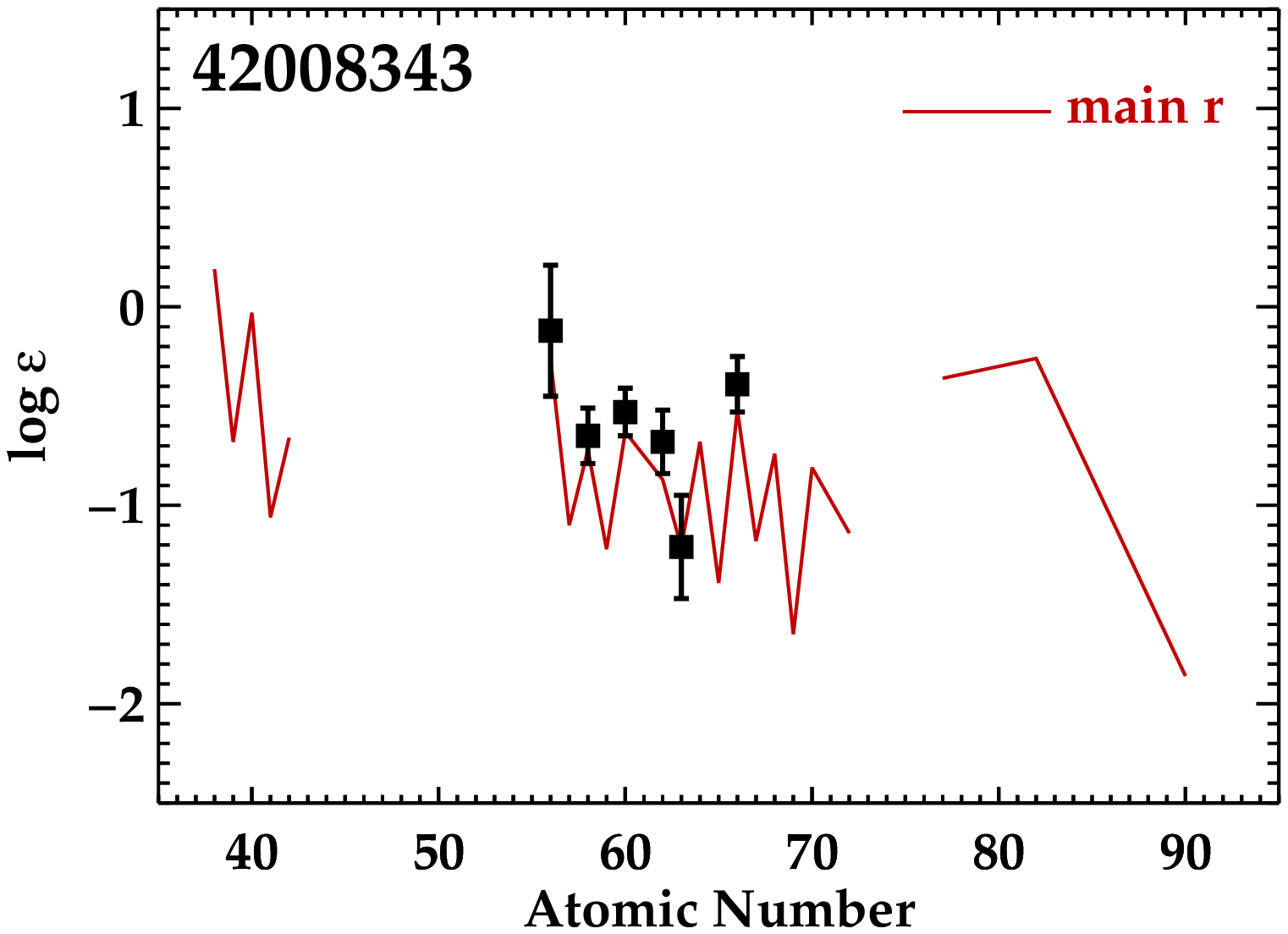}
\includegraphics[angle=0,width=1.625in]{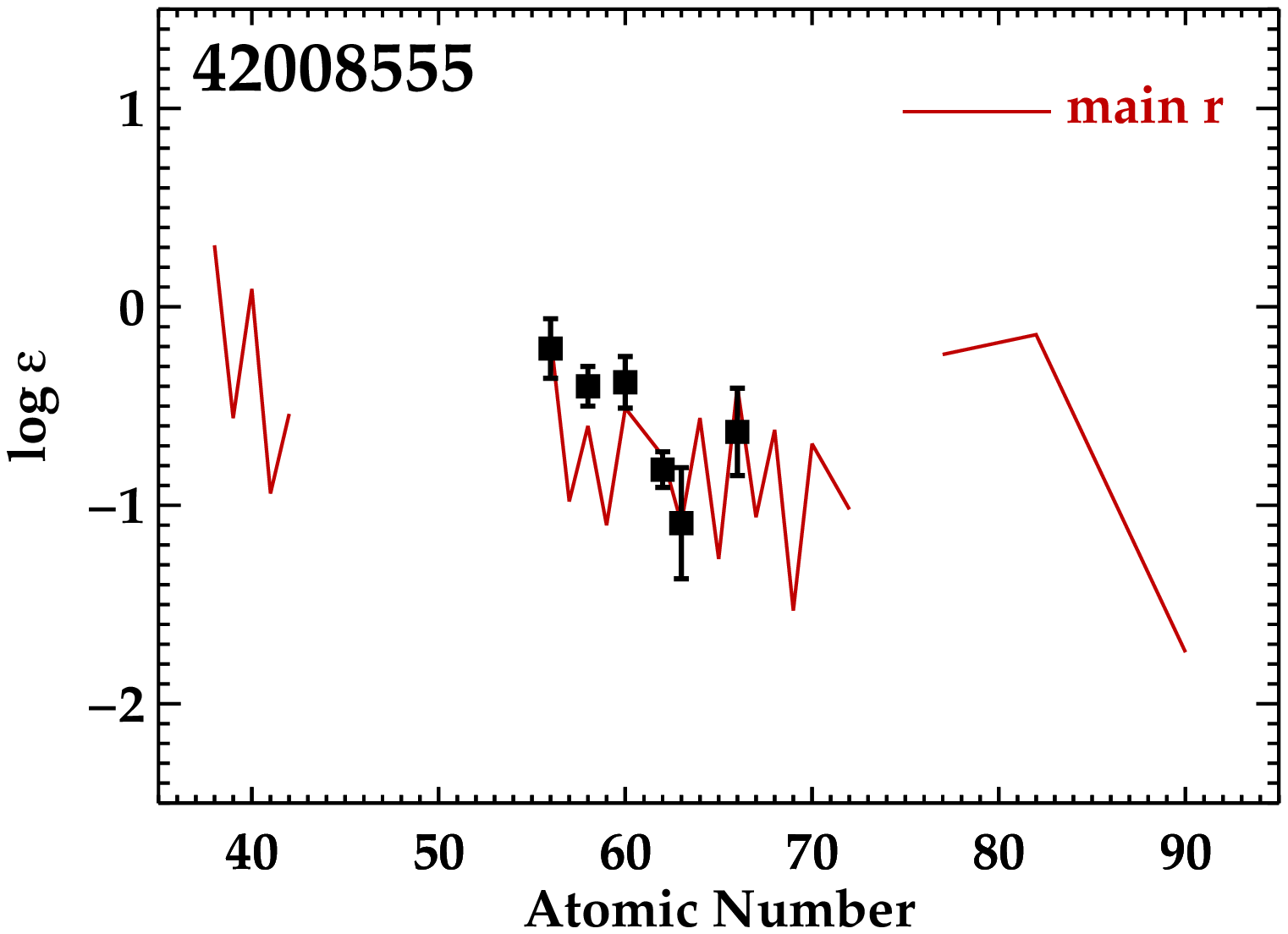}
\includegraphics[angle=0,width=1.625in]{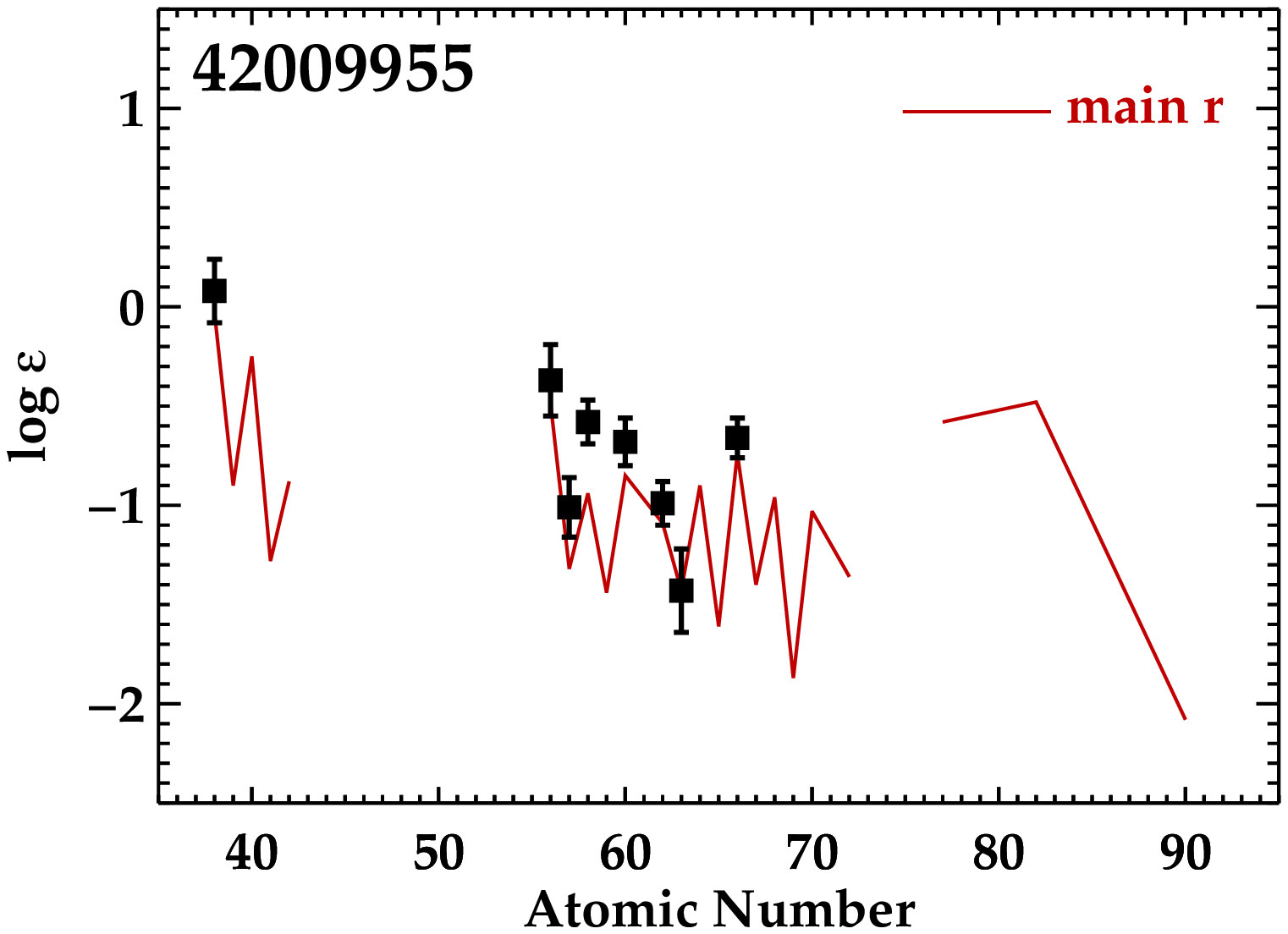}
\includegraphics[angle=0,width=1.625in]{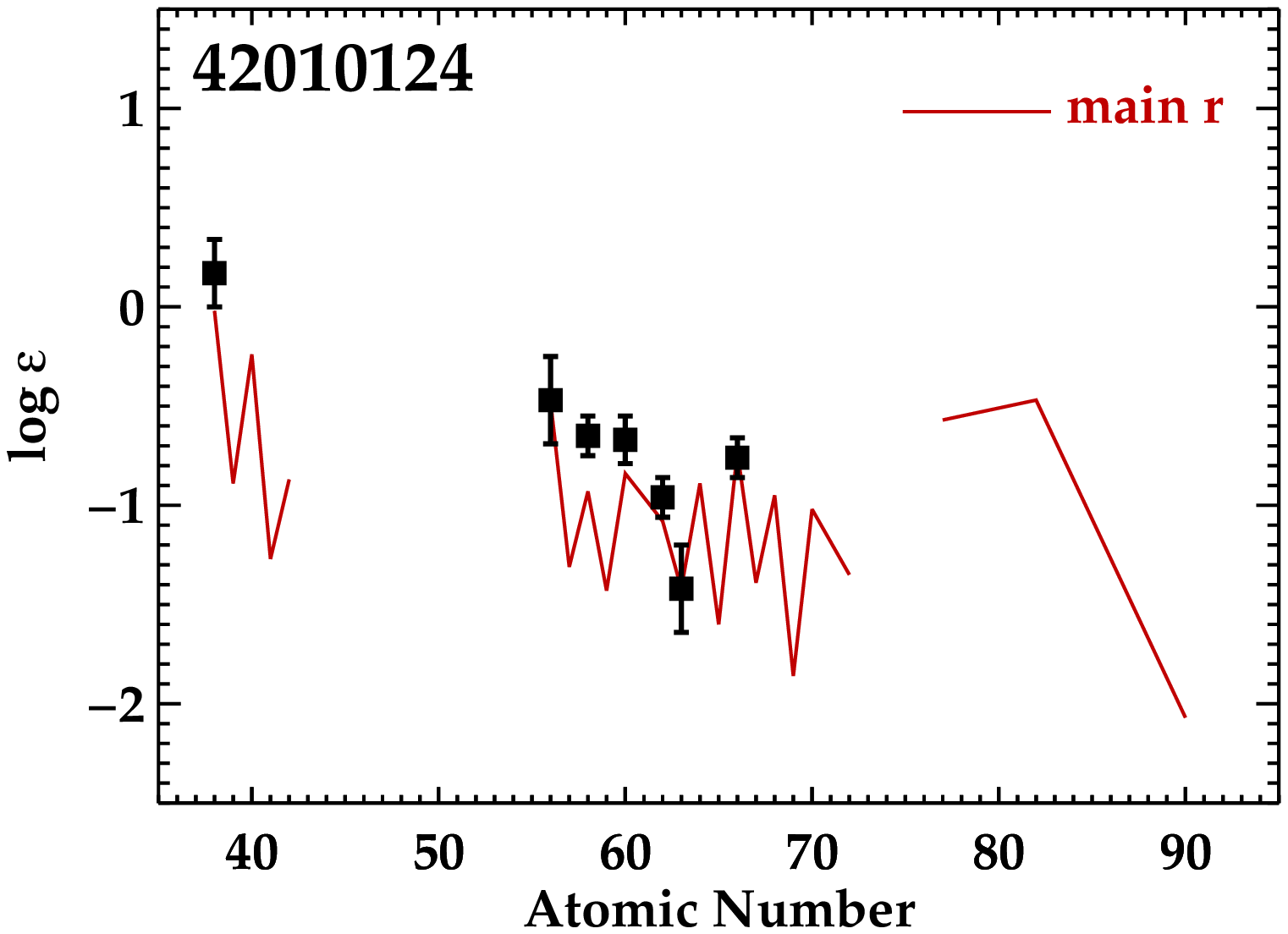} \\
\vspace*{0.05in}
\includegraphics[angle=0,width=1.625in]{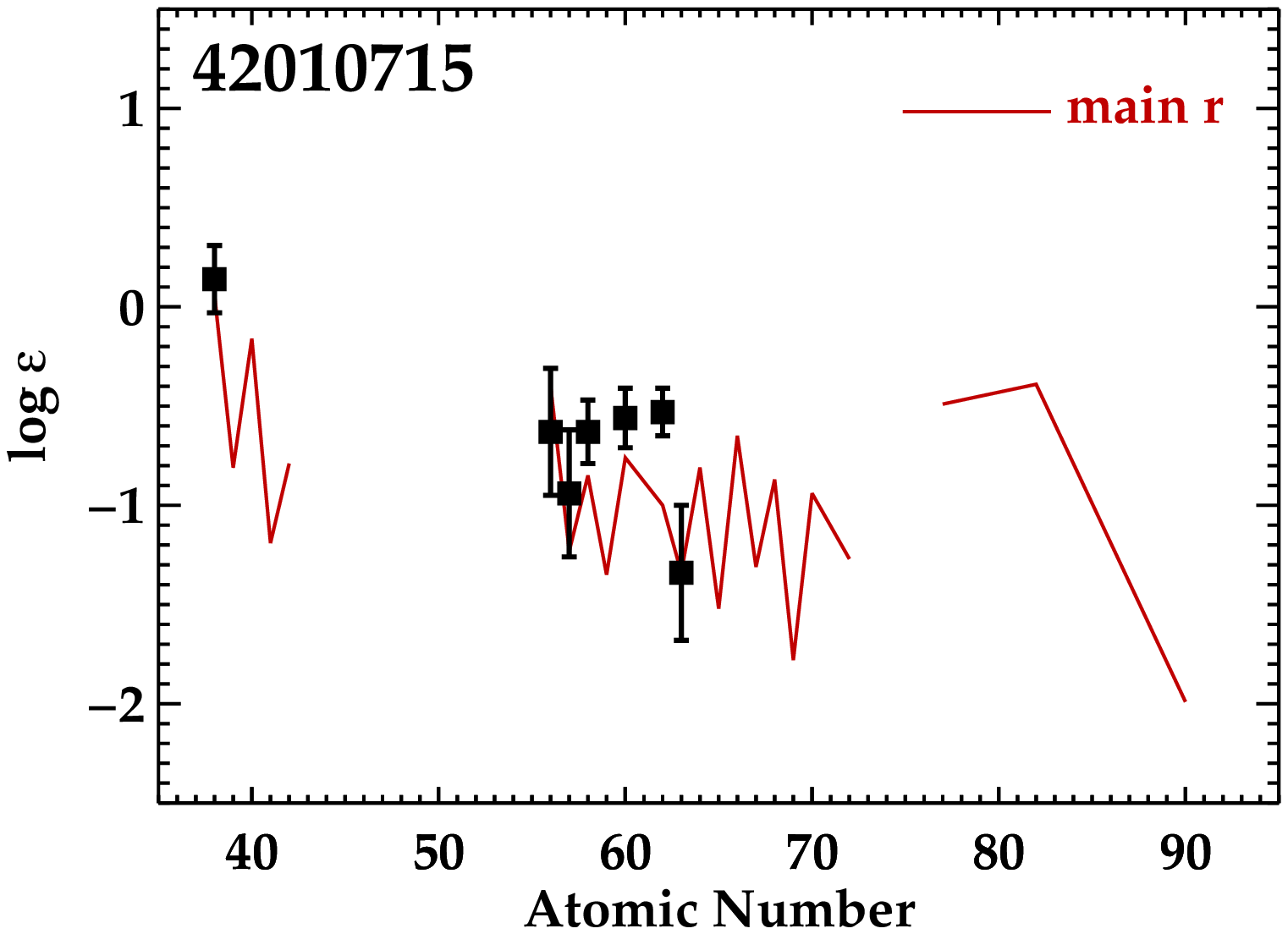}
\includegraphics[angle=0,width=1.625in]{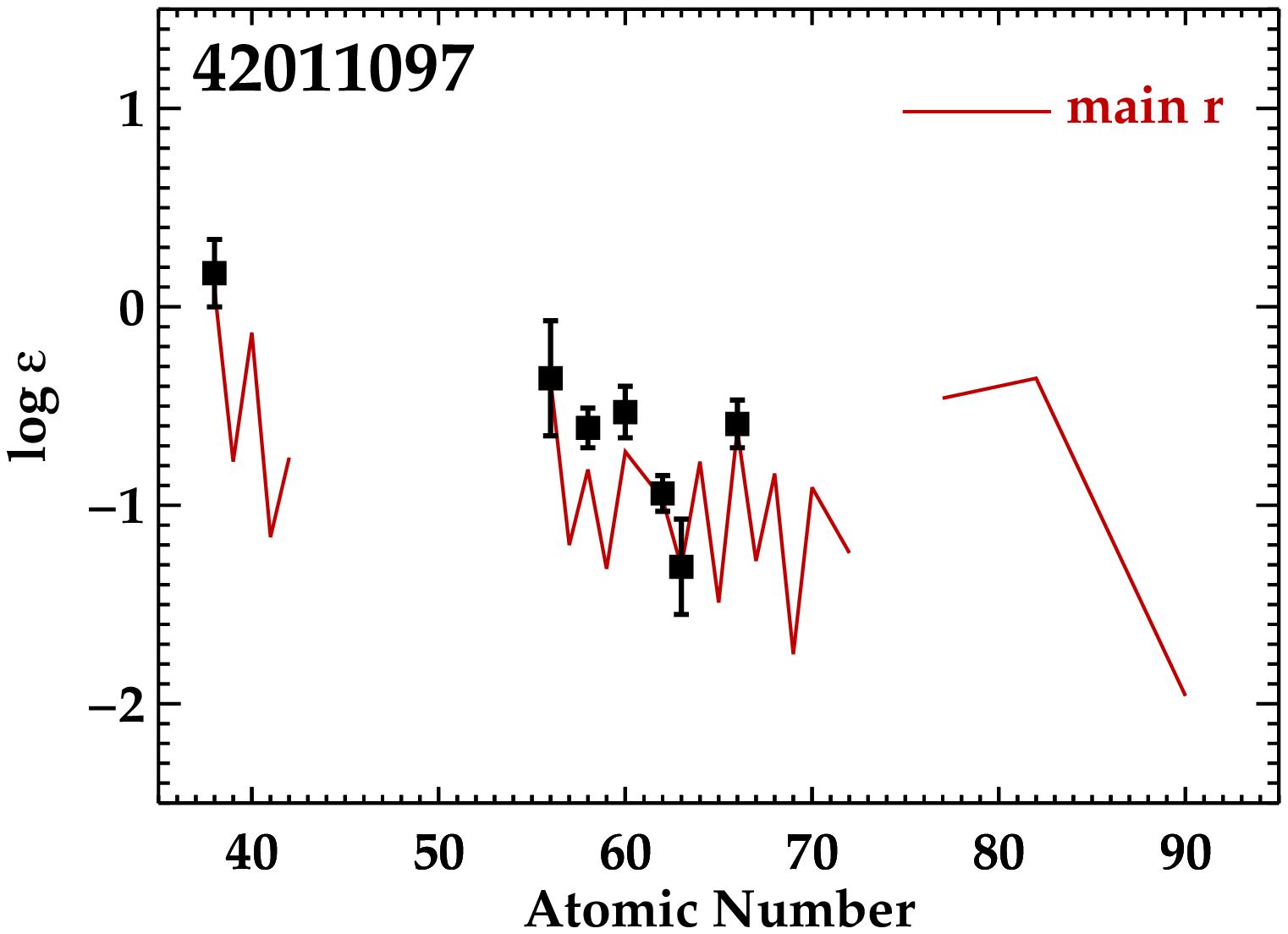}
\includegraphics[angle=0,width=1.625in]{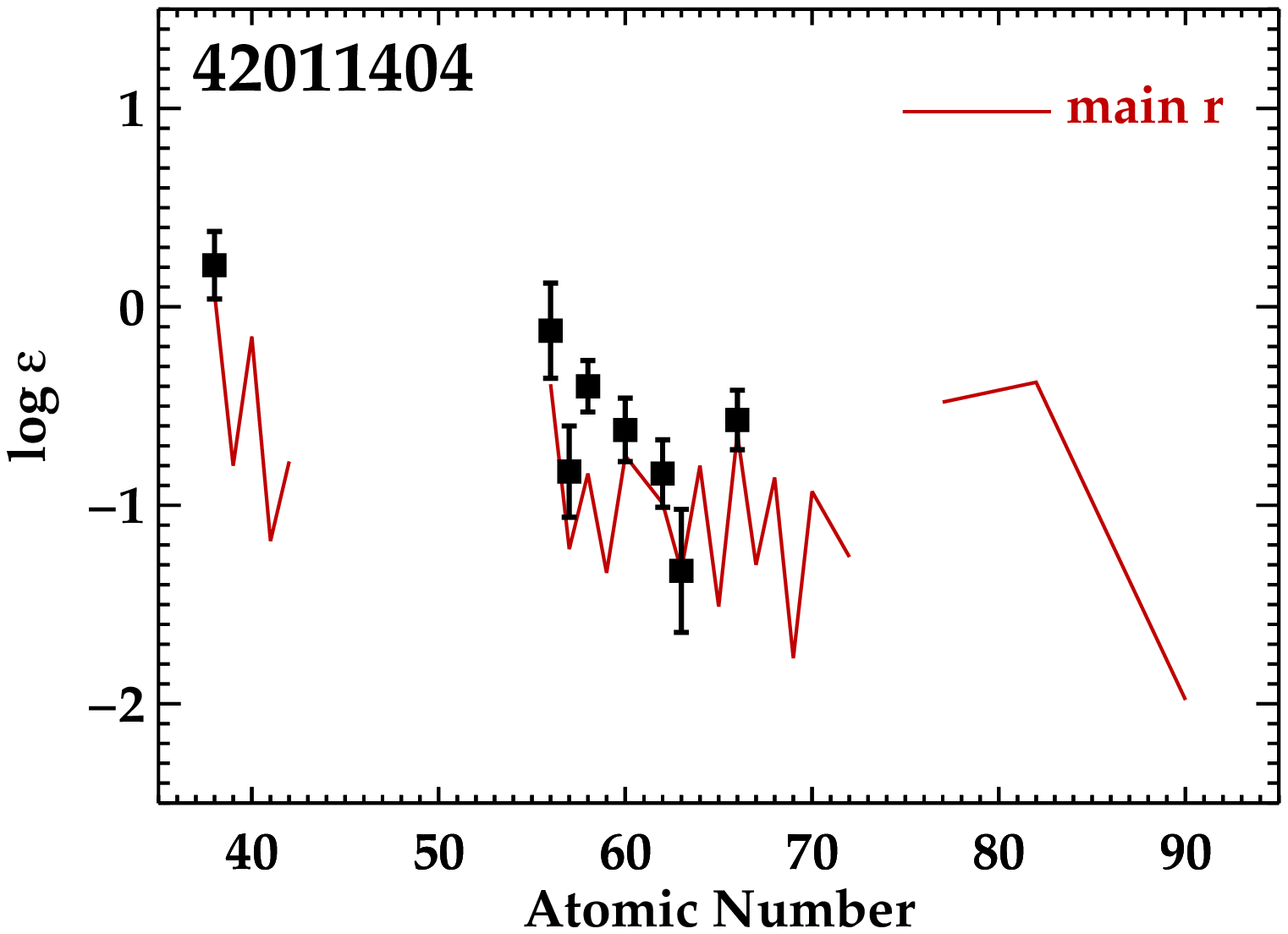}
\includegraphics[angle=0,width=1.625in]{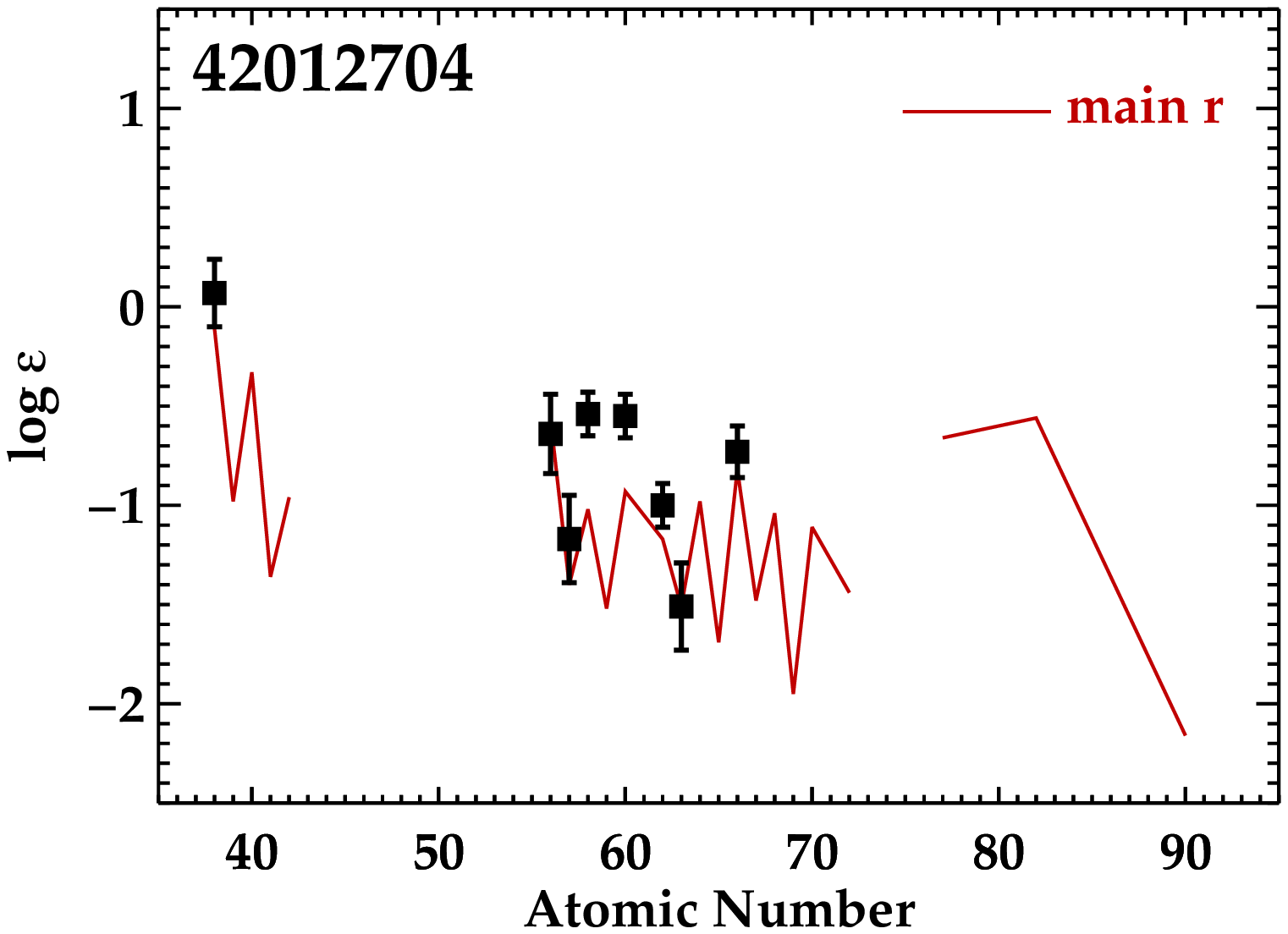} \\
\vspace*{0.05in}
\includegraphics[angle=0,width=1.625in]{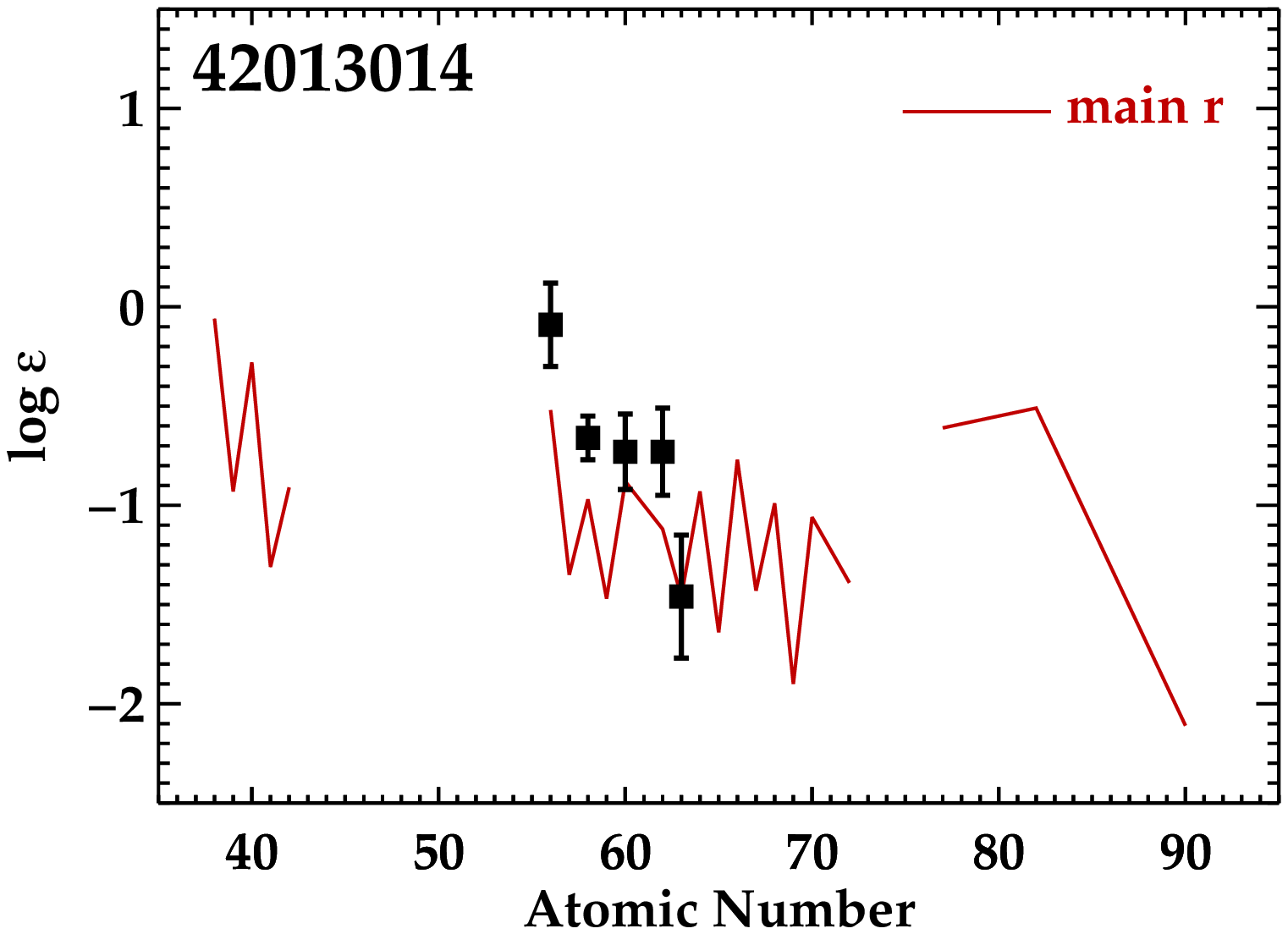}
\includegraphics[angle=0,width=1.625in]{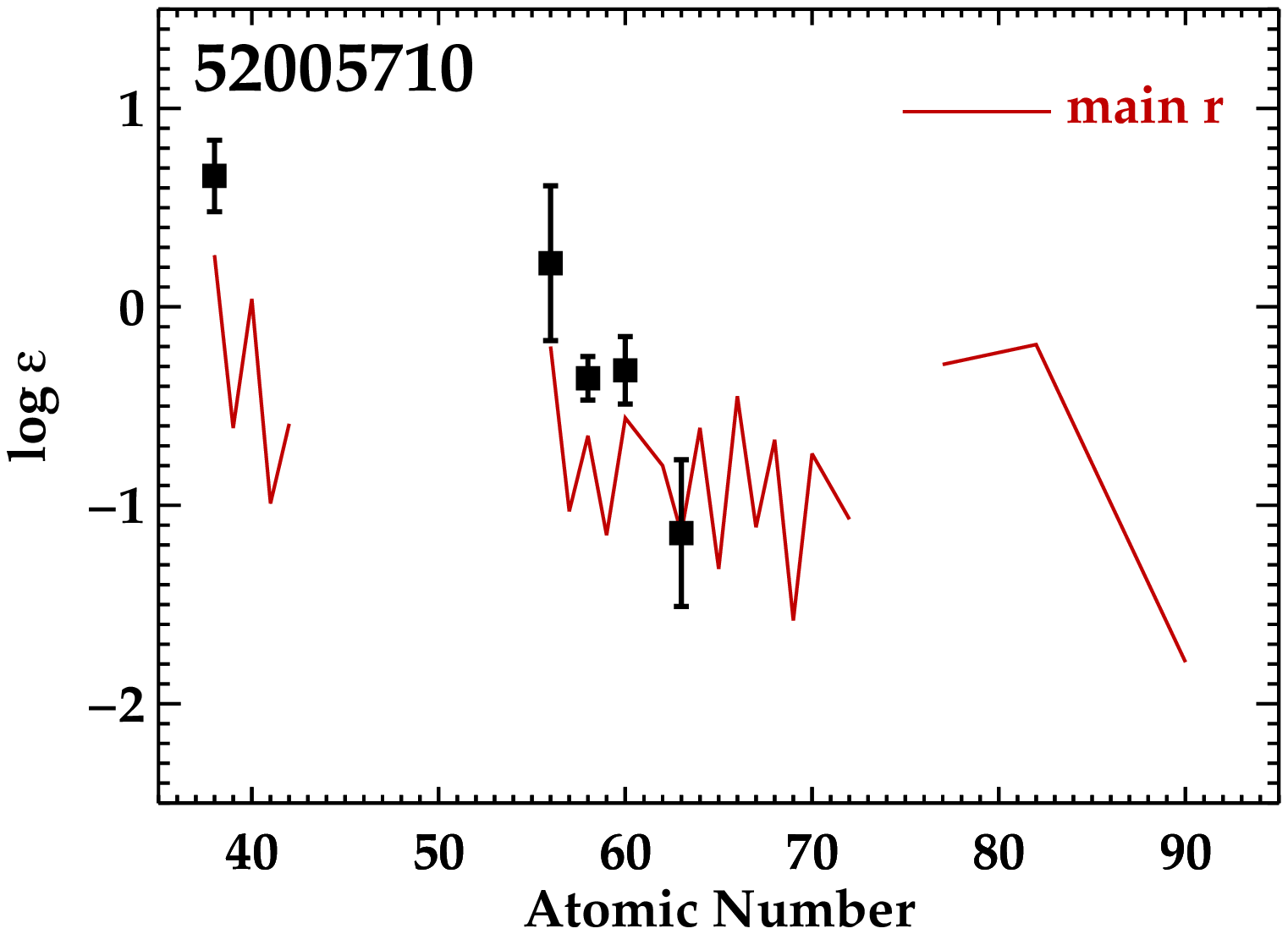}
\includegraphics[angle=0,width=1.625in]{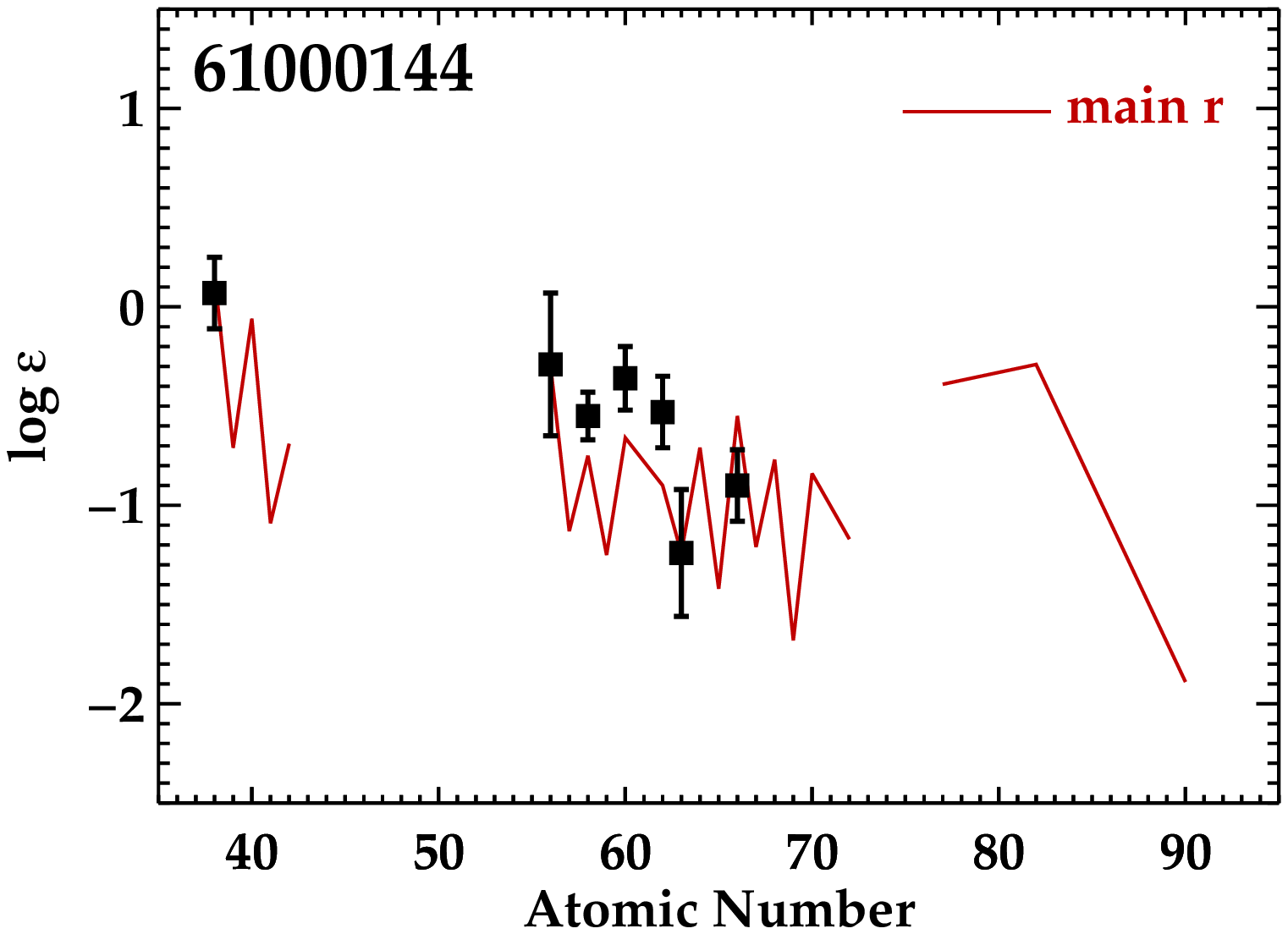}
\includegraphics[angle=0,width=1.625in]{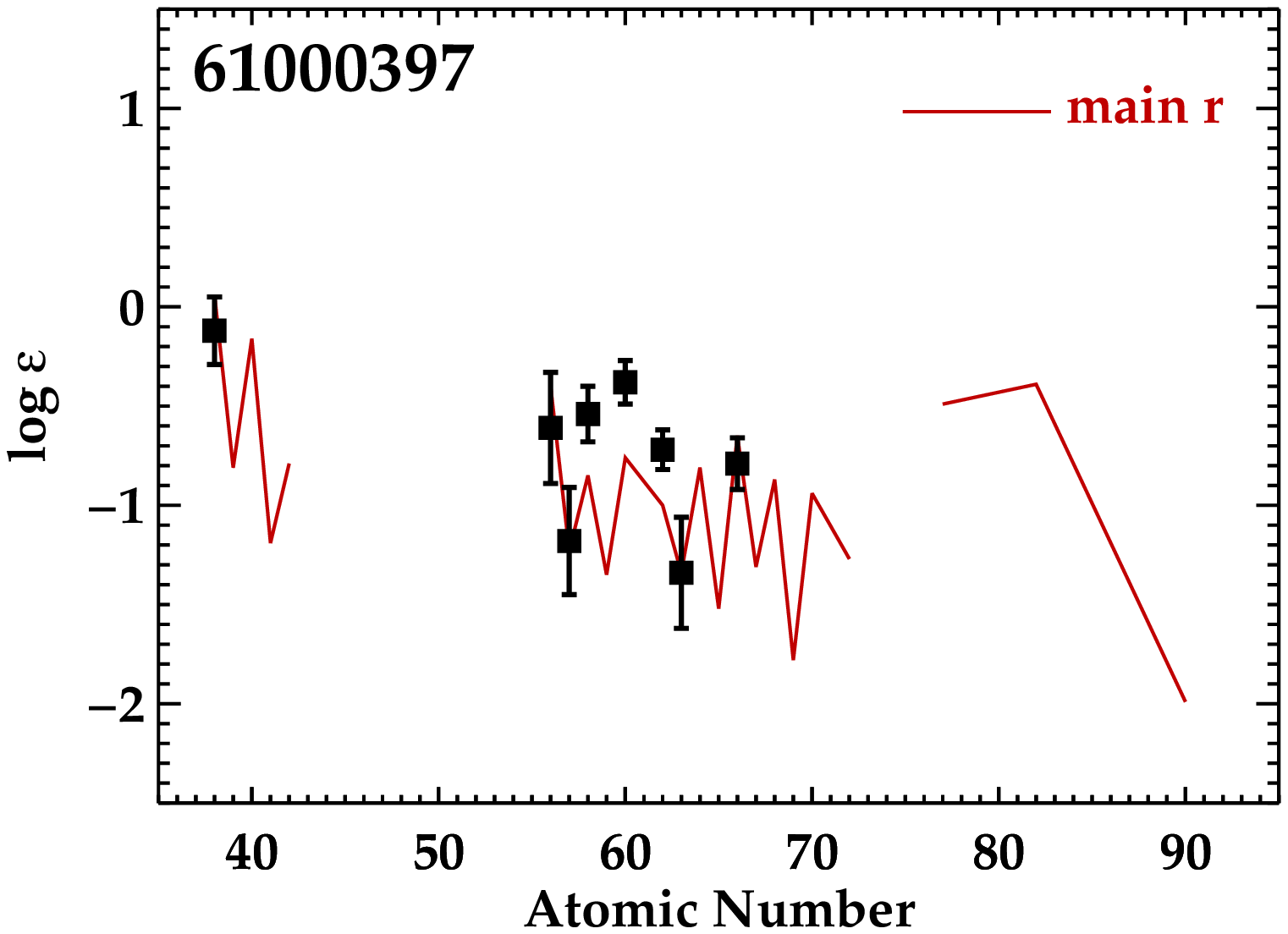} \\
\vspace*{0.05in}
\includegraphics[angle=0,width=1.625in]{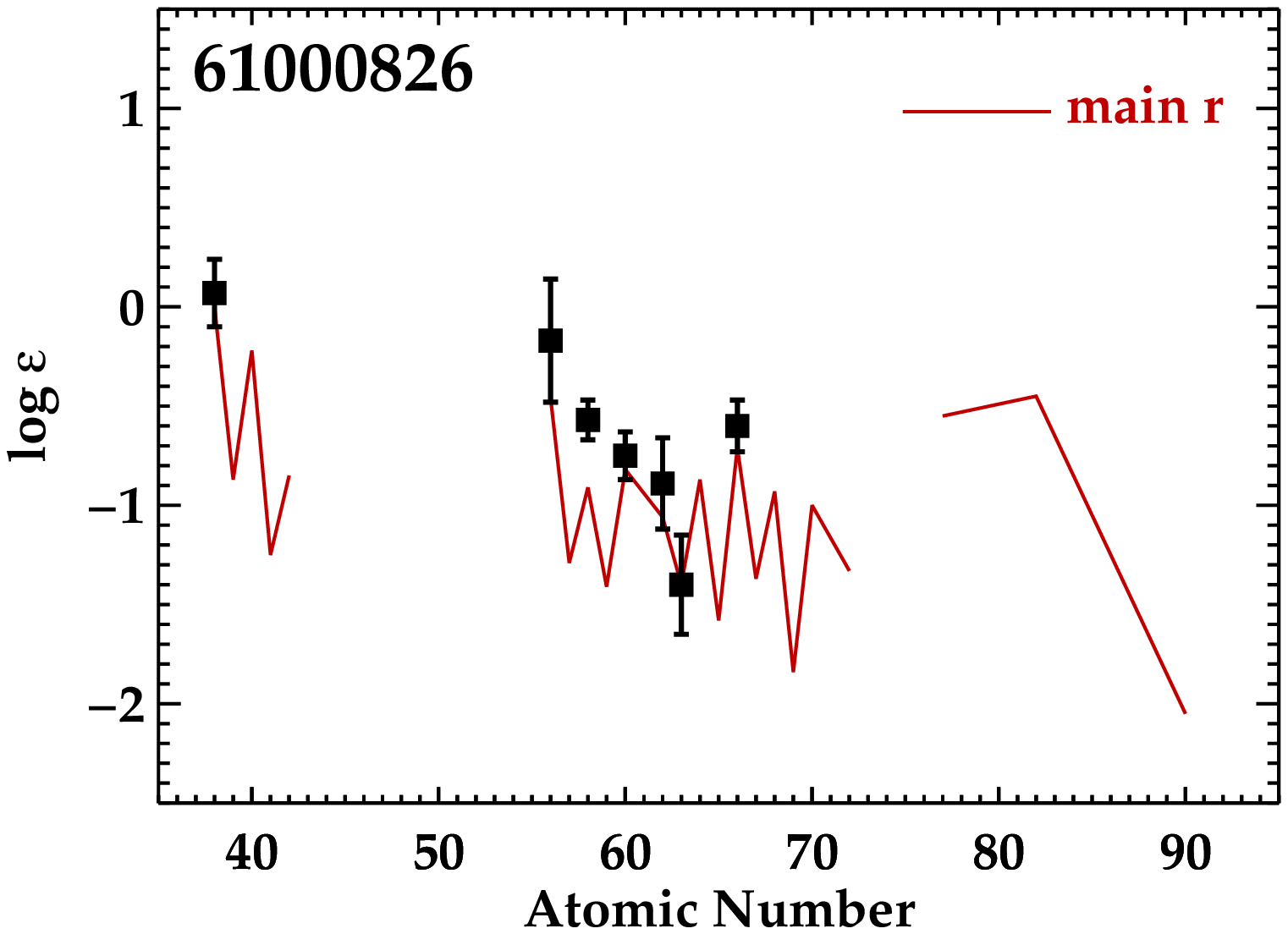}
\includegraphics[angle=0,width=1.625in]{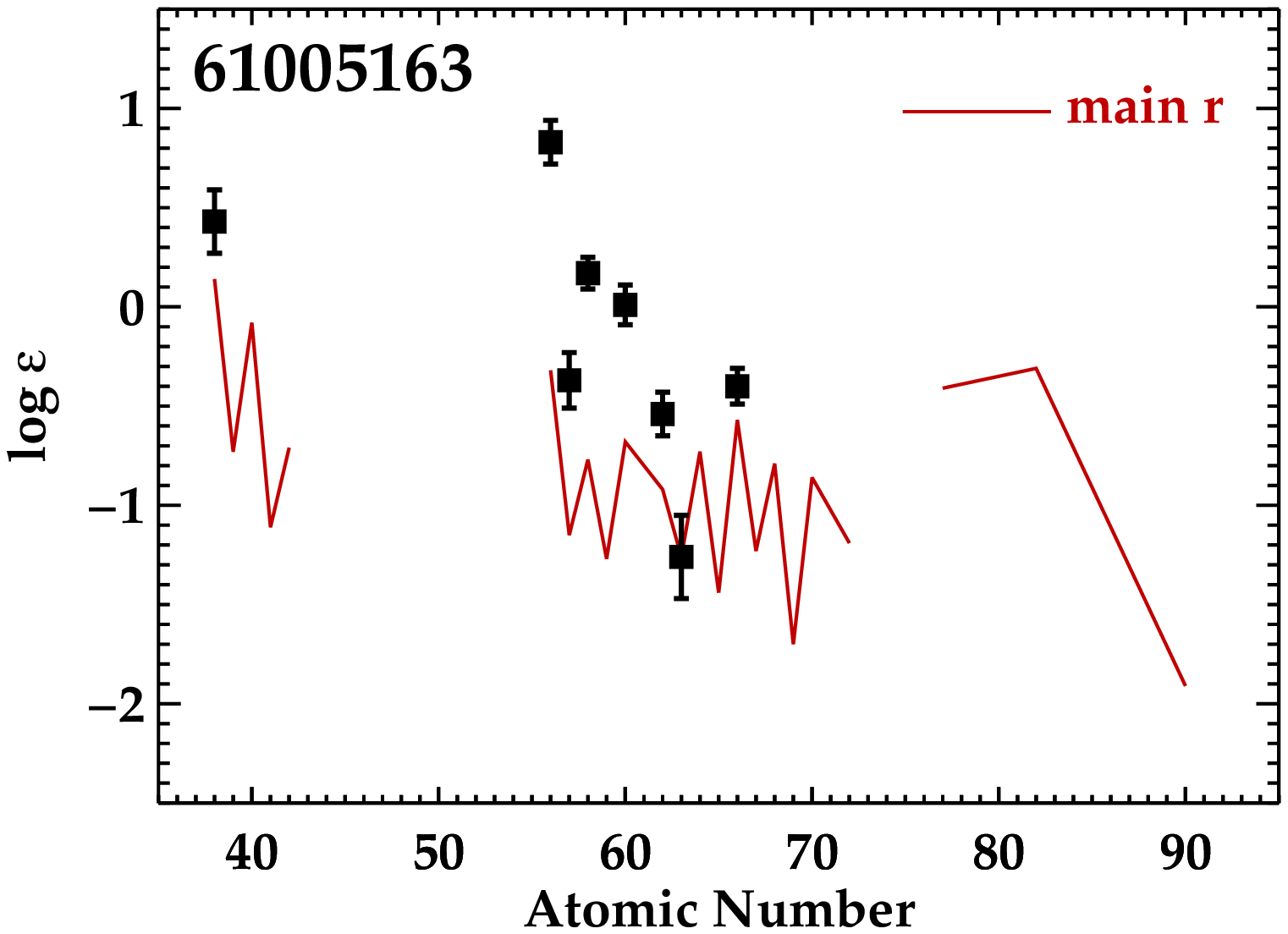}
\includegraphics[angle=0,width=1.625in]{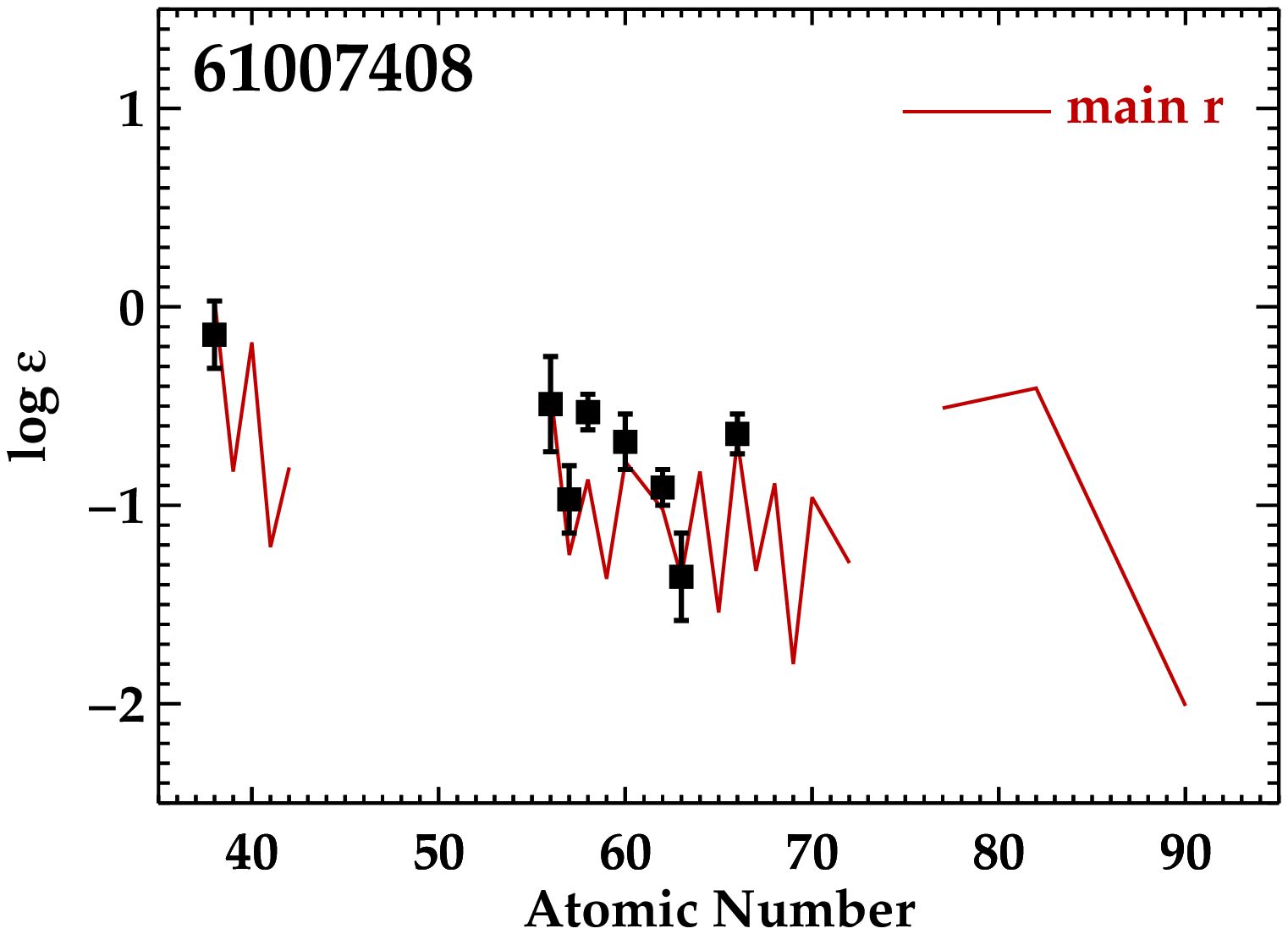}
\includegraphics[angle=0,width=1.625in]{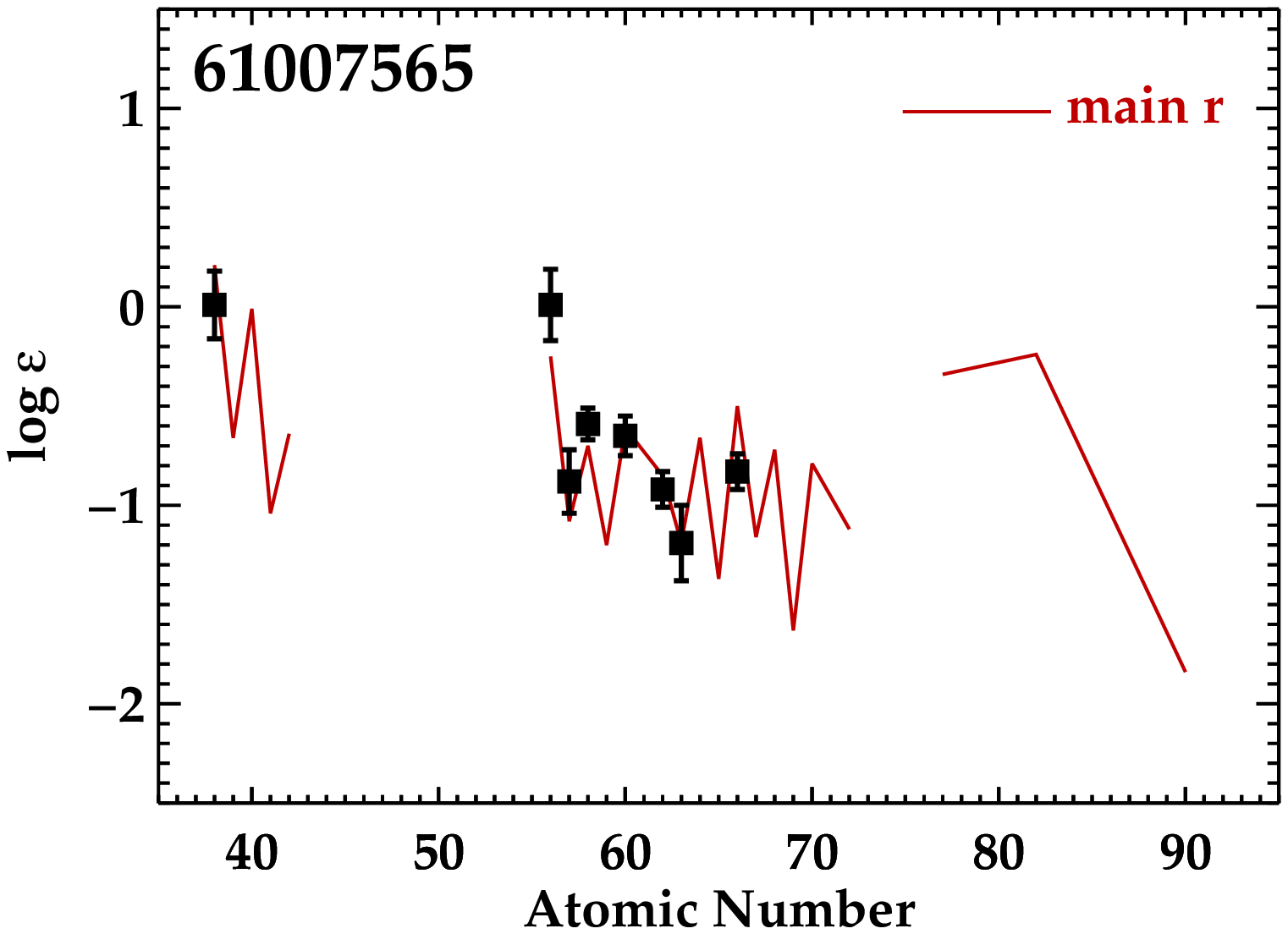} \\
\vspace*{0.05in}
\hspace*{-3.35in}
\includegraphics[angle=0,width=1.625in]{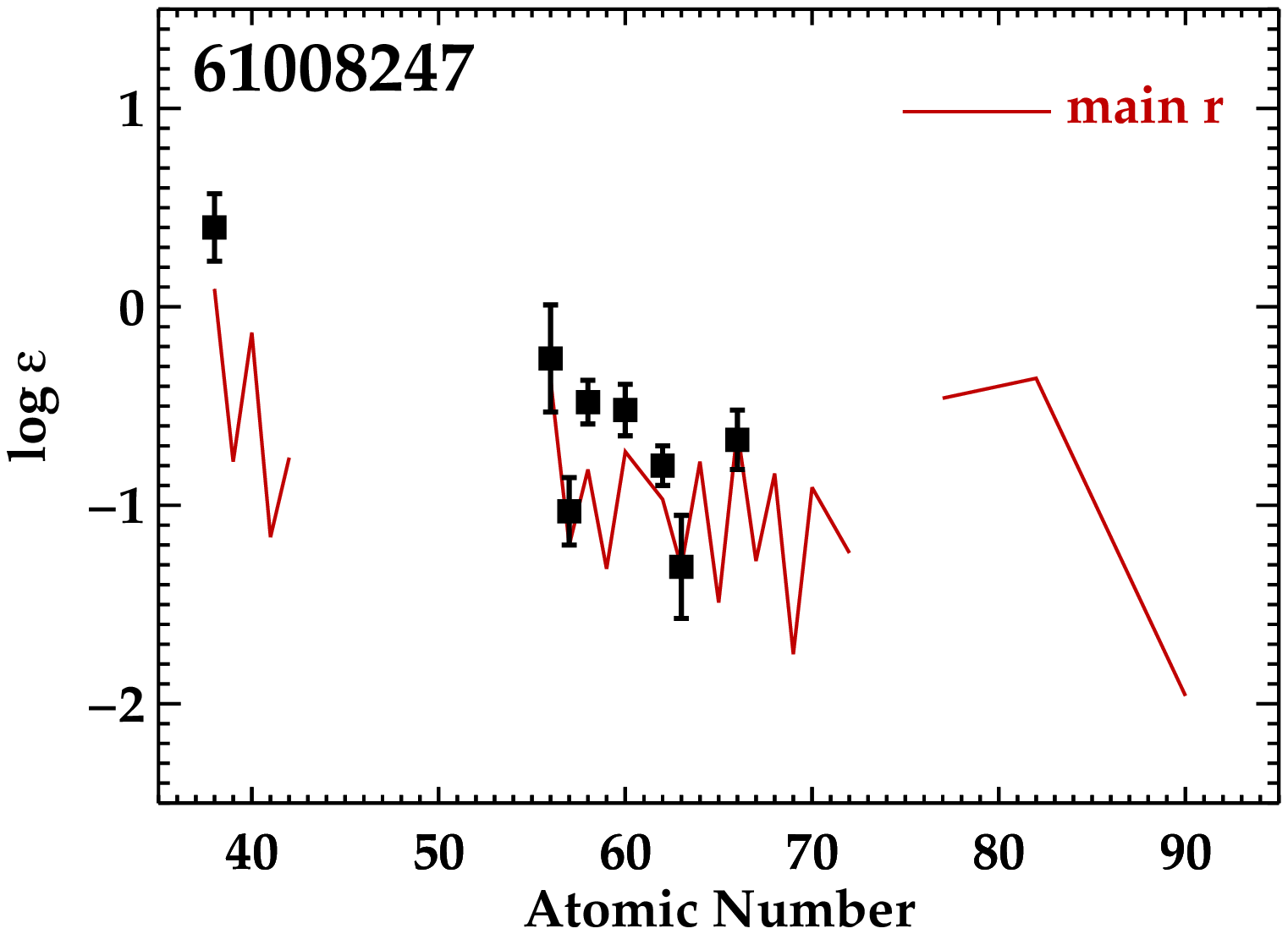}
\includegraphics[angle=0,width=1.625in]{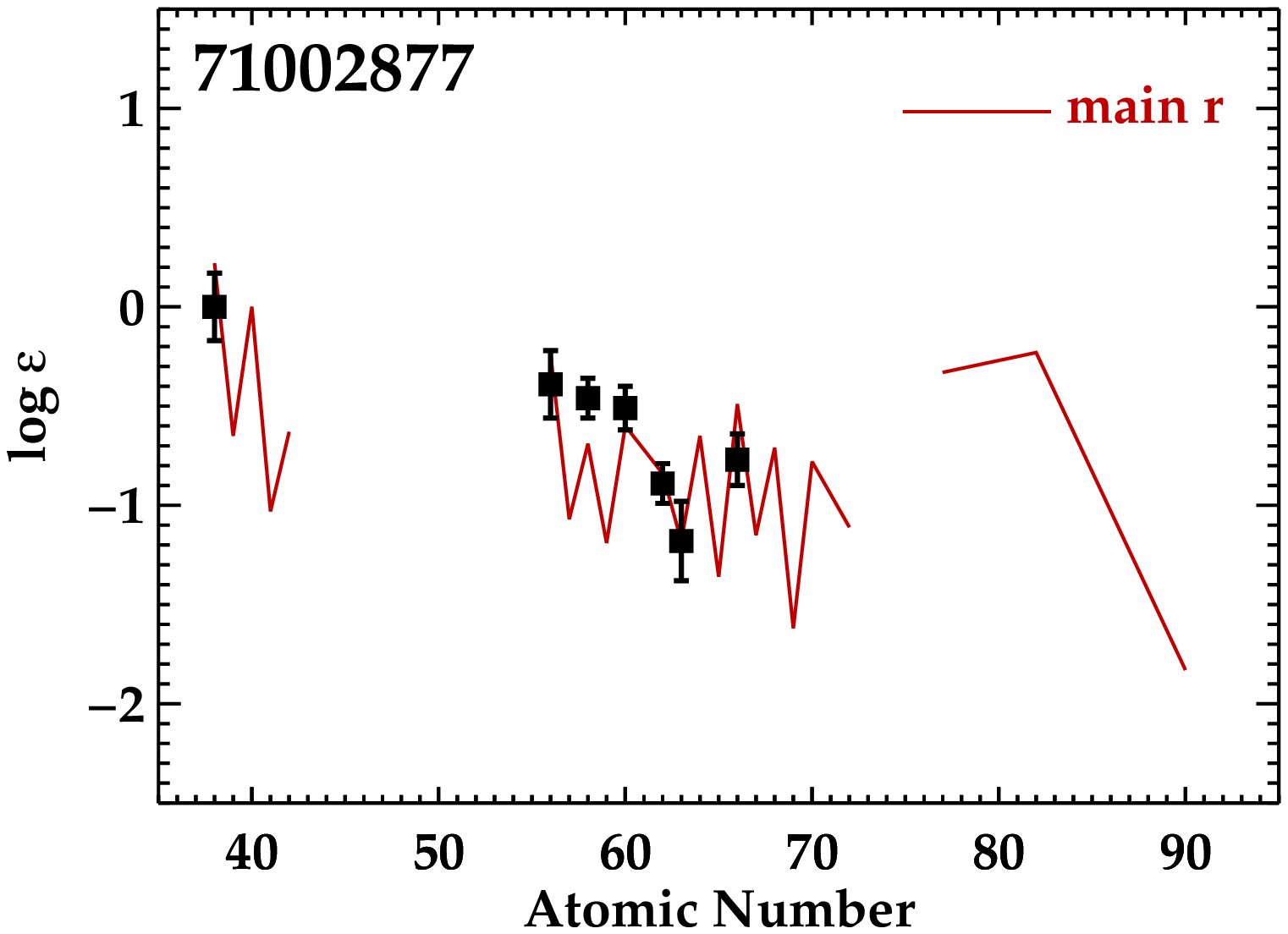} \\
\caption{
\label{ncapplots}
Heavy element abundance patterns derived from M2FS spectra
in all 26~stars examined.
The red line marks a template for the \rpro\ abundance pattern
(the star \mbox{CS~22892--052}; \citealt{sneden03,sneden09,roederer09}),
which is normalized to the Eu abundance in each star.
}
\end{figure*}

Eight elements heavier than the Fe-group are detectable 
in our M2FS spectra, and 17 are detectable
in our MIKE spectra.
The \ncap\ abundance
patterns are virtually identical
in 25 of the 26 stars examined,
as shown in Figure~\ref{ncapplots}.
Each panel in Figure~\ref{ncapplots} shows the 
logarithmic abundances in one star.
For comparison, a template for the ``main'' component
\citep{truran02}
of the \rpro\ pattern is shown.
Overall, the patterns observed in \ngc\ are a close match
to the \rpro\ pattern.
The [Ba/Eu] ratio is commonly used to quantify
the relative contributions of the 
$r$- and \spro.
In \ngc, $\langle$[Ba/Eu]$\rangle = -$0.66~$\pm$~0.05, 
in good agreement with the value found
in highly \rpro-enhanced stars, $-$0.71~$\pm$~0.05
\citep{roederer14b}.

The mean [Eu/Fe] ratio derived from the M2FS spectra, 
$\langle$[Eu/Fe]$\rangle = +$0.11~$\pm$~0.12,
is somewhat lower than that found in other low-metallicity
clusters, $\approx +$0.5~$\pm$~0.2 \citep{roederer11a}.
This ratio is commonly used to assess the overall
enhancement of \rpro\ material in a cluster.
\citet{sneden00} found that the \ncap\ elements
in globular cluster M92 are less abundant by $\approx$~0.3~dex
when compared with M15, another cluster at similar metallicity.
That result suggests there may be a genuine, though
small, cluster-to-cluster dispersion in 
the overall abundance of \ncap\ elements.
\ngc\ may lie at the low end of this dispersion.

Slight overabundances relative to this \rpro\ pattern
are detectable for Sr ($Z =$~38), Ba ($Z =$~56), 
Ce ($Z =$~58), and Nd ($Z =$~60).
It is well-known (e.g., \citealt{mcwilliam98,johnson02})
that the ratios between the light (e.g., Sr) and heavy (e.g., 
Ba or Eu) \ncap\ elements vary in low-metallicity field stars
even when little or no \spro\ contributions are present.
This phenomenon is also observed in globular clusters
(e.g., \citealt{yong14a,roederer15}),
and \ngc\ appears to be no exception in this regard.
The Ba abundances are derived from a single, saturated 
line at 4554~\AA, and they are highly sensitive to the
adopted \vt\ value.
Furthermore, the slight Ba, Ce, and Nd overabundances
can be attributed to contributions from both the
``weak'' and ``main'' components of the \rpro.
We would also expect to see significant Pb enhancement
if these overabundances come from \spro\ nucleosynthesis
(e.g., \citealt{roederer10b}),
since low-metallicity AGB stars are prodigious producers
of Pb (e.g., \citealt{gallino98,vaneck01}).
The low $\log\epsilon$(Pb/Eu) ratio observed in
star 42009955 confirms that little or no
\spro\ material is present.
We conclude that the \ncap\ elements in these 25~stars
were produced by some form of \rpro\ nucleosynthesis.
The \rpro\ is thought to occur in explosive 
environments like core-collapse supernovae 
or neutron-star mergers,
so this enrichment almost certainly occurred
before the present-day stars of \ngc\ formed.

\begin{figure*}
\centering
\includegraphics[angle=0,width=6.5in]{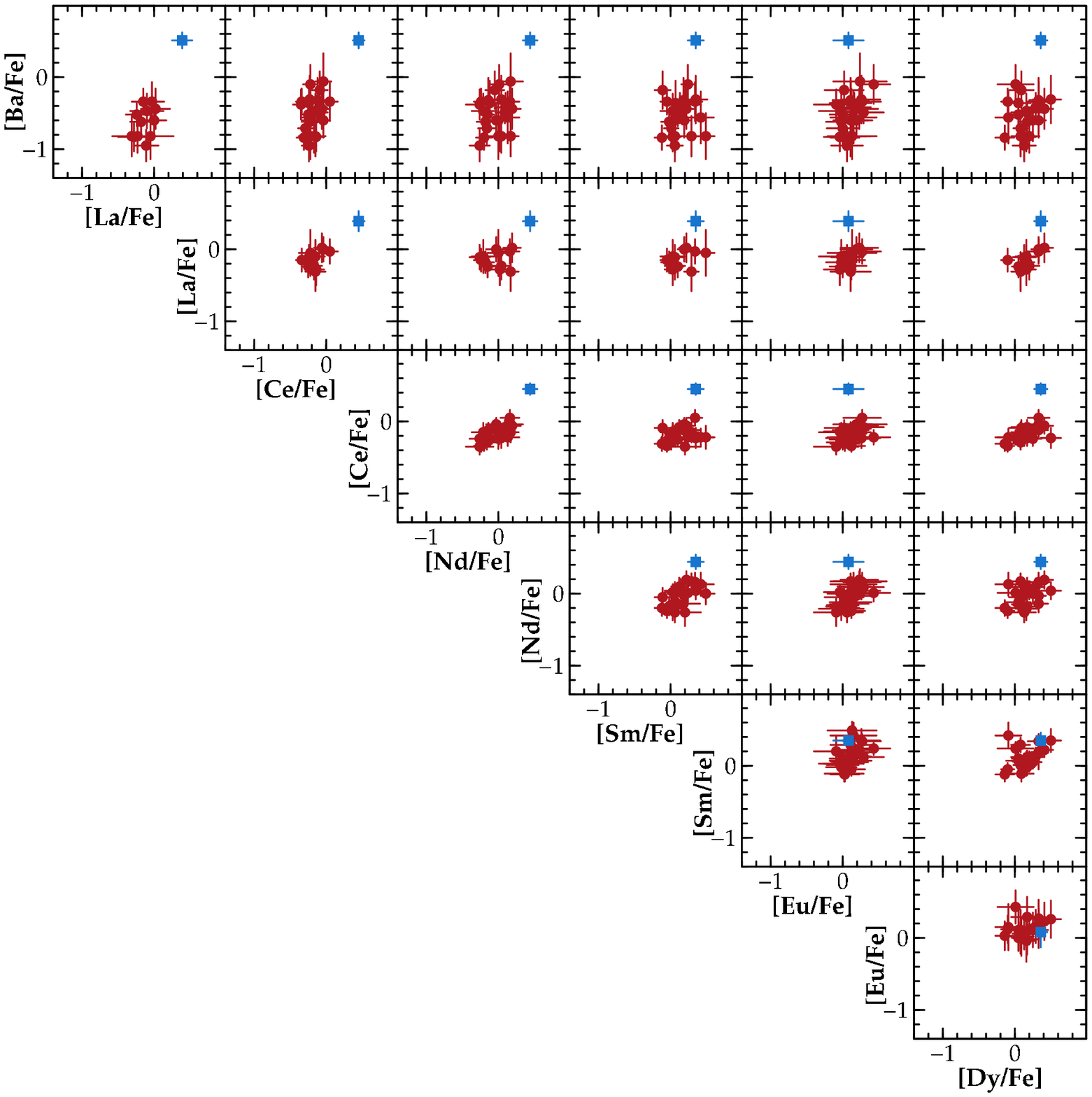}
\caption{
\label{multiabundplot}
Abundances of heavy \ncap\ elements in \ngc.
Red circles mark stars in the $r$-only group, 
and the blue square marks star 61005163.
}
\end{figure*}

Figure~\ref{multiabundplot} illustrates the
relationships between heavy \ncap\ elements found in \ngc.
The 25~stars with only \rpro\ products are shown with red circles.
These form a well-defined locus in each panel,
with the exception of the Ba abundances, which we 
have previously noted are
especially sensitive to the adopted \vt\ values.
Subtle correlations are apparent among some ratios, 
like [Ce/Fe] and [Nd/Fe].
This does not signal cosmic star-to-star dispersion within \ngc\
(cf.\ \citealt{roederer11a,cohen11}).
Rather, these correlations
are likely due to random uncertainties in the model atmosphere
parameters that impact the derived abundances
similarly for each element
\citep{roederer15}.
We conclude that the \ncap\ abundance patterns
within these 25~stars in \ngc\ are effectively identical.

\begin{figure*}
\centering
\includegraphics[angle=0,width=4.0in]{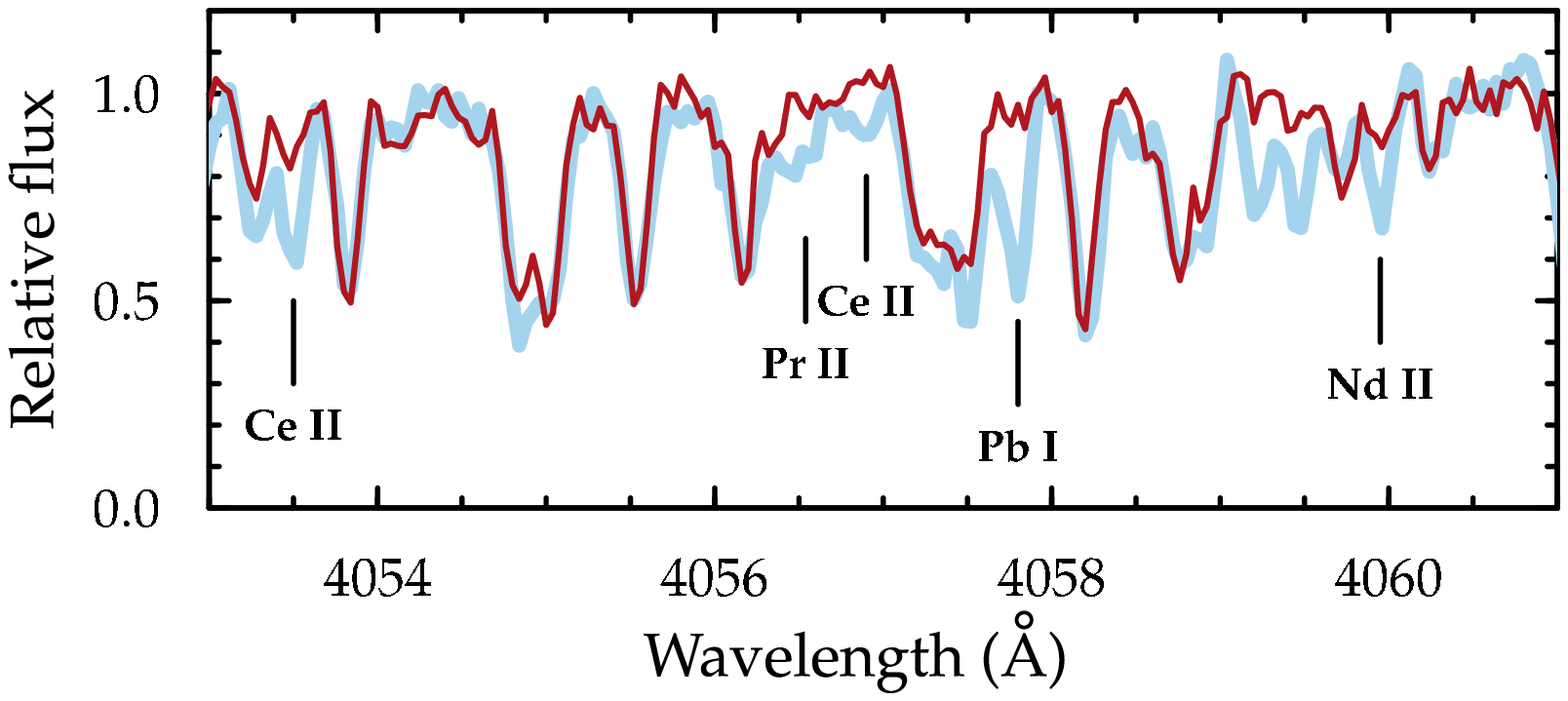} \\
\vspace*{0.05in}
\includegraphics[angle=0,width=4.0in]{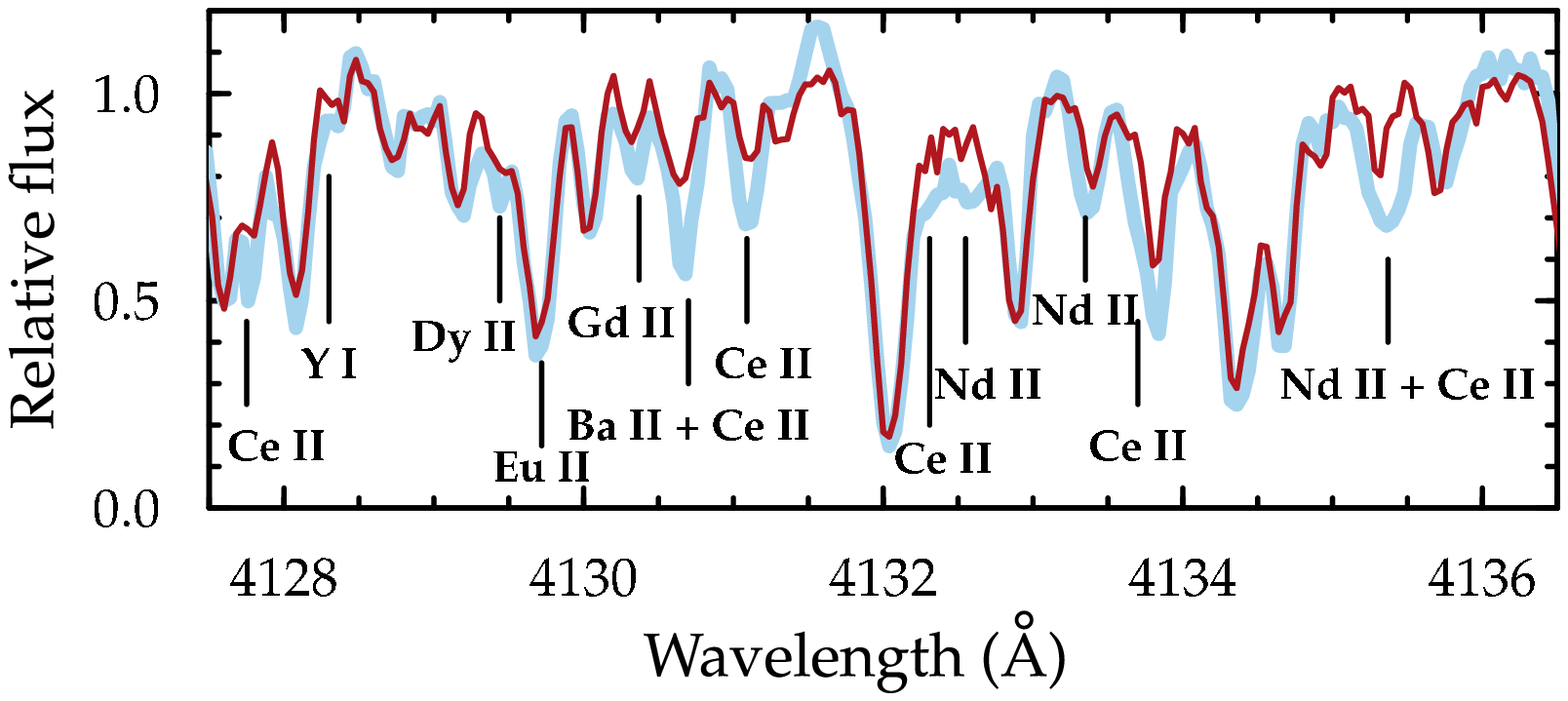} \\
\vspace*{0.05in}
\includegraphics[angle=0,width=4.0in]{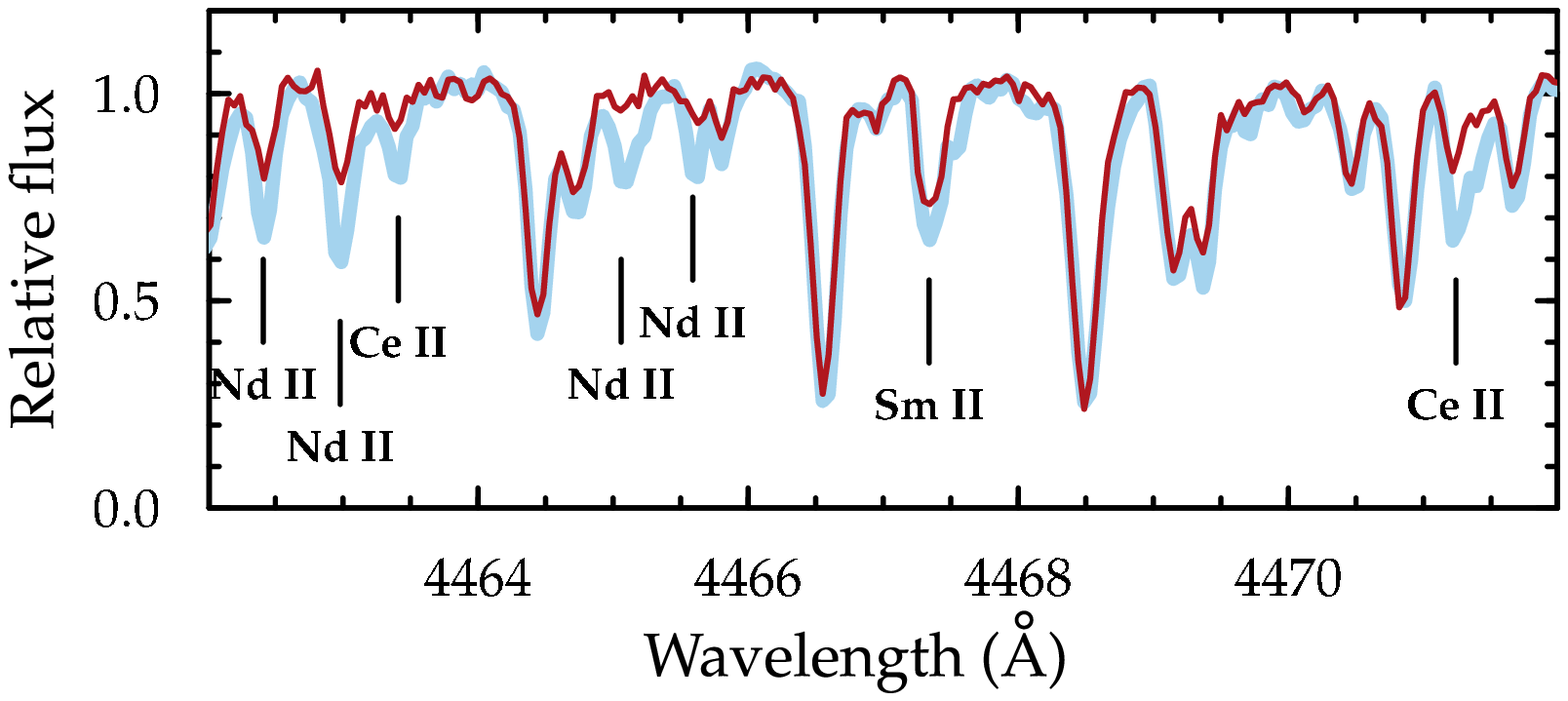} \\
\vspace*{0.05in}
\includegraphics[angle=0,width=4.0in]{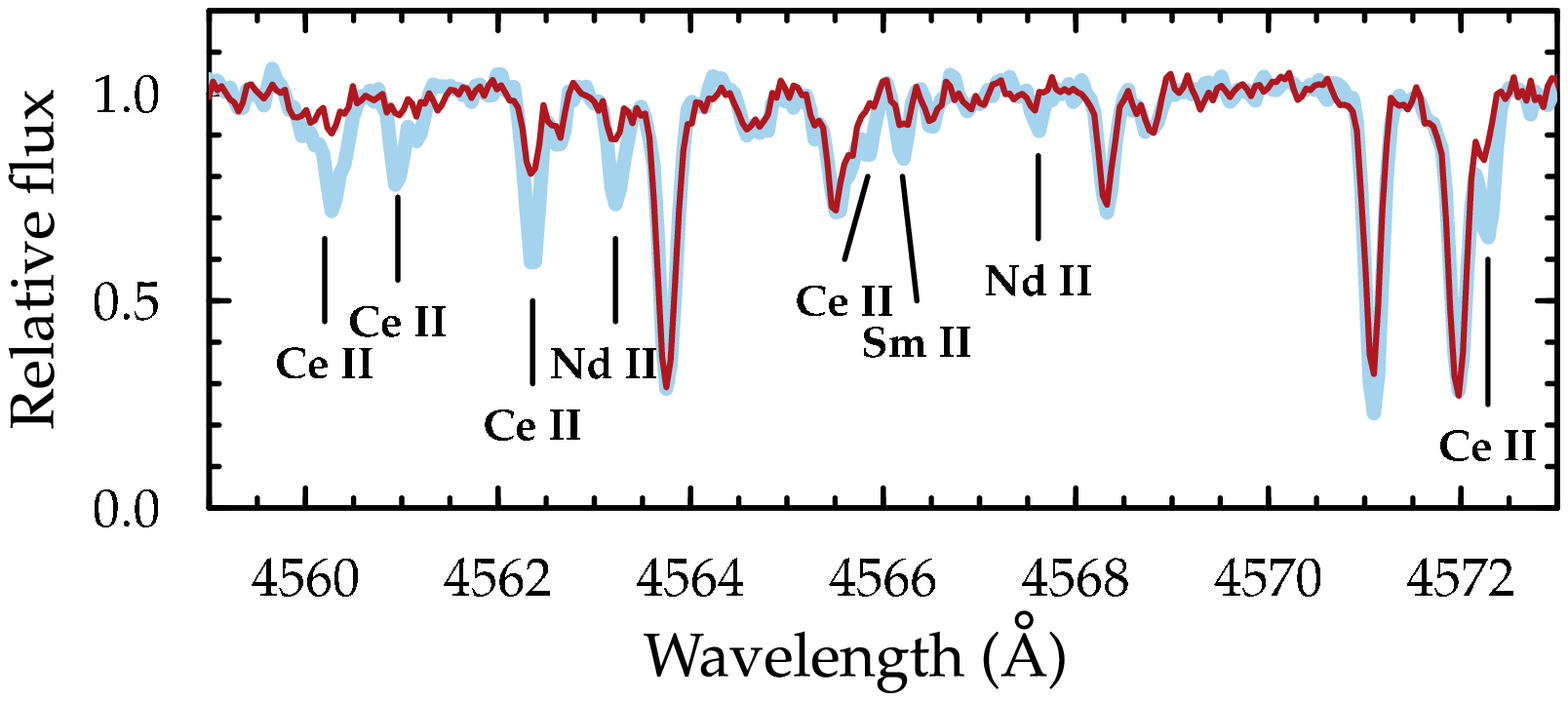} \\
\caption{
\label{heavyspecplot}
Selections of the MIKE spectra of stars 42009955 (thin red line)
and 61005163 (bold blue line) around various lines of \ncap\ elements.
The few differences in the top panel 
that do not correspond to \ncap\ element lines 
are due to differences in the CH abundance.
}
\end{figure*}

The one exception, star 61005163, shows large excesses of
Ba, La, Ce, and Nd in the M2FS spectrum.
This star is shown by the blue square in Figure~\ref{multiabundplot}.
Its [Sm/Fe], [Eu/Fe], and [Dy/Fe] ratios
are indistinguishable from the other stars.
These differences are immediately apparent from
visual inspection of the spectra.
Figure~\ref{heavyspecplot} shows sections of the MIKE spectra
of stars 42009955 and 61005163
surrounding some of the \ncap\ lines.
These two stars have identical stellar parameters
and metallicities,
so differences in their spectra can be attributed
to unequal abundances.
Most of the absorption lines are formed by
Fe-group elements whose abundances
are identical in the two stars (Section~\ref{fegroup}),
and their line strengths are identical, too.
Lines of \ncap\ elements are marked, and
these lines account for nearly all of the differences in the spectra.

\begin{figure}
\centering
\includegraphics[angle=0,width=3.25in]{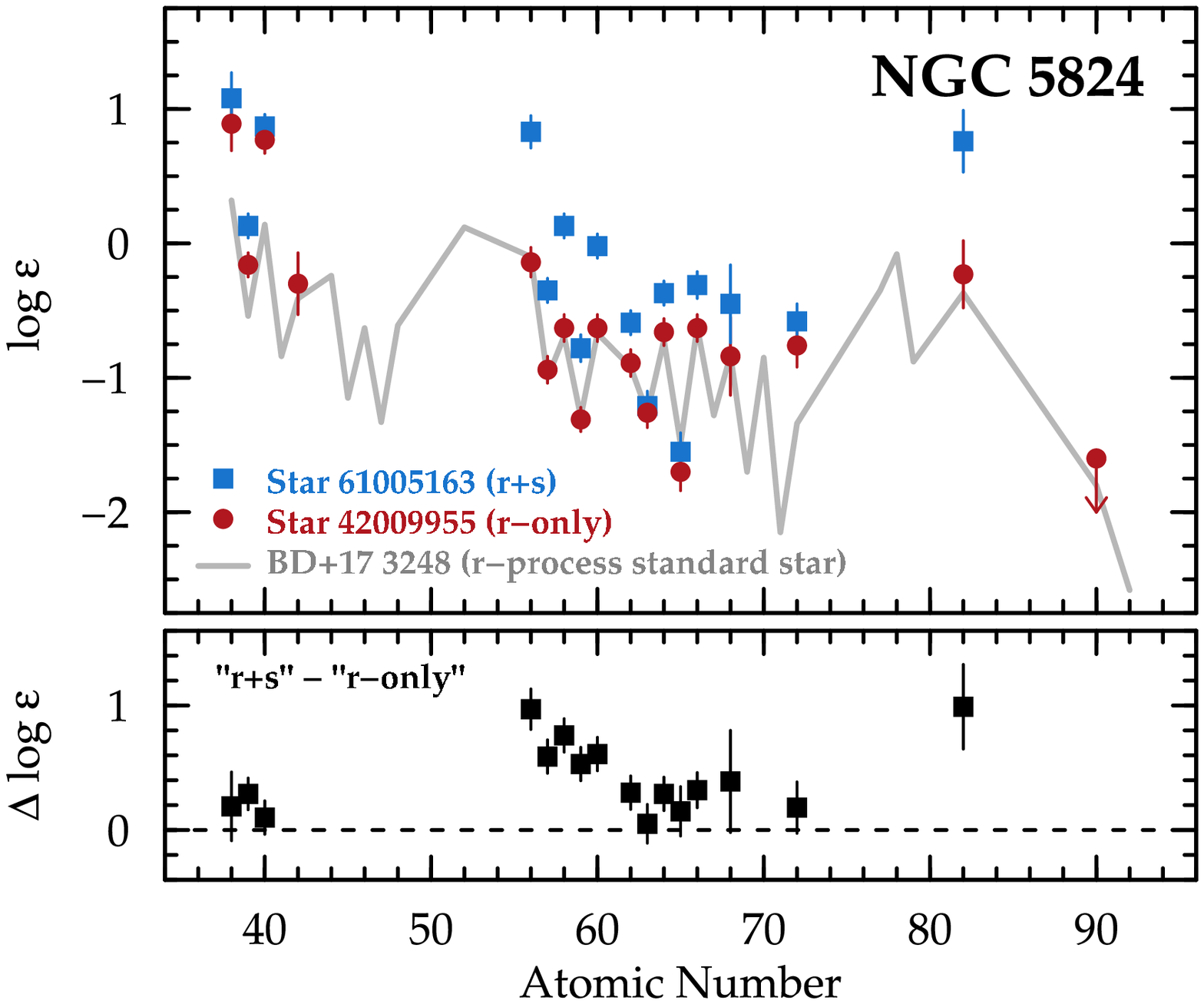}
\caption{
\label{logepsplot}
Logarithmic abundances of the \ncap\ elements in \ngc\
as derived from the MIKE spectra.
In the top panel, 
blue squares mark star 61005163, and the
red circles mark star 42009955.
The gray line marks the \ncap\ element abundance
pattern observed in the metal-poor 
\rpro-enhanced standard star \mbox{BD$+$17~3248}
\citep{cowan02,cowan05,sneden09,roederer09,roederer10a,roederer12a}.
This pattern is normalized to the Eu abundance.
In the bottom panel,
the squares mark the offset between
the two stars in \ngc,
characterized as ``$r+s$ minus $r$-only.''
The dashed line represents a difference of zero.
}
\end{figure}

Figure~\ref{logepsplot} shows the \ncap\ abundance 
patterns derived from the MIKE spectra of stars 42009955 and 61005163.
The metal-poor ([Fe/H]~$= -$2.1)
\rpro-enhanced standard star
\mbox{BD$+$17~3248} is shown for comparison.
The abundances of most \ncap\ elements 
are different between the two stars.
No elements have a lower abundance in star 61005163
than in star 42009955.
\mbox{BD$+$17~3248} is a good match for the elements
with $Z \geq$~56 in star 42009955,
but not for these elements in star 61005163.
We refer to the abundance pattern in star 61005163 as the 
``$r+s$'' pattern, and we refer to the abundance pattern
in star 42009955 as the ``$r$-only'' pattern.

Following the procedure outlined in \citet{roederer11c} for
the two \ncap\ patterns in globular cluster M22,
we subtract the abundance pattern in the $r$-only
group from the abundance pattern in the $r+s$ group
to reveal the ``\spro\ residual'' pattern.
This method assumes that the (\rpro) ``foundation''
is identical in all stars in \ngc,
but it makes no assumptions about the 
nucleosynthesis origins of the foundation.
The difference between the two stars---the \spro\ residual---is
illustrated in the bottom panel of Figure~\ref{logepsplot}.
Note that the abundance differences are
largely insensitive to non-LTE effects that affect line formation
(for, e.g., Pb; \citealt{mashonkina12}),
since stars 42009955 and 61005163 have identical stellar parameters
and metallicities.

\begin{figure}
\centering
\includegraphics[angle=0,width=2.86in]{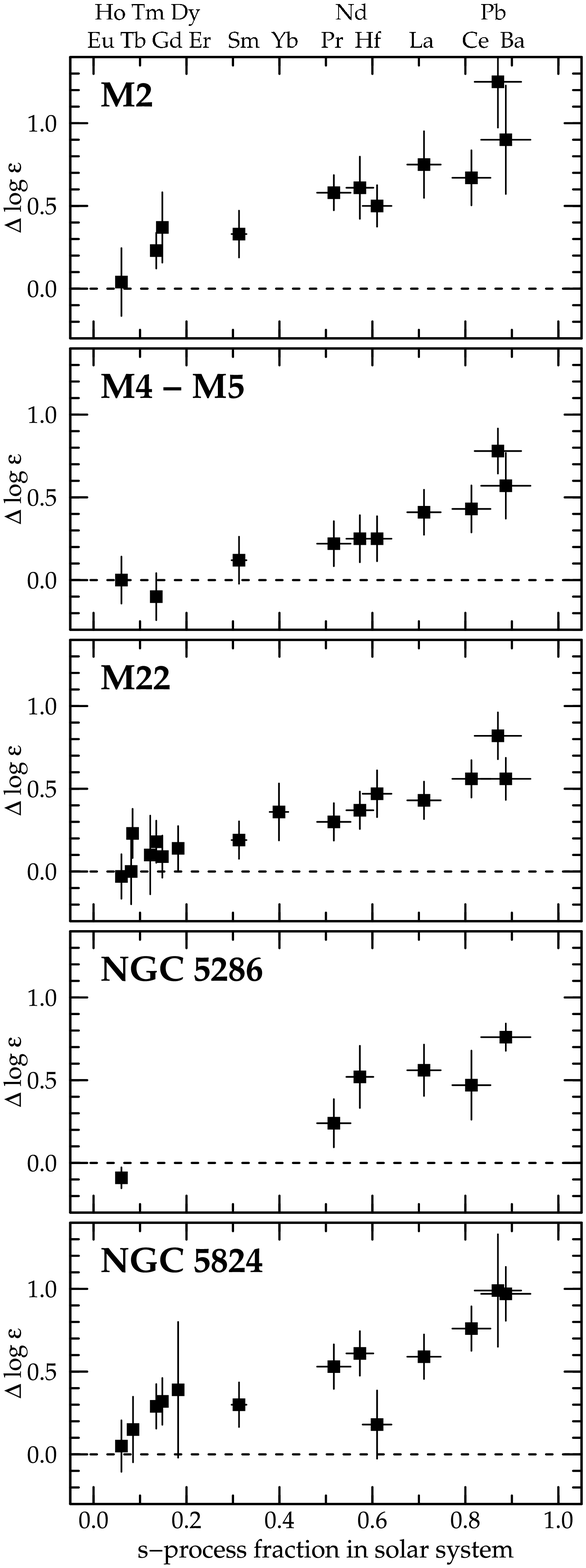}
\caption{
\label{multideltasplot}
Abundance enhancement due to the \spro\ 
in globular clusters M2 \citep{yong14b}, 
M4 \citep{ivans99,ivans01,yong08a,yong08b}, 
M22 (Roederer et al.\ \citeyear{roederer11c}),
\mbox{NGC~5286} \citep{marino15}, and \ngc.
In M2, M22, \mbox{NGC~5286}, and \ngc, 
the abundance differences 
are computed in the sense of ``$r+s$ minus $r$-only.''
In M4, the difference is computed in the sense of 
``M4 minus M5.''
The \spro\ solar system fractions are taken from 
the models of \citet{bisterzo11}.
Only elements with $Z \geq$~56 are shown,
since processes other than the $r$- and \spro\ may 
contribute to the lighter \ncap\ elements.
}
\end{figure}

An unmistakable correlation emerges 
when these differences are plotted as a function
of the \spro\ contribution
to the solar system abundance of each element,
as shown in the bottom panel of Figure~\ref{multideltasplot}.
Elements with minimal \spro\ contribution ($<$~10~per cent; Eu, Tb)
to their solar abundances show no difference
between the $r$-only and $r+s$ groups.
Elements with major \spro\ contributions ($>$~80~per cent; Ba, Ce, Pb)
show the largest differences.
Other \ncap\ elements fall between these two extremes.
The lone exception, Hf, only deviates from the mean trend by
$<$~2~$\sigma$.
We regard this as compelling evidence that \spro\ nucleosynthesis is 
responsible for the differences in the abundance patterns
between stars 42009955 and 61005163.

To summarize,
we find that 25 of the 26~stars 
observed in \ngc\
share a common \ncap\ element abundance pattern
consistent with a nucleosynthetic origin in
some form of \rpro.
One of these 26~stars 
shows substantial amounts of \spro\ material, as well.

\subsection{Natal Enrichment, Self-Enrichment, or 
Mass Transfer from a Companion?}
\label{natal}

We have found only one star with an unusual abundance pattern
in \ngc, which raises the possibility that
the composition of the atmosphere of 
this star does not reflect its natal composition.
In this section, we consider, and dismiss, the 
possibilities that star 61005163 received its \spro\ enhancement
by self-enrichment or 
by mass-transfer from a companion star that passed through the 
AGB phase of evolution.
As a reminder, in this context ``self-enrichment'' refers to 
the mechanism by which the composition of the 
surface layers of a star are changed
through internal nucleosynthesis and/or mixing episodes,
not polluters from one stellar generation 
enriching other stars in a later generation.

Star 61005163 ($M_{\rm bol} = -$2.76) 
is not luminous enough to have 
produced the observed \spro\ material through self-enrichment.
The latest update to the Full-network Repository of Updated
Isotopic Tables \& Yields (FRUITY, version 4; \citealt{cristallo15})
database presents physical parameters and 
\spro\ yields for AGB stars with initial masses
from 1.3 to 6.0~\msun.
The models for low-metallicity stars have
$-$5.2~$< M_{\rm bol} < -$3.5 
during the thermally-pulsing AGB (TP-AGB) phase of evolution,
when \spro\ nucleosynthesis occurs.
Several stars in our sample are more luminous
than star 61005163, and none shows evidence of 
\spro\ enhancement.
We therefore discard the possibility that
star 61005163 is an intrinsic \spro-enriched AGB star.

Ba enhancement is always accompanied by C enhancement
in low-metallicity field stars
whenever the Ba enhancement does not originate from \rpro\
nucleosynthesis
(e.g., \citealt{aoki07}, Sneden et al.\ \citeyear{sneden08}, 
\citealt{allen12}).
Such stars are classified as carbon-enhanced metal-poor stars
with \spro\ enhancement, or \mbox{CEMP-$s$}
\citep{ryan05,beers05}.
Frequently, these stars are found in binary (or multiple)
star systems, as revealed by radial velocity variations.
They are presumed to have acquired their carbon and \spro\ 
enhancement from a more massive companion that 
evolved through the AGB phase of evolution
and transferred the C and \spro\ material
to the longer-lived star, which we now observe.

Star 61005163 does not exhibit radial velocity variations
in our two observations spaced $\approx$~2~months apart.
This information alone does not exclude the possibility 
of a companion, since the system
could be observed face-on or have a very long ($\ga$~10$^{3}$~d)
orbital period.

Star 61005163 is not C-enhanced,
and it probably was not C-enhanced 
before evolving up the RGB (Section~\ref{light}).
It is also not C-enhanced 
when evaluated by its C/O ratio,
which is $\ll$~1, even if the LTE O abundance
is overestimated by 1~dex.
Both models and observations support this view.
All of the low-metallicity models of \citet{cristallo15}
end the TP-AGB phase with C/O~$>$~1.
The compilation of \citet{masseron10} lists 11~\mbox{CEMP-$s$}
stars with [Fe/H]~$> -$3,
and all have C/O~$>$~1.

The [Ba/Fe], [La/Fe], [Ce/Fe], [Pb/Fe], and other ratios
in \mbox{CEMP-$s$} stars
are usually several dex higher than the solar ratios
and significantly higher than the 
ratios observed in star 61005163 in \ngc\
(e.g., \citealt{vaneck01,sneden03b,aoki08,allen12}).
The lack of radial velocity variations,
lack of C enhancement, and 
relatively low [\ncap/Fe] ratios 
collectively support our assertion that star 61005163 is not
related to the stars in the \mbox{CEMP-$s$} class.

Searches for Ba-enhanced stars in normal globular clusters
have found that the fraction of such stars
(5 out of 1205, or 0.4~per cent; \citealt{dorazi10})
is lower than in the field
($>$~8~per cent; \citealt{aoki15}).
Only two such stars in clusters
have been confirmed observationally
from the analysis of many elements
\citep{kacharov13,cordero15}.
The \ncap\ abundance pattern in one of them,
\mbox{Lee 4710} in cluster \mbox{47~Tuc},
is consistent with \spro\ enrichment from an AGB companion
with an initial mass of $\sim$~1.3~\msun\ (\citeauthor{cordero15}) 
No such comparison was performed for the other star, 
in globular cluster M75.
One simple explanation for the lack of Ba-enhanced stars
could be the dense stellar environments 
of globular clusters,
which would disrupt most wide binary systems and
produce a lower binary fraction than in the field
(cf., e.g., \citealt{cote96,mayor96,milone12}).

One star out of 26 (4~per cent) analyzed
by us shows an anomalous abundance pattern.
Finding additional stars with abundance patterns like star 61005163
in \ngc\ would greatly alleviate the concerns that
the composition of this particular star 
does not reflect its natal composition.
At first glance, this percentage is much lower
than the percentage of spectroscopically-confirmed members 
of the $r+s$ groups
in other clusters:\
M2 (40~per cent, 10~stars total), 
M22 (40~per cent, 35~stars total), and 
\mbox{NGC~5286} (43~per cent, 7 stars total).
However, it is important to recognize that those samples
are highly biased in favor of stars in the 
$r+s$ groups.
More representative percentages can be estimated
from photometric analyses of the multiple
subgiant or red giant sequences (which correspond to the
$r$-only and $r+s$ groups), for example.
The populations associated with the $r+s$ groups in 
M2, M22, and \mbox{NGC~5286} comprise only
3~per cent, 40~per cent, and 14~per cent of stars
\citep{milone15c,piotto12,marino15}.
\ngc\ is not an outlier with respect to other
complex clusters in this regard.
It would be of great interest to
obtain new ultraviolet broadband photometry of 
\ngc\ with the \textit{Hubble Space Telescope}
(cf.\ \citealt{piotto15})
to search for the presence of multiple sequences in this cluster.

\subsection{The Origin of the $s$-process Material}
\label{sprocess}

Stars in most low-metallicity globular clusters
only show enrichment patterns like
that found in the $r$-only group in \ngc.
Three other low-metallicity clusters 
host groups of stars resembling the 
distinct $r$-only and $r+s$ patterns:\
M2 \citep{yong14b},
M22 (\citealt{marino09,marino11}; Roederer et al.\ \citeyear{roederer11c}),
and
\mbox{NGC~5286} \citep{marino15}.
All stars in another low-metallicity cluster, M4,
resemble the $r+s$ pattern
\citep{ivans99,yong08a,yong08b},
thus distinguishing this cluster
from other low-metallicity ones.
In this section we examine the \ncap\ elements in M2, M4, M22, and 
\mbox{NGC~5286}, along with \ngc,
to attempt to constrain the 
mass range of AGB stars that 
may have polluted the cluster ISM
from which star 61005163 formed.

Figure~\ref{multideltasplot} illustrates the differences
between the $r+s$ and $r$-only groups in these five clusters.
In all cases, $\Delta \log \epsilon$ is calculated as
``$r+s$ minus $r$-only''
(or ``$s$-rich minus $s$-poor'' in other nomenclature; 
e.g., \citealt{marino11}).
We use the globular cluster M5, whose \ncap\ elements
originated mainly in some form of \rpro\ nucleosynthesis
\citep{ivans01,yong08a,yong08b,lai11},
as a foil for M4 to perform the subtraction ``M4 minus M5.''
Qualitatively, the abundance patterns in these five 
physically-unrelated clusters
are remarkably similar,
which could indicate that they share
similar chemical enrichment histories.

\begin{figure}
\centering
\includegraphics[angle=0,width=3.25in]{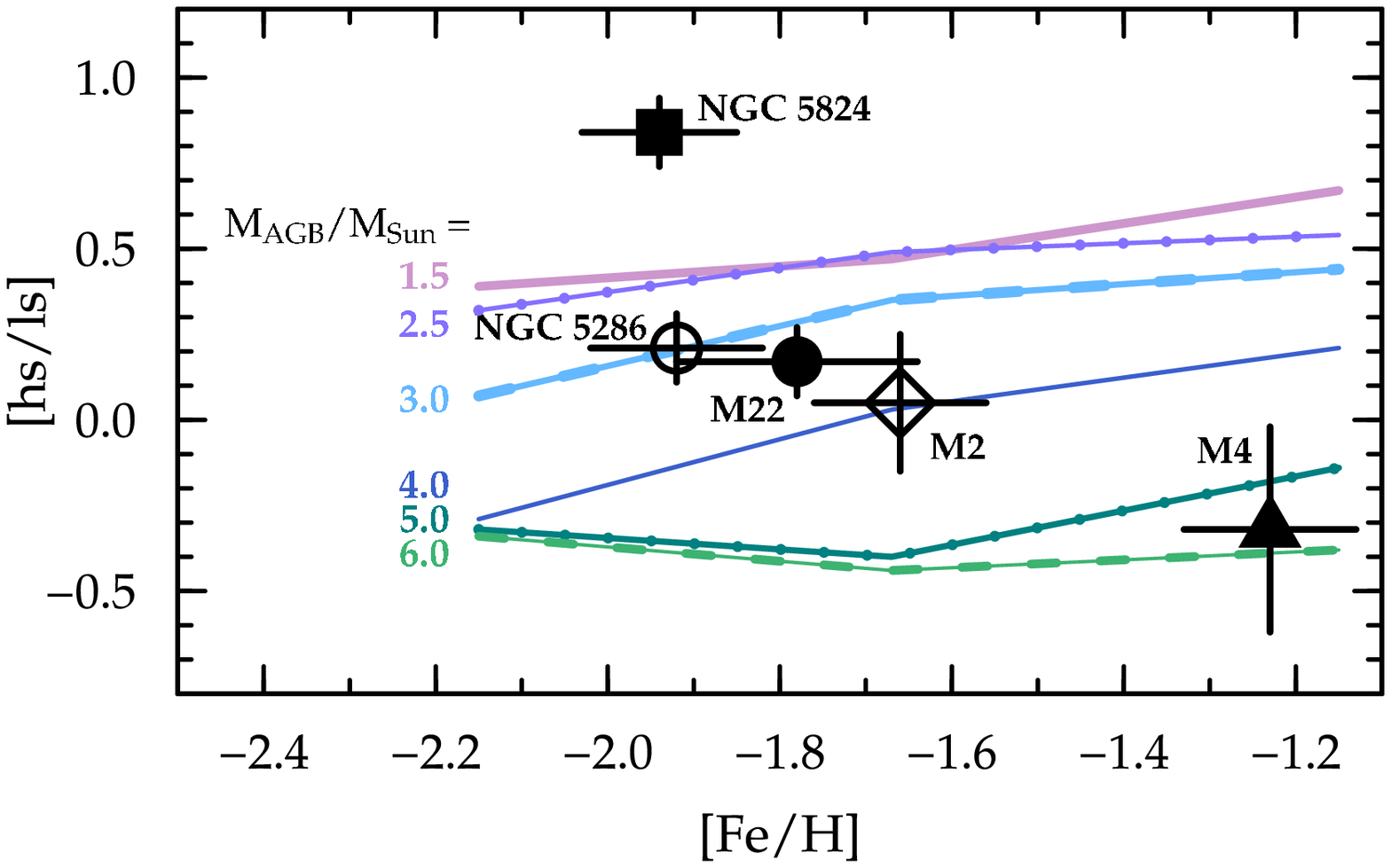} \\
\vspace*{0.1in}
\includegraphics[angle=0,width=3.25in]{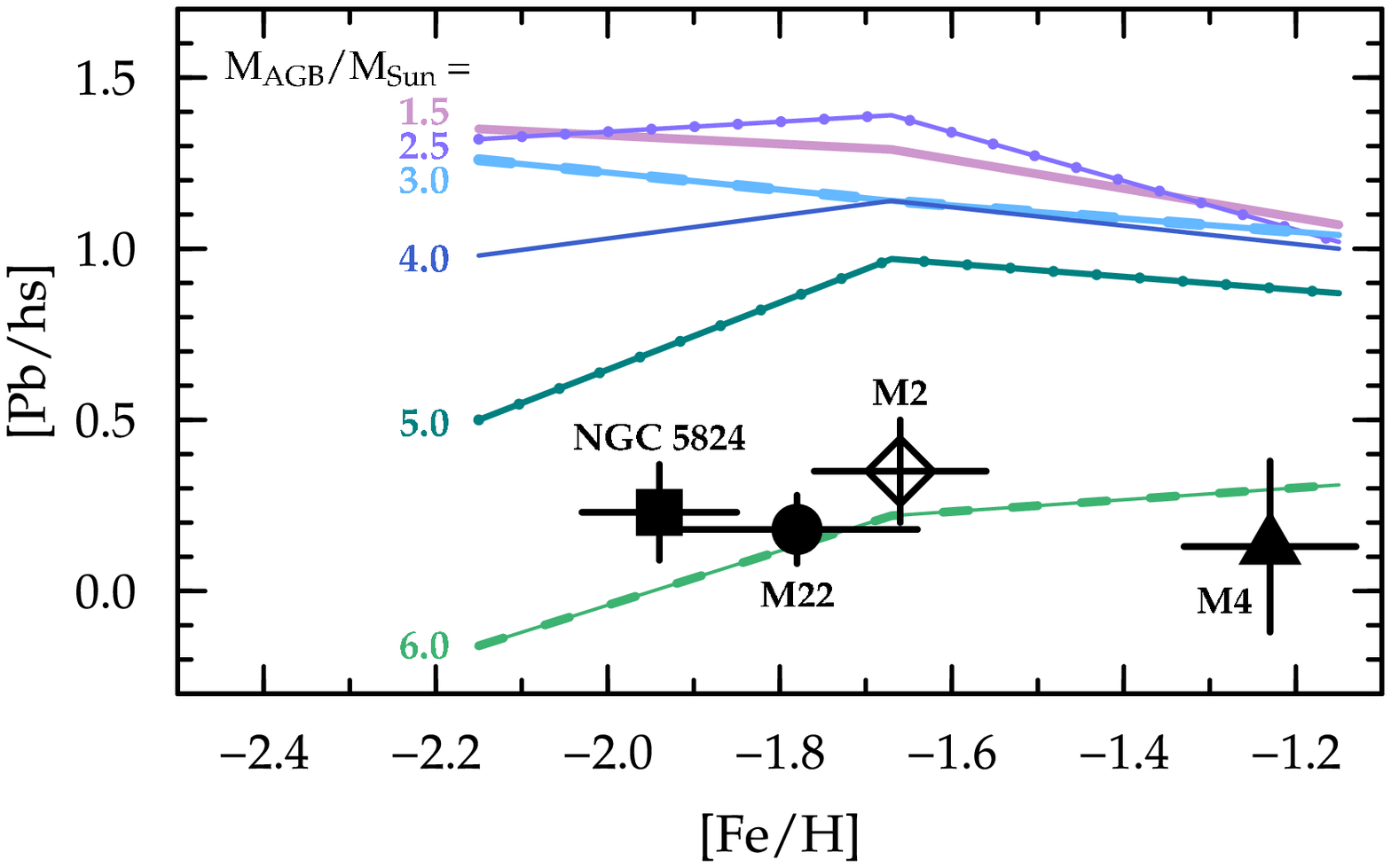}
\caption{
\label{fruityplot}
Comparison of the \hsls\ and \pbhs\ indices
in five globular clusters and 
AGB model predictions from the FRUITY database \citep{cristallo15}.
In M4, the \spro\ residual is computed using the 
abundance pattern in the 
(physically unrelated) globular cluster
M5 as the $r$-only group.
In \ngc, the difference is computed in the sense of
``star 61005163 minus 42009955.''
We recompute the \hsls\ and \pbhs\
indices using our definitions and the
yields of the FRUITY database.
}
\end{figure}

We compute a few indices from well-measured
abundance ratios 
to serve as indicators of the \spro\ nucleosynthesis patterns.
We define
[$ls$/Fe] $=$ ([Y/Fe]$+$[Zr/Fe])/2,
[$hs$/Fe] $=$ ([La/Fe]$+$[Ce/Fe]$+$[Nd/Fe])/3,
\hsls\ $=$ [$hs$/Fe]$-$[$ls$/Fe], and
\pbhs\ $=$ [Pb/Fe]$-$[$hs$/Fe].
These definitions are constructed so that all ratios
are measured in each of the five clusters,
and thus they differ slightly from some
versions of [$ls$/Fe] and [$hs$/Fe] in the literature.
These indices are computed
from the $\spro$ residual in each cluster,
not the derived ratios in the $r+s$ groups.
The indices \hsls\ and \pbhs,
unlike [X/Fe] ratios,
are insensitive to the details of how material
was acquired by the star observed today.
For \ngc,
\hsls~$= +$0.84~$\pm$~0.10, and
\pbhs~$= +$0.23~$\pm$~0.14.
Qualitatively speaking, the elements at the
second \spro\ peak (Ba, La, Ce, Pr, Nd) are significantly
overproduced relative to the elements at the first \spro\ peak
(Sr, Y, Zr)
and mildly underproduced relative to one of the elements
at the third \spro\ peak (Pb).

The AGB models in the FRUITY database
are computed at several metallicities
that bracket the metallicities of the clusters of interest.
The differing neutron fluxes,
ambient physical conditions,
and subsequent AGB evolution 
affect the \spro\ yields
to produce a substantial dependence on 
the mass of the AGB progenitor.
We overlay these predictions on the cluster
values shown in Figure~\ref{fruityplot}.
A few points are notable.
First, the \hsls\ index in \ngc\ is higher by $\approx$~0.6~dex
than the next-highest cluster,
although the \hsls\ indices in the other four clusters
span a range of $\approx$~0.5~dex.
No single model from the FRUITY database can reproduce
the \hsls\ index in \ngc,
and the models would predict AGB masses ranging from
$\approx$~3 to 6~\msun\ for the other four clusters.
Second, the \pbhs\ index in \ngc\ is in very good agreement
with the other three clusters where Pb has been detected.
The models consistently imply AGB masses $\approx$~6~\msun\
based on the \pbhs\ index.
The models predict 
AGB masses of $\approx$~6~\msun\ produced the \spro\ material
in M4;
however, for all other clusters, no single mass model
can self-consistently account for the observed indices.

In principle, the timescale for star formation and
chemical enrichment in each cluster
matches the lifetime of the lowest-mass stars
known to contribute to the metals in the present-day stars.
A 6~\msun\ star has a lifetime of $\approx$~60~Myr,
while a 3~\msun\ star has a lifetime of $\approx$~300~Myr.
Roederer et al.\ (\citeyear{roederer11c}) 
examined tables of AGB nucleosynthesis yields and
concluded that AGB stars with $M >$~3~\msun\
were responsible for producing the \spro\ material in M22,
implying a formation timescale of $\approx$~300~Myr or less.
\citet{straniero14} re-examined the M22 abundances
derived by Roederer et al.\
using a self-consistent approach to the AGB models,
integrating the yields of all stars above a certain mass
and weighting by the initial mass function.
They confirmed the suspicions of Roederer et al.,
deriving a timescale of 
144~$\pm$~49~Myr for M22,
requiring the yields from AGB stars with masses $>$~4~\msun\ or so.
\citet{shingles14} adopted a different set of models
and found a minimum enrichment timescale of 
$\approx$~300~$\pm$~60~Myr,
with slightly lower-mass AGB stars contributing.
The difference between the two results
can be attributed to the prescription for
the formation of the $^{13}$C pocket in AGB
stars with 3~$< M/\msun <$~4.5.
Both timescales are consistent with the
maximum age spread allowed ($\sim$~300~Myr)
by the split subgiant branch in M22 \citep{marino12}.

The weak \spro, which operates in massive
($\sim$~25~\msun),
rapidly-rotating, low-metallicity stars,
is another proposed mechanism for producing \spro\ material
in the early universe
(e.g., \citealt{the00,pignatari08}).
These rapidly-evolving stars leave little time for
the \spro\ to flow to the highest-mass nuclei.
Using Figure~1 of \citet{frischknecht12}, we estimate
that \hsls\ can take a wide range of values, but 
\pbhs~$< -$0.6 in all of their models.
The \pbhs\ index is slightly super-solar
in the four clusters shown in Figure~\ref{fruityplot},
so this is not a likely candidate for the \spro\ material
found in these clusters.

Detailed comparisons with AGB models, 
like the work of Straniero et al.\ (\citeyear{straniero14})
or \citet{shingles14},
are beyond the
scope of the present study.
The similarity of the \pbhs\ ratios between \ngc\ and M22
argues for a relatively short (few hundred Myr or less)
enrichment timescale for \ngc.
We have no reason to discount the measurements that
form the \hsls\ index in \ngc, but its value
is quite different from that found in the other four clusters
and predicted using low-metallicity AGB models.
We recommend the issue be revisited 
should more stars in the $r+s$ group in \ngc\ be found.

\subsection{The $^{232}$Th Nuclear Chronometer in NGC~5824}
\label{age}

We derive an upper limit from the non-detection of the
Th~\textsc{ii} line at 4019.13~\AA\ in the MIKE spectrum of
star 42009955.  
The $^{232}$Th isotope is radioactive, and it can only
be produced by \rpro\ nucleosynthesis.
We calculate an age of the \rpro\ material in \ngc\
by comparing the current ratio of $^{232}$Th and a stable
element produced by the \rpro\ (Eu)
with their initial production ratio (Table~9, \citealt{roederer09}).
The current ratio, $\log\epsilon$(Th/Eu)~$< -$0.34~$\pm$~0.08,
implies an age greater than 0.3~Gyr.
This is not a useful age constraint.

Alternatively, if we assume a relatively old age for \ngc\
(e.g., \citealt{carretta10a}),
we can use the $\log\epsilon$(Th/Eu) ratio
to exclude the
presence of a rare, poorly-understood 
aspect of \rpro\ nucleosynthesis, the so-called
``actinide boost'' \citep{hill02,schatz02}.
This phenomenon is characterized by enhanced Th abundances,
and it is found in a handful of low-metallicity field stars.
If \ngc\ is 13~Gyr old, the radioactive decay from
Th produced in an actinide boost would 
yield a current ratio of 
$\log\epsilon$(Th/Eu)~$-$0.17~$\pm$~0.07,
which is significantly higher than our upper limit.
Younger ages for \ngc\ would imply even higher 
$\log\epsilon$(Th/Eu) ratios, in sharper
tension with our observed limit.

\citet{roederer15} summarized the
Th abundances or upper limits reported for
stars in six clusters.
To this list we add \ngc\ 
(and M75, \citealt{kacharov13}, 
which was inadvertently omitted).
None of these clusters, including \ngc,
show evidence of an actinide boost,
hinting that it may not occur
or its signature is diluted beyond recognition
in globular cluster environments.
This observation
may help in the hunt to identify or exclude candidate
sites for the actinide boost phenomenon.

\section{Summary}
\label{summary}

We have examined the first sets of high-resolution
spectroscopic observations obtained for the luminous,
metal-poor globular cluster \ngc.
Fifty stars have been observed using the M2FS spectrograph at Magellan,
and 26 of them have S/N sufficient to perform
a detailed abundance analysis of 20~species of 17~elements.
Two stars were re-observed using the MIKE spectrograph
to achieve broader wavelength coverage.

We derive $\langle$[Fe/H]$\rangle = -$1.94~$\pm$~0.02 (statistical)
$\pm$~0.10 (systematic).
We exclude the presence of 
an intrinsic metallicity spread 
among these 26~stars
at the 0.08~dex level.
Previous work by \citet{dacosta14} 
did not detect a metallicity spread among these
26~stars, either, and only detected 
a possible spread among the fainter stars in their sample.
Other elements in the Fe group show no significant star-to-star
dispersion, and their ratios are well-matched to
those in field stars with similar stellar parameters and metallicities.

The [Mg/Fe] ratios in \ngc\ have a large 
star-to-star dispersion, 0.28~dex,
which is 
only found among some luminous,
metal-poor clusters
like \mbox{$\omega$~Cen},
M3, M13, M15, \mbox{NGC~2419},
\mbox{NGC~2808}, \mbox{NGC~4833}, and \mbox{NGC~6752}.
Our limited data on O, Na, and Al do not permit us to 
examine the details of the light-element correlations
and anti-correlations produced by $p$-capture reactions.
However, these elements vary with each other, and Mg, 
in the usual manner
in the two stars observed with MIKE.

The \ncap\ abundance patterns in 25 of the 26~stars
observed with M2FS are effectively identical
and consistent with \rpro\ nucleosynthesis
patterns found in field stars with 
similar stellar parameters and metallicities.
One star shows a significant enhancement of \spro\ material,
as well.
We consider, and dismiss, the possibilities 
that this star self-enriched or obtained the
\spro\ material from a companion that 
passed through the AGB phase of evolution.
The \spro\ pattern resembles that found in
groups of stars in the other low-metallicity clusters
M2, M22, and \mbox{NGC~5286}.
Mounting evidence suggests that
intermediate-mass AGB stars ($\approx$~3--6~\msun)
were responsible for producing the
\spro\ material during the time of star formation
in these clusters.
The percentage of \spro-enhanced stars in \ngc\
is within the range of that found in other
clusters in this peculiar class.

We close by noting that several other
low-metallicity clusters
show evidence for internal
dispersion among the \ncap\ elements.
M2, M22, and \mbox{NGC~5286} have been discussed
extensively in Section~\ref{sprocess}.
Heavy element dispersion
in clusters $\omega$~Cen \citep{smith00,dorazi11},
M15 (e.g., \citealt{sneden97,otsuki06,worley13}), and
\mbox{NGC~1851} \citep{yong08c,villanova10,carretta11}
has been confirmed by multiple investigators.
Compelling evidence for variations in M19 
has been presented recently by \citet{johnson15b}.
Other clusters with more subtle variations or less secure evidence
include
\mbox{47~Tuc} \citep{cordero15},
M75 \citep{kacharov13},
M80 \citep{carretta15},
\mbox{NGC~362} \citep{carretta13a},
\mbox{NGC~4372} \citep{sanroman15},
and
\mbox{NGC~5897} \citep{koch14}.
We have not included these clusters in the discussion here
due to their enormous complexity,
small sample sizes, 
or small numbers of \ncap\ elements studied.
Spectroscopic followup observations
of the \ncap\ elements in these clusters
would be greatly welcomed.


\section*{Acknowledgments}

I.U.R.\ thanks
G.\ Da Costa for thoughtful, encouraging discussions
throughout the course of this study;
D.\ Kelson and E.\ Villanueva for their assistance
installing and maintaining the CarPy MIKE reduction pipeline;
A.\ Koch for checking the list of clusters with
possible \ncap\ variations;
E.\ Olszewski, I.\ Thompson, and M.\ Walker
for helping to make M2FS a reality; 
and
V.\ Placco for sharing insights on CEMP star binaries.
We appreciate our referee's thoughtful comments 
that have substantially improved this manuscript.
This research has made use of NASA's 
Astrophysics Data System Bibliographic Services, 
the arXiv preprint server operated by Cornell University, 
the SIMBAD and VizieR databases hosted by the
Strasbourg Astronomical Data Center,  
the Atomic Spectra Database \citep{kramida14} hosted by
the National Institute of Standards and Technology, and
the \textsc{r} suite of software \citep{rsoftware}.
\textsc{iraf} is distributed by the National Optical Astronomy Observatories,
which are operated by the Association of Universities for Research
in Astronomy, Inc., under cooperative agreement with the National
Science Foundation.
M.M.\ and J.I.B.\ gratefully acknowledge support
from the U.S.\ National Science Foundation to develop M2FS
(AST-0923160).

\label{lastpage}

\end{document}